\documentclass[preprint,aps,nofootinbib]{revtex4}

\usepackage{dcolumn}
\usepackage{bm}
\usepackage{graphicx,subfigure}
\usepackage[bookmarks=true]{hyperref}

\usepackage{amsmath}
\usepackage{amsfonts}
\usepackage{amssymb}
\usepackage{slashed}
\usepackage{hyperref}
\usepackage{cprotect}
\usepackage{color}

\def\be{\begin{equation}}
\def\ee{\end{equation}}
\def\bea{\begin{eqnarray}}
\def\eea{\end{eqnarray}}

\renewcommand\Im{\operatorname{Im}}
\DeclareMathOperator{\arcsinh}{arcsinh}

\def\ss2l{SS2$\ell$}
\def\3l{3$\ell$}

\def\slashchar#1{\setbox0=\hbox{$#1$}           
   \dimen0=\wd0                                 
   \setbox1=\hbox{/} \dimen1=\wd1               
   \ifdim\dimen0>\dimen1                        
      \rlap{\hbox to \dimen0{\hfil/\hfil}}      
      #1                                        
   \else                                        
      \rlap{\hbox to \dimen1{\hfil$#1$\hfil}}   
      /                                         
   \fi}

\begin{document}
\title{What's in the Loop? \\
The Anatomy of Double Higgs Production}
\vspace*{1cm}

\author{\vspace{1cm} S.~Dawson$^{\, a}$, A.~Ismail$^{\, b,c}$ and Ian Low$^{\, b,d}$ }

\affiliation{
\vspace*{.5cm}
  \mbox{$^a$ Physics Department, Brookhaven National Laboratory, Upton, NY 11973}\\
 \mbox{$^b$ High Energy Physics Division, Argonne National Laboratory, Argonne, IL 60439}\\
\mbox{$^c$ Department of Physics, University of Illinois, Chicago, IL 60607}\\
\mbox{$^d$ Department of Physics and Astronomy, Northwestern University, Evanston, IL 60208} \\
\vspace*{1cm}}

\begin{abstract}
\vspace*{0.5cm}
Determination of Higgs self-interactions through the double Higgs production from gluon fusion is a major goal of current and future collider experiments.   We point out  this channel could help disentangle and resolve the nature of  ultraviolet contributions to Higgs couplings to two gluons. Analytic properties of the double Higgs amplitudes near kinematic threshold are used to study features resulting from scalar and fermionic loop particles mediating the interaction. Focusing on the $hh$ invariant mass spectrum, we consider the effect from anomalous top and bottom Yukawa couplings, as well as from scalar and fermionic loop particles. In particular, the spectrum at high $hh$ invariant mass is sensitive to the spin of the particles in the loop.  
\end{abstract}

\maketitle

\section{Introduction}

Now that the Higgs boson has been discovered at $m_h=125$ GeV, the next important task is
a detailed exploration of the Higgs properties. The measured Higgs boson production rates and the extracted
values of the Higgs couplings are close to the Standard Model (SM) predictions, but at the ${\cal O}(10-20)\%$ level, there
is room for new physics effects in the Higgs sector.  
The structure of the Higgs potential is completely determined in the SM and measuring  
the Higgs self-interactions is an important step in determining if the observed boson is 
identical to the Higgs boson predicted by the SM.
The Higgs self-interactions are most directly probed by double Higgs ($2h$) production at the Large Hadron Collider (LHC), $gg\rightarrow hh$,
which has a very small rate,
$\sigma\sim 34~\mathrm{fb}$ at $\sqrt{S}=13$ TeV \cite{Plehn:1996wb,Glover:1987nx,Dawson:1998py,deFlorian:2013jea,Frederix:2014hta,Maltoni:2014eza,Grigo:2014jma}, making this measurement only feasible at high luminosity \cite{Moretti:2004wa,Binoth:2006ym,Grober:2010yv,Shao:2013bz,Goertz:2013kp,Barr:2013tda,Barger:2013jfa,deLima:2014dta,Wardrope:2014kya,Papaefstathiou:2015iba,Dolan:2012rv}.  In the SM the cross section receives contributions from 
both box and triangle diagrams, and the large cancellation between the diagrams at threshold makes
the $gg\rightarrow hh$ process particularly sensitive to new physics contributions \cite{Arhrib:2009hc,Baglio:2012np,Dolan:2012ac,Cao:2013si,Nhung:2013lpa,Han:2013sga,Haba:2013xla,Slawinska:2014vpa,Goertz:2014qta,Chen:2014xra,Li:2015yia,Dicus:2015yva}. 

Beyond-the-SM  (BSM) physics can contribute to $2h$ production in a variety of ways, including anomalous $ t {\overline t} h$ ($b {\overline b}h$) and $t {\overline t}hh$  $(b{\overline b}  hh)$ couplings \cite{Delaunay:2013iia,Nishiwaki:2013cma,Chen:2014xwa,Azatov:2015oxa,Contino:2012xk,Contino:2010mh,Gillioz:2012se,Liu:2014rba},
 resonant enhancements \cite{Christensen:2013dra,Gouzevitch:2013qca,Liu:2013woa,No:2013wsa,Baglio:2014nea,Kumar:2014bca,Hespel:2014sla,Barger:2014taa,Chen:2014ask,vanBeekveld:2015tka}, exotic decays \cite{Ellwanger:2014hca,vanBeekveld:2015tka}
and  new colored scalar \cite{Belyaev:1999mx,BarrientosBendezu:2001di,Asakawa:2010xj,Kribs:2012kz,Enkhbat:2013oba,Heng:2013cya} or fermonic \cite{Liu:2004pv,Dib:2005re,Pierce:2006dh,Ma:2009kj,Han:2009zp,Dawson:2012mk,Edelhaeuser:2015zra} particles contributing to the loop amplitudes. Some of these effects, for example the modified $ t {\overline t} h$ couplings or new colored particles in the loop, also affect the single Higgs ($1h$) production in the 
$gg\to h$ channel. However, it is difficult to disentangle new physics effects in $1h$ production because of the limited number of kinematic observables in the final state. Using the higher-order process of $gg\to h+j$ may help with measuring the top Yukawa coupling \cite{Grojean:2013nya,Azatov:2013xha}, but is unlikely to resolve the nature  of the 
colored particle mediating the loop.  In this work we are interested in the question of whether $2h$ production is sensitive to the underlying ultraviolet source of new physics and can potentially differentiate between different sources of new physics. We will see that many of the aforementioned new physics effects can significantly change the rate as well as the kinematic distributions in $2h$ production. (See Refs. \cite{Asakawa:2010xj,Chen:2014xra} for previous studies of new physics effects in the $2h$ kinematic distributions.) In some cases, the changes are severely restricted by the (close to SM predicted)  measurements of $1h$ production.

This work is organized as follows. In Section \ref{basic}, we review the basics of $1h$ and $2h$ production
to set our notation.  One of our major new results is  in Section \ref{analytic}, where we discuss the analytic structure of the $2h$ amplitude near threshold
in the case where the new physics arises from heavy fermions or from heavy colored scalars in the loops.
  Section \ref{example}  contains numerical results for $2h$ production at $\sqrt{S}=13$ and $100$ TeV.  The amplitude for $2h$ production
  from intermediate colored scalar loops is reviewed in an appendix.

\section{Basics of $1h$ and $2h$ cross-sections}
\label{basic}
In this section, we review the lowest order results for $1h$ and $2h$ production from gluon fusion  in order to fix our notation. 
We begin by presenting an effective Lagrangian,  and then consider the specific contributions from heavy fermions and heavy colored scalars.
\subsection{Non-SM Interactions}
We consider the following effective Lagrangian, where we are only interested in new physics affecting Higgs rates, assume
SM kinetic energy terms ($L_{KE}$),
and assume no light particles other than those of the SM.  Including only third generation fermions, 
\begin{eqnarray}
L_{eff}
 &=&L_{KE}
 -\biggl(1+\delta_t\biggr)
 {m_t\over v}{\overline t} t h
+c_{2h}^{(t)}{\overline t}t h^2
-\biggl(1+\delta_b\biggr){m_b\over v}{\overline b} b h
+c_{2h}^{(b)}{\overline b} b h^2
\nonumber \\
&&
-\biggl(1+\delta_3\biggr) {m_h^2\over 2 v} h^3 -
\biggl(1+\delta_4\biggr) {m_h^2\over 8 v^2}  h^4
+{c_g\alpha_s\over 12 \pi v}G^{A,\mu\nu}G_A^{\mu\nu}h
-{c_{gg}\alpha_s\over 24 \pi v^2}G^{A,\mu\nu}G_A^{\mu\nu}h^2
 \, ,
\label{effective}
\end{eqnarray}
where  in the SM,  $\delta_t=\delta_b=c_g=c_{gg}=c_{2h}^{(t)}=c_{2h}^{(b)}=\delta_3=\delta_4=0$.  Global fits to Higgs production rates at the LHC
limit the deviations of $\delta_t$ and $c_{g}$ from $0$ in a correlated fashion, as described below in Eq. \ref{singans}.  
Deviations of  the $b$-Yukawa coupling from the SM prediction,  $\delta_b$, are less constrained \cite{atlasfits,Khachatryan:2014jba}.  

\subsection{Colored scalars}
The contributions from colored scalars
depend on the parameters of the scalar potential. 
We use the following Lagrangian for an $SU(2)_L$ singlet,  $SU(3)_c$ complex scalar, $s$,
\begin{eqnarray}
{\cal L}_{s,c}&=&(D_\mu s)^*(D^\mu s)-m_0^2 s^*s -{\lambda_s\over 2} (s^*s)^2-\kappa s^*s \left|H^\dagger H\right|
\\ &\rightarrow &
(D_\mu s)^*(D^\mu s)-m_0^2 s^*s -{\lambda_s\over 2}(s^*s)^2
-\kappa s^*s \biggl|
{(h+v)\over \sqrt{2}}\biggr|^2\, ,
\label{eq:lsc}
\end{eqnarray}
where $H$ is the SM $SU(2)_L$ doublet with $\langle H\rangle =(0, v/\sqrt{2})^T$. In this normalization,
 the Fermi constant $G_F=1/(\sqrt{2}v^2)$ and $v\approx 246$ GeV.
If the scalar, $s$, is real,
\begin{eqnarray}
{\cal L}_{s,r}&=&{1\over 2} (D_\mu s)(D^\mu s)
-{m_0^2\over 2}s^2 -{\lambda_s\over 4} s^4
-{\kappa\over 2}s^2 \left|H^\dagger H\right|\, .
\end{eqnarray}
The physical mass for either a real or complex scalar is, 
\begin{equation}
m_s^2=m_0^2+{\kappa v^2\over 2}\, ,
\end{equation}
where  $m_0=0$ is the limit where the scalar gets all of its mass from electroweak symmetry breaking.  
The cubic and quartic scalar couplings are,
\begin{eqnarray}
L&\sim & -g_{h2 s} s^*s h-{g_{2h2 s}\over 2}s^* s h^2\nonumber \\
g_{h2s} & =& \kappa v \ , \qquad g_{2h2s}= \kappa\ ,
\end{eqnarray}
and similarly for a real scalar.

\subsection{$1h$ Production from Scalars and Fermions}

The leading-order (LO) $gg\rightarrow 1h$ production rates due to virtual scalars and fermions are well-known. It is convenient to introduce the loop functions
\begin{eqnarray}
\label{eq:f12}
F_{1/2}(\tau)&=&-2\tau[1+(1-\tau)f(\tau)] \, \\
\label{eq:f0}
F_0(\tau)&=& \tau[1-\tau f(\tau)]\ ,
\end{eqnarray}
where $\tau_i=4m_i^2/m_h^2$  and  
\begin{eqnarray}
f(\tau)&=& \biggl[
\sin^{-1}
(1/\sqrt{\tau})
\biggr]^2
\qquad \rm{if} ~\tau\ge 1
\nonumber \\
&=& -{1\over 4}
\biggl[ \ln\biggl(
{1+\sqrt{1-\tau}
\over 1-\sqrt{1-\tau)}}
\biggr)-i\pi\biggr]^2\qquad {\rm if} ~\tau < 1 \ .
\end{eqnarray}
Then including colored scalars and non-SM fermion interactions as defined in the previous subsections, 
\bea
\label{eq:sigmalo}
\sigma^{\rm (LO)} (gg\to h)&=& \frac{\alpha_s^2}{1024\pi} \left| \sum_{f_i} 
T(f_i){2 (1+\delta_{f_i})\over v}  F_{1/2}(\tau_{f_i}) +c_g \biggl(-{4\over 3v} \biggr)\right.
\nonumber \\ && \left.
+ \sum_{s_i} \delta_R T(s_i) \frac{g_{h2s_i}v}{m_{s_i}^2} F_{0}(\tau_{s_i})\right|^2\ ,
\eea 
where $T(\cdot)$ is the Dynkin index for the corresponding representation under $SU(3)_c$ defined as $Tr(T^AT^B)=\delta_{AB}T(\cdot)$, and 
$\delta_R=1/2$ for real scalars  and 1 for complex scalars. The Dynkin index is $1/2$ for fundamental representations and $3$ for 
adjoint representations, respectively. For SM  fermions, $T(f)=1/2$.

Neglecting the $b$-quark  contribution and noting that $F_{1/2}$ and $F_0$ are well approximated by their large mass limits, 
 $F_{1/2}(\tau_t\rightarrow \infty)=-{4\over 3}$  and $F_0(\tau_s\rightarrow \infty) \rightarrow -{1\over 3}$,
\begin{eqnarray}
R_h 
&\equiv & 
{\sigma(gg\rightarrow h)\over \sigma(gg\rightarrow h)\mid_{SM}}
\nonumber \\
&\rightarrow & 
1+2\biggl(
\delta_t+c_g+\sum_{s_i} \delta_R T(s_i)
{g_{h2s_i}v\over 4 m_{s_i}^2}
\biggr) \, .
\label{singans}
\end{eqnarray}
Eq. \ref{singans} is the well-known result that $1h$ production has little discriminating power between $\delta_t$
and $c_g$ \cite{Banfi:2013yoa,Dawson:2014ora,Azatov:2014jga,Azatov:2013hya}.  The coefficients of $c_g$ and $c_{gg}$
from heavy colored scalars can be found in Refs.~\cite{Dawson:2015gka,Bonciani:2007ex,Gori:2013mia}.
In the SM,   $gg\rightarrow 1h$ production receives significant QCD corrections beyond LO QCD. The NNLO contributions  from arbitrary fermions \cite{Anastasiou:2010bt}
 and scalars \cite{Boughezal:2011mh,Boughezal:2010ry} are significant.  However, since we are typically concerned with ratios relative to the SM, we work at leading order.

\subsection{$2h$ Production for Fermions and Scalars}

\begin{figure}[t]
\includegraphics[height=0.5in]{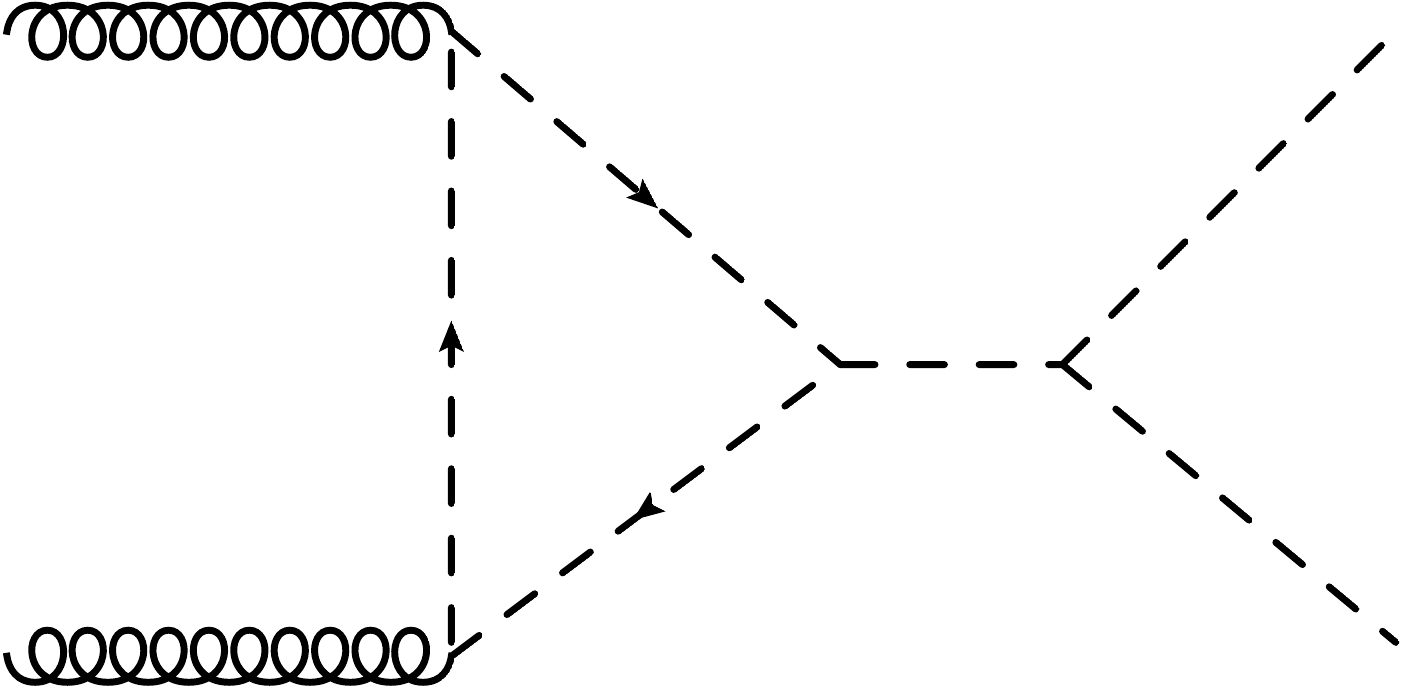} \hspace{0.25in}
\includegraphics[height=0.5in]{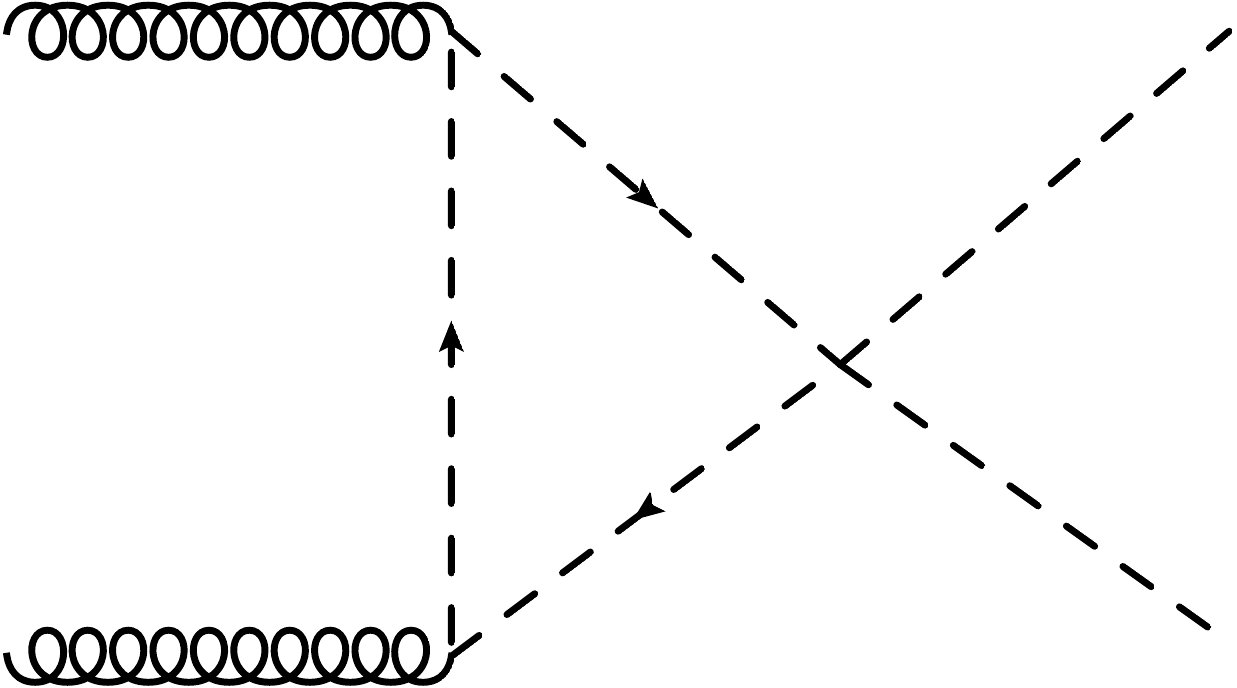} \hspace{0.25in}
\includegraphics[height=0.5in]{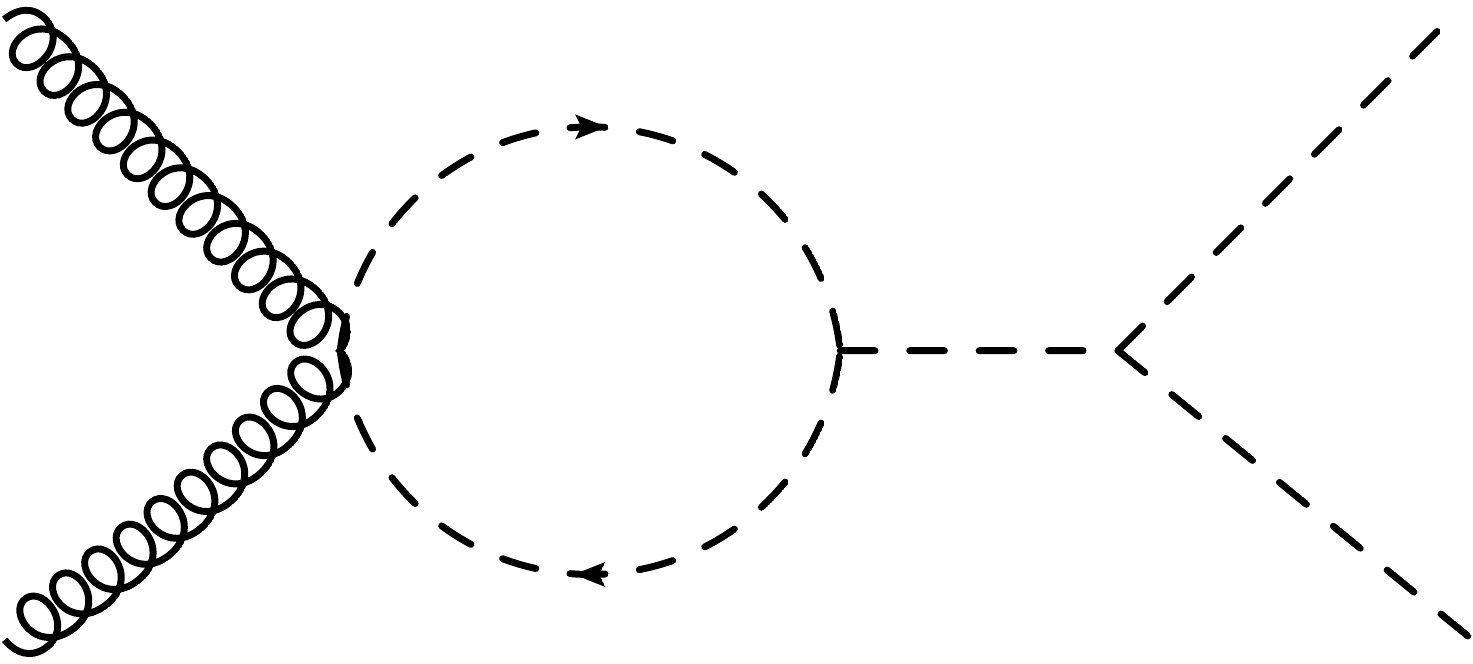} \hspace{0.25in}
\includegraphics[height=0.5in]{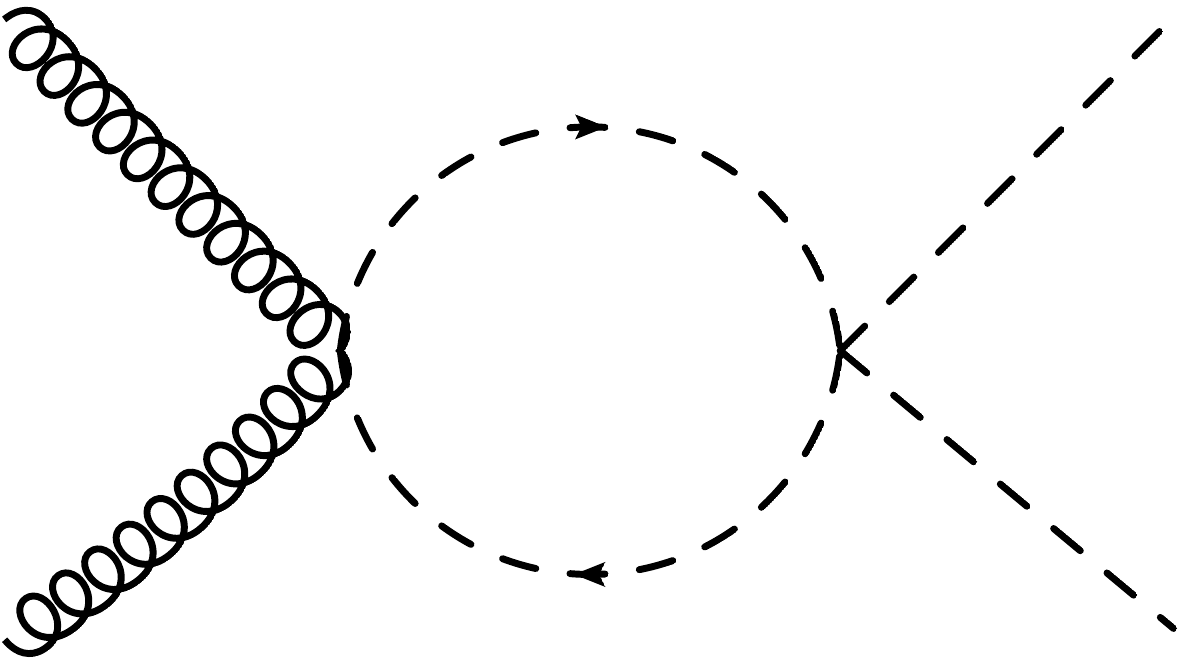} \\[0.25in]
\includegraphics[height=0.5in]{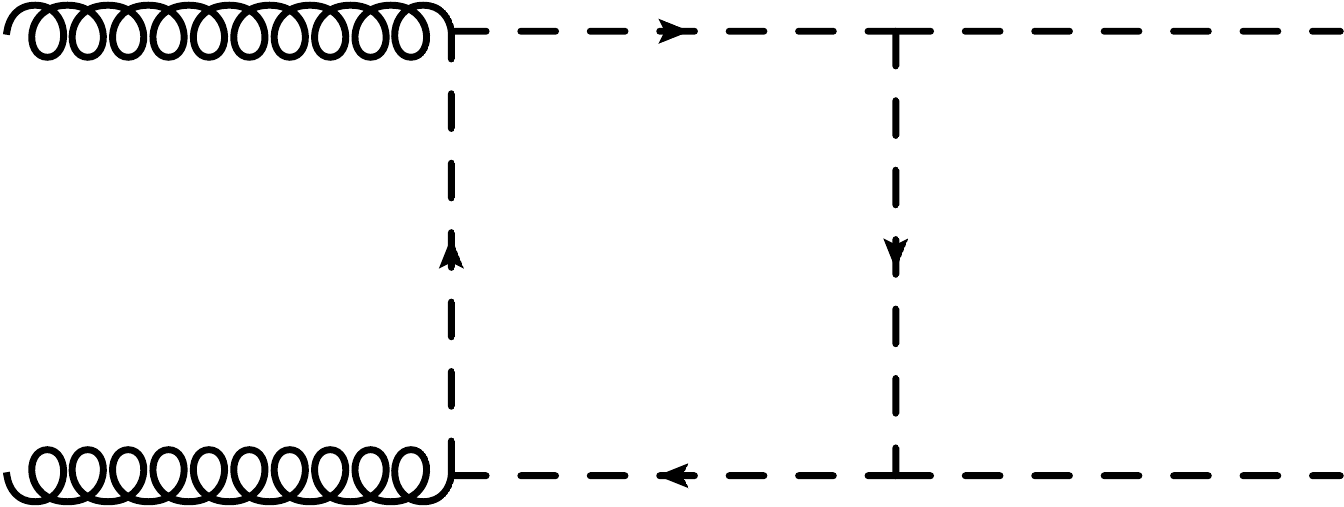} \hspace{0.25in}
\includegraphics[height=0.5in]{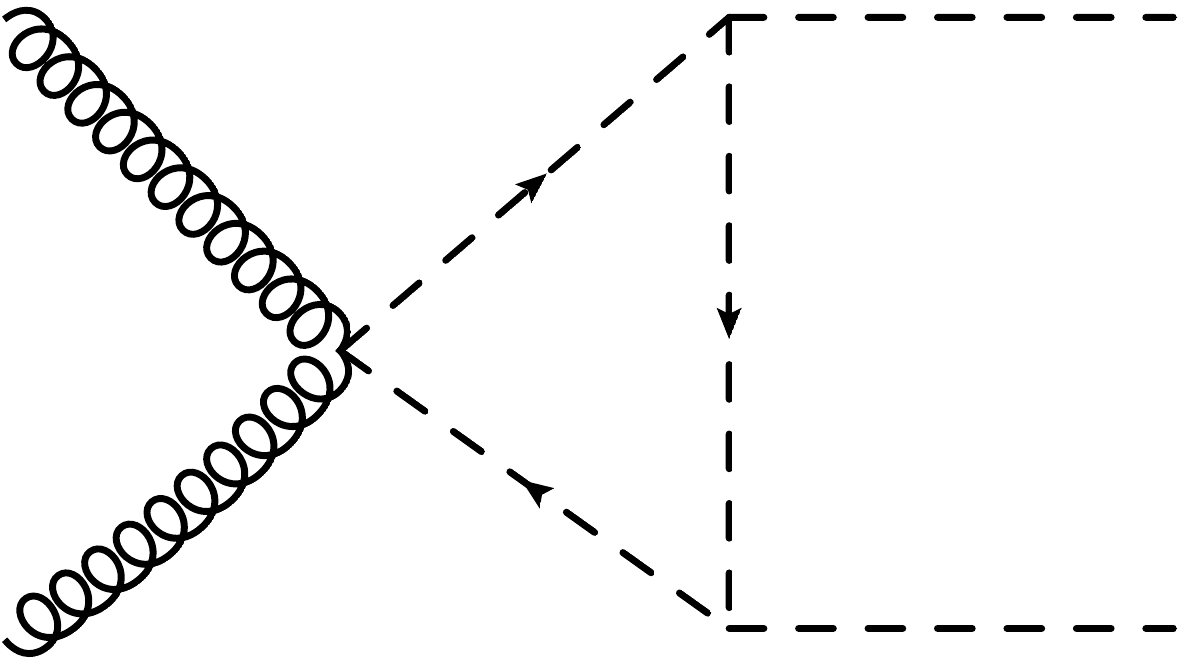}
\caption{\em Top row: Triangle diagrams for $g g \to h h$. Bubble diagrams with quartic scalar-gluon vertices are included
with the triangle diagrams in our results.  Bottom row: Box diagrams for $g g \to h h$. Triangle diagrams with quartic scalar-gluon vertices are included with the box diagrams.}
\label{fg:scdiag}
\end{figure}

The LO $gg\rightarrow 2h$ production rates from fermion and scalar
loops can be found in Refs.~\cite{Glover:1987nx,Plehn:1996wb} and Refs.~\cite{Asakawa:2010xj, Kribs:2012kz}, respectively. The LO partonic cross-section for $g(p_1) g(p_2)\to h(k_1) h(k_2)$ is given by
\be
\sigma^{\rm (LO)} (gg\to 2h) = \int  dt\ \frac1{2^2} \frac1{8^2} \frac1{2!} \frac1{16\pi {\hat s}^2} \left|{\cal M}\right|^2  \ ,
\ee
where ${\hat s}=(p_1+p_2)^2,~{{\hat t}} = (p_1 - k_1)^2$. In the above $1/2^2$ comes from averaging over the initial gluon helicities, $1/8^2$ from color averaging and $1/2!$ from 
the identical final state particles. The amplitude-squared can be written as
\begin{eqnarray}
\label{eq:ampgghh}
\left| {\cal M}\right|^2 &=& (N_c^2-1)\frac{\alpha_s^2}{8\pi^2}\frac{{\hat s}^2}{v^4}  
\left[ \left|   {3m_h^2\over {\hat s}-m_h^2} (1+\delta_3)
\biggl\{\sum_{f_i}
(1+\delta_{f_i}) F_\bigtriangleup({\hat s},{\hat t},m_h^2, m_{f_i}^2)
+{2\over 3} c_g\biggr\} 
\right.\right.\nonumber \\ && \left.\left.
+\biggl\{1+{3m_h^2 (1+\delta_3) \over {\hat s-m_h^2}}\biggr\}\sum_{s_i} \delta_R T(s_i) \frac{g_{h2s_i}v}{m_{s_i}^2} F_\bigtriangleup({\hat s},{\hat t},m_h^2, m_{s_i}^2)
\right . \right .\nonumber \\  &&
 \left . \left . -2v^2 \sum_{f_i}{c_{2h}^{(f_i)}\over m_{f_i}}
 F_\bigtriangleup({\hat s},{\hat t},m_h^2, m_{f_i}^2)
 + \sum_{f_i}(1+\delta_{ f_i})^2 F_\Box({\hat s}, {\hat t}, m_h^2, m_{f_i}^2)
 \right.\right. \nonumber \\ &&
 \left .
 +\sum_{s_i} \delta_R T(s_i) {g_{h2 s_i}^2v^2\over  m_{s_i}^4} F_\Box({\hat s}, {\hat t}, m_h^2, m_{s_i}^2)
 -{2c_{gg}\over 3}   \right|^2
 \nonumber \\ && \left. 
 + \left|
 \sum_{f_i} (1+\delta_{f_i})^2 
 G_\Box({\hat s},{\hat t}, m_h^2, m_i^2)
 +\sum_{s_i} \delta_R T(s_i)
  {g_{h2s_i}^2 v^2 \over  m_{s_i}^4} G_\Box({\hat s}, {\hat t}, m_h^2, m_{s_i}^2)
 \right|^2\right]  ,
 \label{eq:hhdif}
\end{eqnarray} 
where $N_c^2-1$ comes from summing over the gluon color index. The form factors $F_\bigtriangleup({\hat s},{\hat t}, m_h^2, m_i^2)$,
 $F_\Box({\hat s},{\hat t}, m_h^2, m_i^2)$, and $G_\Box{(\hat s},{\hat t}, m_h^2, m_i^2)$ resulting from the SM 
 top quark are given in the appendix of Ref.~\cite{Plehn:1996wb}\footnote{ In the SM, including only the top quark contribution Eq. \ref{eq:hhdif}
differs from Eq. 13 in Ref.~\cite{Plehn:1996wb} as well as Eq. 4 in Ref.~\cite{Baglio:2012np}, but agrees with Eq. 6 in Ref.~\cite{Glover:1987nx},  Eq. 5 in Ref.~\cite{Dawson:2012mk} and Eq. 4 in \cite{Chen:2014xra}, after plugging in $G_F =1/(\sqrt{2} v^2)$ and taking account  differences in the normalization of 
the form factors employed.}. 
The Feynman diagrams for the case of scalar particles are shown in Fig.~\ref{fg:scdiag}, and the corresponding form factors are given in Appendix \ref{appendix:2hamp}\footnote{
We disagree with the overall normalization of the corresponding expressions in Refs. \cite{Asakawa:2010xj,Kribs:2012kz}. In addition, the first $2$ arguments of the last $D$ function in Eq. 16 of Ref. \cite{Kribs:2012kz} should be swapped.}. 
 In the large mass limits,
 \begin{eqnarray}
 F_\bigtriangleup({\hat s},{\hat t},m_h^2, m_{f_i}^2)&\rightarrow &{2\over 3}\rightarrow 
 -{1\over 2}F_{1/2}(\tau_f\rightarrow\infty)
 \nonumber \\
 F_\bigtriangleup({\hat s},{\hat t},m_h^2, m_{s_i}^2)&\rightarrow &{1\over 6}
 \rightarrow -{1\over 2}F_{0}(\tau_s\rightarrow\infty)
 \nonumber \\
 F_\Box({\hat s},{\hat t},m_h^2, m_{f_i}^2)&\rightarrow & -{2\over 3}
 \rightarrow {1\over 2}F_{1/2}(\tau_f\rightarrow\infty)
 \nonumber \\
 F_\Box({\hat s},{\hat t},m_h^2, m_{s_i}^2)&\rightarrow & -{1\over 6}
\rightarrow {1\over 2}F_{0}(\tau_s\rightarrow\infty)
\nonumber \\
G_\Box({\hat {s}}, {\hat{t}}, m_h^2, m^2)&\rightarrow & {\cal O}\biggl({p_T^2\over m^2}\biggr)
 \, .
\end{eqnarray}
 
 In the 
large mass limit,    only the the spin-0 contribution survives,
\begin{eqnarray}
\label{eq:ampgghh}
\left| {\cal M}\right|^2 &\rightarrow & (N_c^2-1)\frac{\alpha_s^2}{18\pi^2}
\frac{{\hat s}^2}{v^4}  
 \left|   {3m_h^2\over {\hat s}-m_h^2}
  \biggl(
  (1+\delta_3)(1+\delta_t)+c_g
  \biggr)
-(1+\delta_t)^2-c_{gg}-
{2c_{2h}^{(t)}v^2\over m_t}
\right.
\nonumber \\ && \left.
+\sum_{s_i} \delta_R T(s_i) \frac{g_{h2s_i}v}{4m_{s_i}^2}
\biggl(1+{3m_h^2\over {\hat s}-m_h^2}(1+\delta_3)-{g_{h2s_i} v\over m_{s_i}^2}\biggr)
   \right|^2\, .
   \label{largemass}
 \end{eqnarray}
 The contributions from the anomalous couplings in Eq. \ref{largemass} are consistent with those in Refs.
 \cite{Gillioz:2012se,Delaunay:2013iia,Chen:2014xwa,Chen:2014xra}.
 Furthermore, the  contributions to Eq. \ref{largemass} which come from the triangle diagrams
are related to the fermionic and scalar contributions to the 1-loop QCD $\beta$ function via the Higgs low-energy theorems \cite{Shifman:1979eb}, which can be used to systematically compute higher order QCD corrections to the triangle loops \cite{Dawson:1996xz,Gori:2013mia}.

\section{Analytic Structure}
\label{analytic}
A closed-form analytic expression for the $2h$ production amplitude at threshold, $\hat{s} = 4m_h^2$, may be obtained from the imaginary
part of the amplitude, combined with a knowledge of the amplitude's limiting behavior as the particle masses in the loops go to either zero or infinity \cite{Li:2013rra}.
Alternatively, the threshold result can be obtained by a direct expansion of the full amplitude. For a colored fermion of mass $m_f$ running in the loops, the separate components of the amplitude arising from the triangle and box diagrams are, at threshold,
\bea
F_\triangle^{(f)}\mid_{th} &\equiv& F_\triangle^{(f)}(\hat{s} = 4m_h^2, \hat{t} = -m_h^2, m_h^2, m_f^2) \nonumber \\
&=& \frac{1}{2} T(f) \tau_f \left( 1 + \left( 1 - \frac{\tau_f}{4} \right) \arcsin^2 \left( \frac{2}{\sqrt{\tau_f}} \right) \right) \nonumber \\
F_\Box^{(f)}\mid_{th} &\equiv& F_\Box^{(f)}(\hat{s} = 4m_h^2, \hat{t} = -m_h^2, m_h^2, m_f^2) \nonumber \\
&=& -\frac{1}{2} T(f) \tau_f \Bigg( -1 + \tau_f \left( 1 - \frac{\tau_f}{4} \right) \arcsin^2 \left(\frac{2}{\sqrt{\tau_f}}\right) + \\
&&\ \left( \frac{\tau_f}{2} - 1 \right) (\tau_f + 1) \arcsin^2 \left(\frac{1}{\sqrt{\tau_f}}\right) + (\tau_f - 3) \sqrt{\tau_f - 1} \arcsin \left(\frac{1}{\sqrt{\tau_f}}\right) + \nonumber \\
&&\ \left( \frac{\tau_f}{2} - 1 \right) (\tau_f + 1) \arcsinh^2 \left(\frac{1}{\sqrt{\tau_f}}\right) - (\tau_f - 3) \sqrt{\tau_f + 1} \arcsinh \left(\frac{1}{\sqrt{\tau_f}}\right) \Bigg) 
\, ,
\nonumber
\label{eq:ferthr}
\eea
where $\tau_f = 4 m_f^2 / m_h^2$ and $T(f)$ is  again the Dynkin index of the $SU(3)$ representation of the fermion. The total amplitude is proportional to the sum of the two expressions above,
\bea
F^{(f)}\mid_{th} &=& F_\triangle^{(f)}\mid_{th} + F_\Box^{(f)}\mid_{th} \, .
\eea
In the heavy mass limit,
\bea
F^{(f)}\mid_{th}
&\approx& -\frac{14}{45} \tau_f^{-1} - \frac{8}{7} \tau_f^{-2} + {\cal O}(\tau_f^{-3})
\label{eq:ffth}
\eea
and agrees with the result of \cite{Li:2013rra}.

For a scalar of mass $m_s$ with $\kappa = \kappa_0 \equiv 2 m_s^2 / v^2$, the triangle and box amplitudes at threshold are found  by analytic
continuation of the imaginary contributions given in Appendix B, 
\bea
F_\triangle^{(s)}\mid_{th}^{\kappa = \kappa_0} &\equiv& F_\triangle^{(s)}(\hat{s} = 4m_h^2, \hat{t} = -m_h^2, m_h^2, m_f^2) \nonumber \\
&=& -\frac{1}{16} T(s) \tau_s \left( 4 - \tau_s \arcsin^2 \left( \frac{2}{\sqrt{\tau_s}} \right) \right) \nonumber \\
F_\Box^{(s)}\mid_{th}^{\kappa = \kappa_0} &\equiv& F_\Box^{(s)}(\hat{s} = 4m_h^2, \hat{t} = -m_h^2, m_h^2, m_f^2) \nonumber \\
&=& \frac{1}{16} T(s) \tau_s^2 \Bigg( -\frac{\tau_s}{2} \arcsin^2 \left(\frac{2}{\sqrt{\tau_s}}\right) + \\
&&\ (2 + \tau_s) \arcsin^2 \left(\frac{1}{\sqrt{\tau_s}}\right) + 2 \sqrt{\tau_s - 1} \arcsin \left(\frac{1}{\sqrt{\tau_s}}\right) + \nonumber \\
&&\ (2 + \tau_s) \arcsinh^2 \left(\frac{1}{\sqrt{\tau_s}}\right) - 2 \sqrt{\tau_s + 1} \arcsinh \left(\frac{1}{\sqrt{\tau_s}}\right) \Bigg) \, , \nonumber
\label{eq:scthr}
\eea
where $\tau_s = 4 m_s^2 / m_h^2$ and $T(s)$ is the Dynkin index of the scalar's $SU(3)$ representation. We have included bubble diagrams with quartic scalar-gluon couplings in $F_\triangle^{(s)}$ and triangle diagrams with such couplings in $F_\Box^{(s)}$.  
Just as for fermions, we may expand the total threshold amplitude as
\bea
F^{(s)}\mid_{th}^{\kappa = \kappa_0} &=& F^{(s)}_\triangle\mid_{th}^{\kappa = \kappa_0} + F^{(s)}_\Box\mid_{th}^{\kappa = \kappa_0} \, ,
\label{eq:fsth}
\eea
where the functions $F^{(s)}_\triangle\mid_{th}, F^{(s)}_\Box\mid_{th}$ are the threshold values of the form factors $F^{(s)}_\triangle, F^{(s)}_\Box$ in Eq. \ref{eq:ampgghh}. In the heavy mass limit, the threshold result is
\bea
F^{(s)}\mid_{th}^{\kappa = \kappa_0} &\approx& -\frac{4}{45} \tau_s^{-1} - \frac{8}{21} \tau_s^{-2} + {\cal O}(\tau_s^{-3})\, .
\eea

\begin{figure}[t]
\begin{centering}
\includegraphics[height=2in]{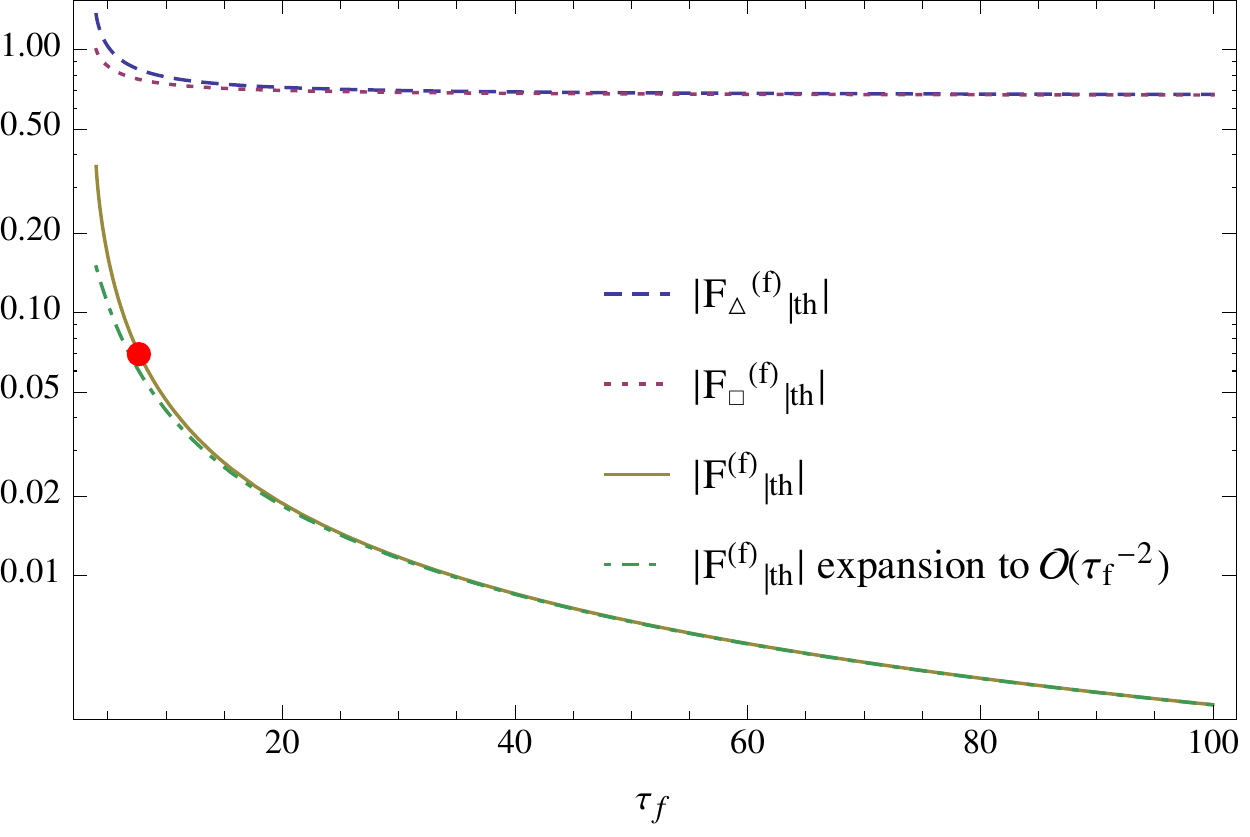}
\includegraphics[height=2in]{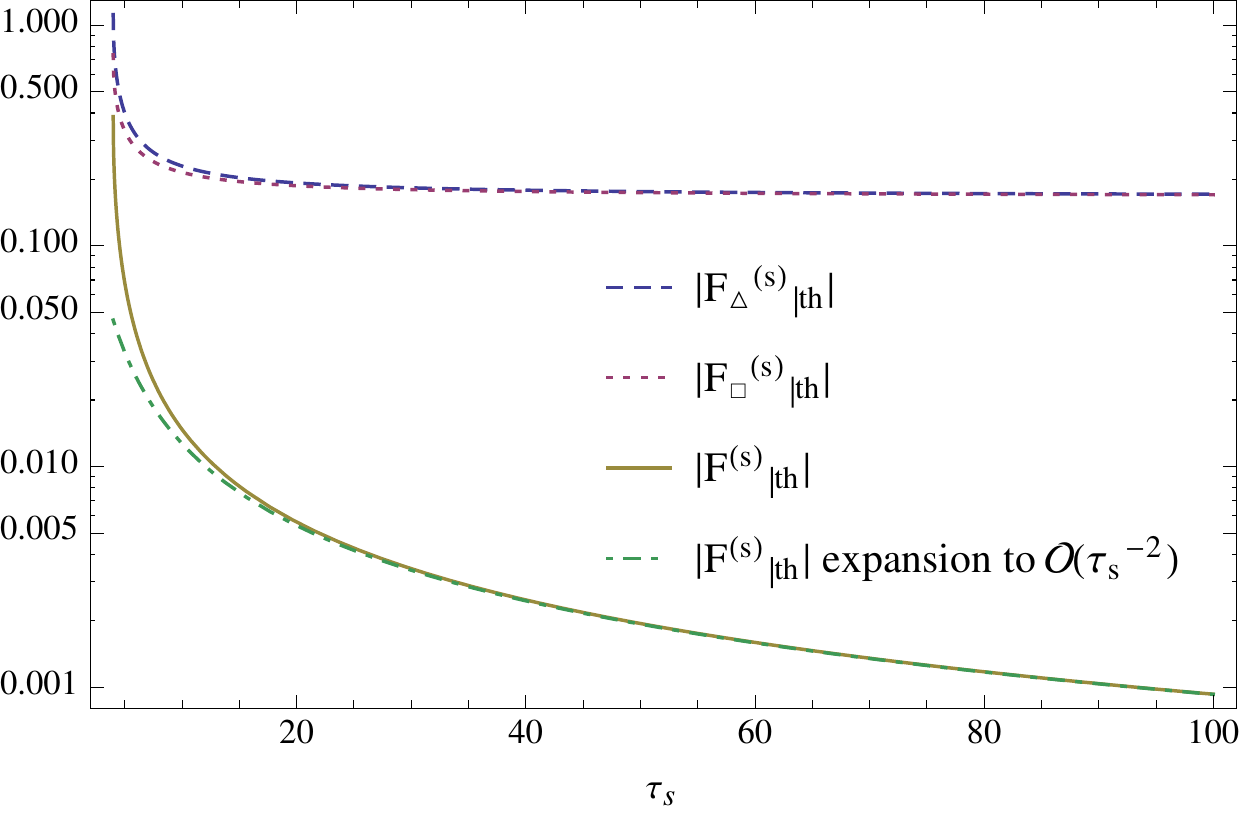}
\par\end{centering}
\caption{\em Cancellation between contributions to the $2h$ amplitude for fermions (left) or scalars (right) that get their mass entirely through couplings to the Higgs doublet. In the left panel, the value of $\tau_f$ corresponding to the top quark is indicated.
}
  \label{fig:cancel}
\end{figure}

The cancellations between the triangle and box functions for fermions and scalars are shown in Fig. \ref{fig:cancel}. In each panel, the cancellation clearly gets more exact
 for heavy loop particles. However, due to the small coefficients in the expansions above, the amplitude at threshold is still significantly suppressed for finite masses. For the SM top, the indicated point in the left panel of Fig. \ref{fig:cancel} shows that the triangle and box functions cancel to ${\cal O}(10\%)$.  This cancellation will be spoiled by a non-SM Higgs self-coupling, additional interactions between the fermions and the Higgs boson, or if the scalar mass receives a contribution that is not from the Higgs (${\it{i.e.}},  \kappa\ne \kappa_0$).

\begin{figure}[t]
\begin{centering}
\includegraphics[width=0.5\textwidth]{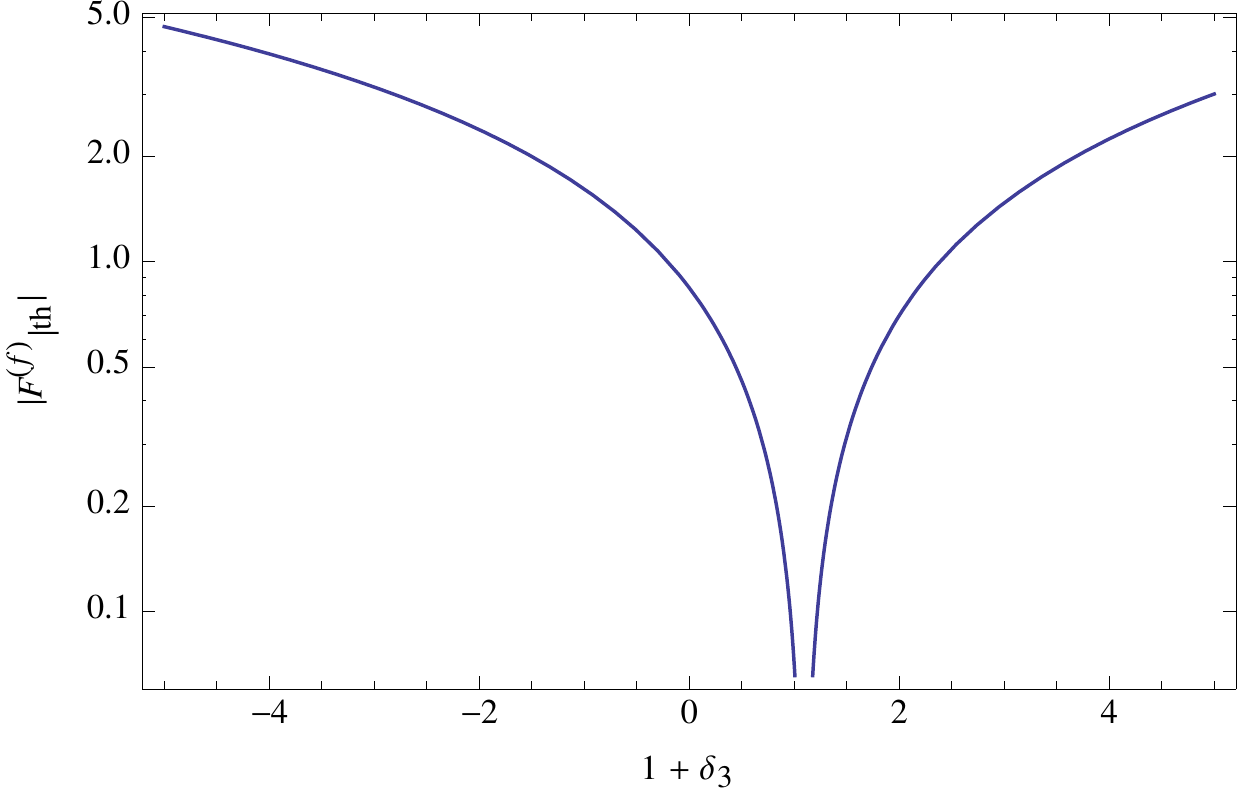}
\par\end{centering}
\caption{\em $2h$ amplitude at threshold in the SM, as a function of the Higgs self-coupling. Only top quarks are included, with $m_t$ = 173 GeV.
}
  \label{fig:toplambda}
\end{figure}

Having established that the cancellation between $F_\triangle\mid_{th}$ and $F_\Box\mid_{th}$ is largely present for finite loop particle masses, we now investigate the behavior of the cancellation in the presence of additional couplings. It is natural to begin by considering the effect of a non-SM Higgs self-coupling on the $2h$ amplitude from top loops at threshold. Such a rescaling would affect $F^{(f)}_\triangle\mid_{th}$ only, since the box diagrams do not involve the Higgs self-coupling. Fig. \ref{fig:toplambda} shows how a modified Higgs tri-linear coupling would significantly alter the threshold amplitude, leading to the well-known result that $2h$ production is a sensitive probe of the Higgs self-coupling. Indeed, for arbitrary loop particle mass, there is a perfect cancellation between the one-loop  triangle and box diagrams for
\bea
1 + \delta_3 &=& -\frac{F^{(f)}_\Box\mid_{th}}{F^{(f)}_\triangle\mid_{th}} \, ,
\label{eq:hhhcancel}
\eea
where the ratio of the box and triangle diagrams at threshold approaches -1 as the loop particle gets infinitely heavy. For the SM top mass, the cancellation is perfect when $\delta_3 \approx 0.09$. The next section considers further new couplings between the SM quarks and the Higgs bosons.

\begin{figure}[t]
\begin{centering}
\includegraphics[height=1.9in]{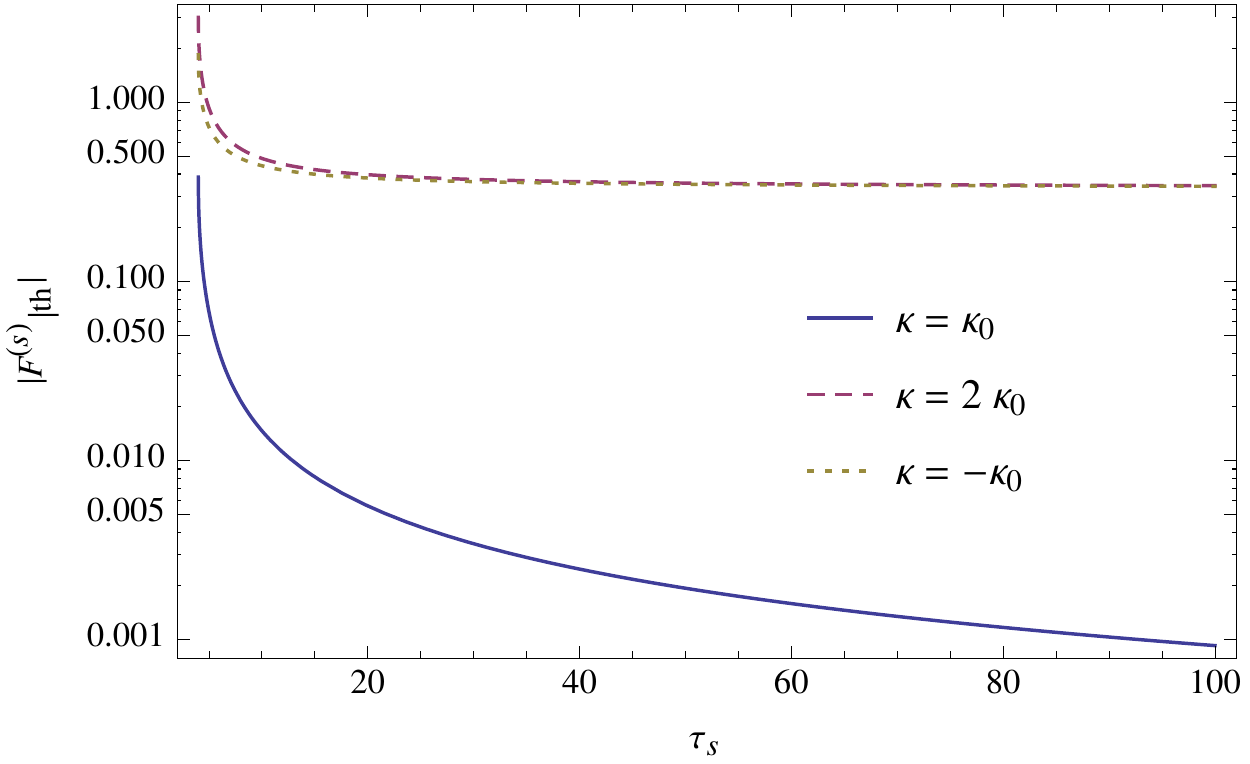}
\includegraphics[height=1.9in]{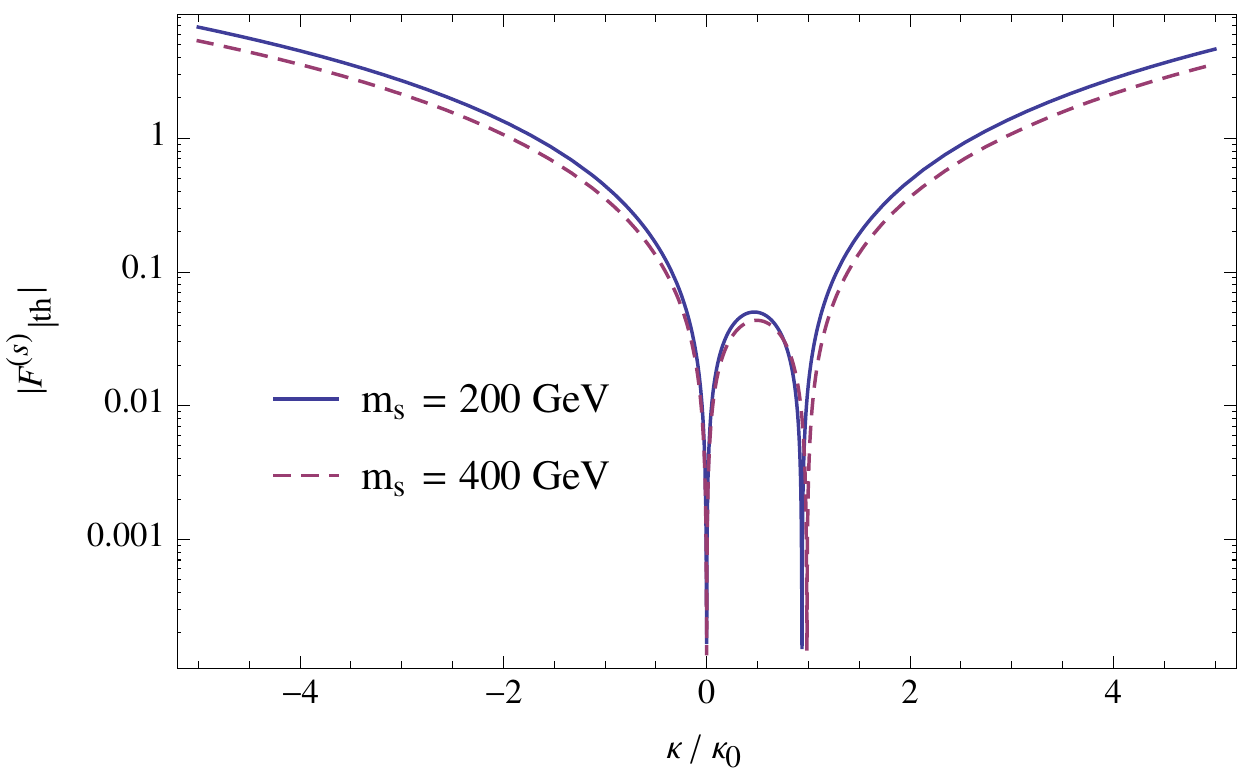}
\par\end{centering}
\caption{\em $2h$ amplitude at threshold for scalars that get different contributions to their masses from electroweak symmetry breaking. The left panel shows curves of constant $\kappa/\kappa_0$, while the right panel shows curves of constant physical scalar mass $m_s$.
}
  \label{fig:scyuk}
\end{figure}

We now turn to the scalar case, including arbitrary soft masses of scalars coupling to the Higgs. For a scalar which does not receive its mass entirely from electroweak symmetry breaking, $m_0 \ne 0$, the amplitude at threshold is
\bea
F^{(s)}\mid_{th} &=& \frac{\kappa}{\kappa_0} F^{(s)}_\triangle\mid_{th}^{\kappa = \kappa_0} + \left( \frac{\kappa}{\kappa_0} \right)^2 F^{(s)}_\Box\mid_{th}^{\kappa = \kappa_0} \, .
\eea
Fig. \ref{fig:scyuk} shows how the cancellation between $F^{(s)}_\triangle\mid_{th}$ and $F^{(s)}_\Box\mid_{th}$ breaks down for $m_0 \ne 0$. We see in the left panel that the triangle and box functions do not cancel when the scalar has a soft mass term, and the total amplitude at threshold tends to a non-zero value in the limit of infinitely heavy scalar mass. The right panel shows how sensitive the cancellation between $F^{(s)}_\triangle\mid_{th}$ and $F^{(s)}_\Box\mid_{th}$ is to the presence of soft scalar mass terms, for a selection of fixed physical scalar masses. In addition to the cancellation at $\kappa = \kappa_0$, the amplitude obviously vanishes when the scalar does not couple to the Higgs, $\kappa = 0$. Discounting this trivial case, the amplitude quickly grows as we move away from the scenario where the scalar gets all of its mass from the Higgs.

\begin{figure}[t]
\begin{centering}
\includegraphics[width=0.5\textwidth]{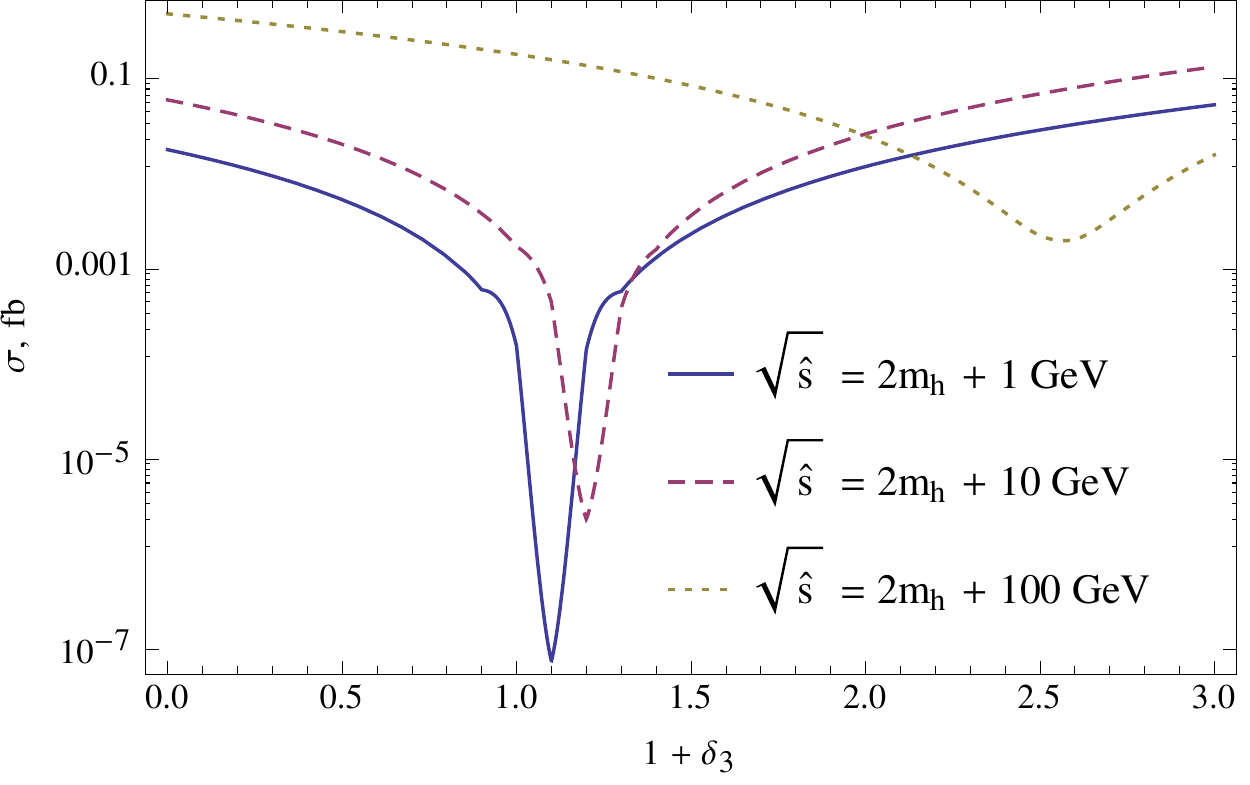}
\par\end{centering}
\caption{\em $2h$ partonic cross section in the SM, as a function of the Higgs self-coupling. Only top quarks are included, with $m_t$ = 173 GeV.
}
  \label{fig:toppartonic}
\end{figure}

Finally, we examine the cancellation away from threshold. In addition to the spin- 0 amplitudes realizing their full functional dependence on the partonic CM energy beyond $\hat{s} = 4 m_h^2$, there are spin-2 
contributions $G_\Box$ to $2h$ production. The full amplitudes are known in terms of loop integrals, and reveal the strong cancellation near threshold when evaluated numerically. In the fermion case, Fig. \ref{fig:toppartonic} shows how the partonic $2h$ cross section changes above threshold for the SM top. Near threshold, there is a pronounced cancellation between the triangle and box diagrams for the value of the Higgs self-coupling predicted by Eq. \ref{eq:hhhcancel}. This dip shifts and becomes much weaker as we move above threshold.

\begin{figure}[t]
\begin{centering}
\includegraphics[width=0.495\textwidth]{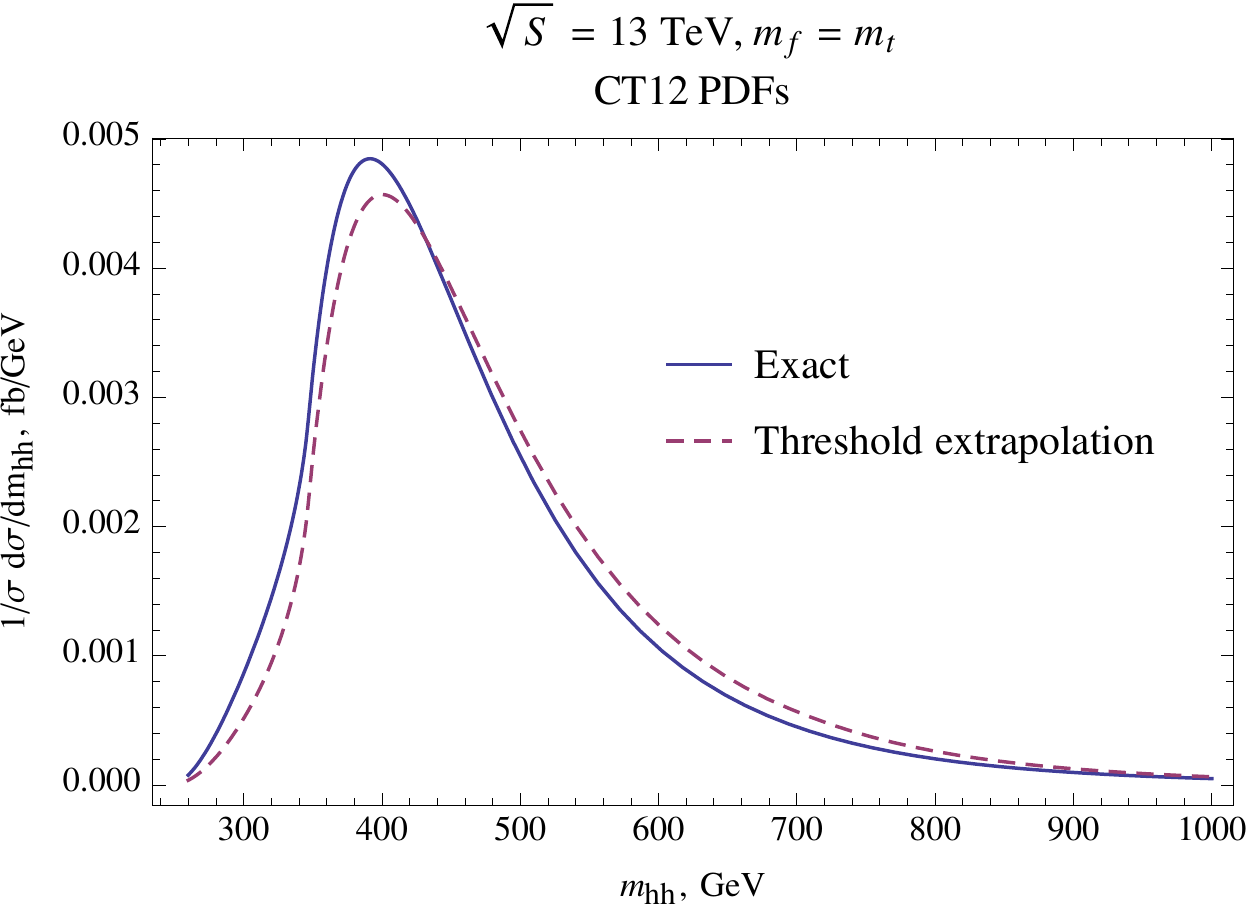}
\includegraphics[width=0.495\textwidth]{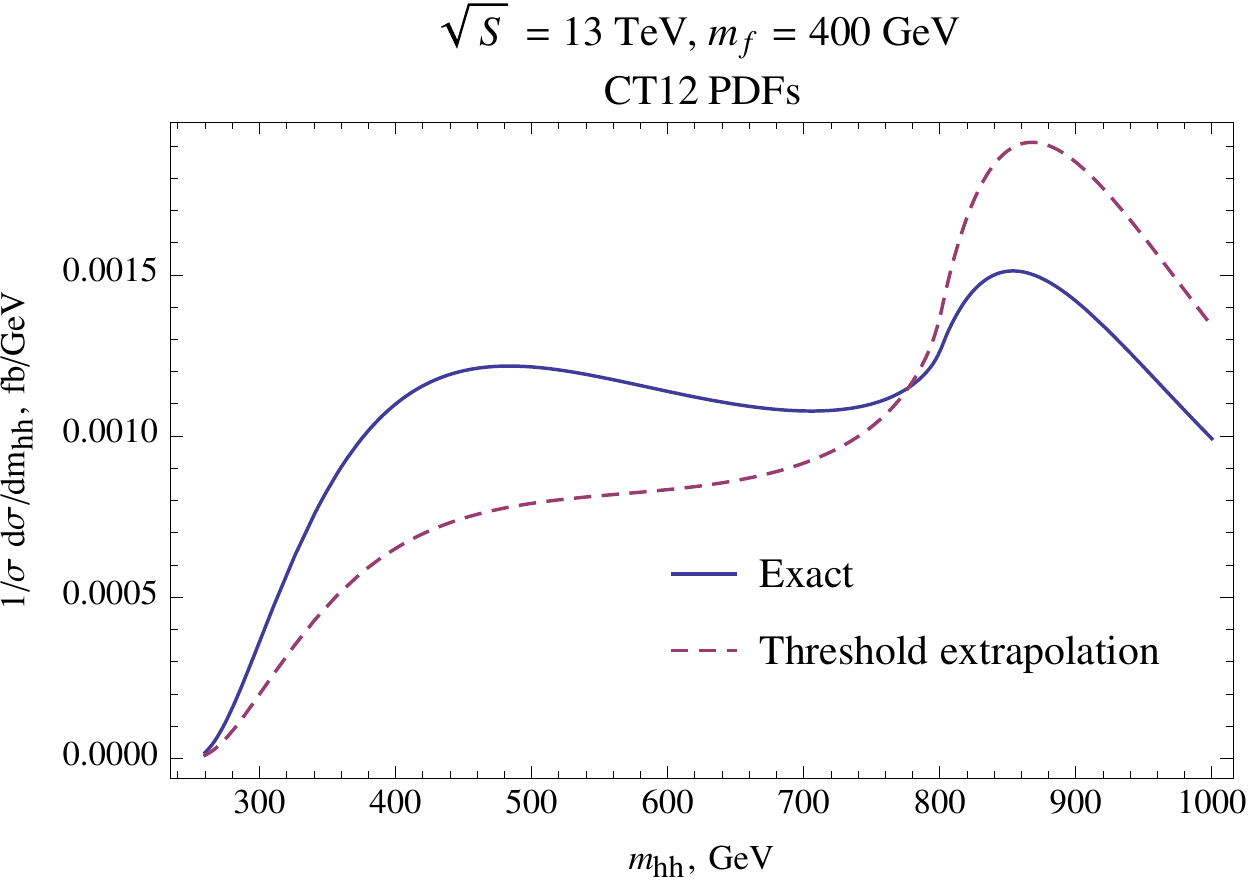}
\includegraphics[width=0.495\textwidth]{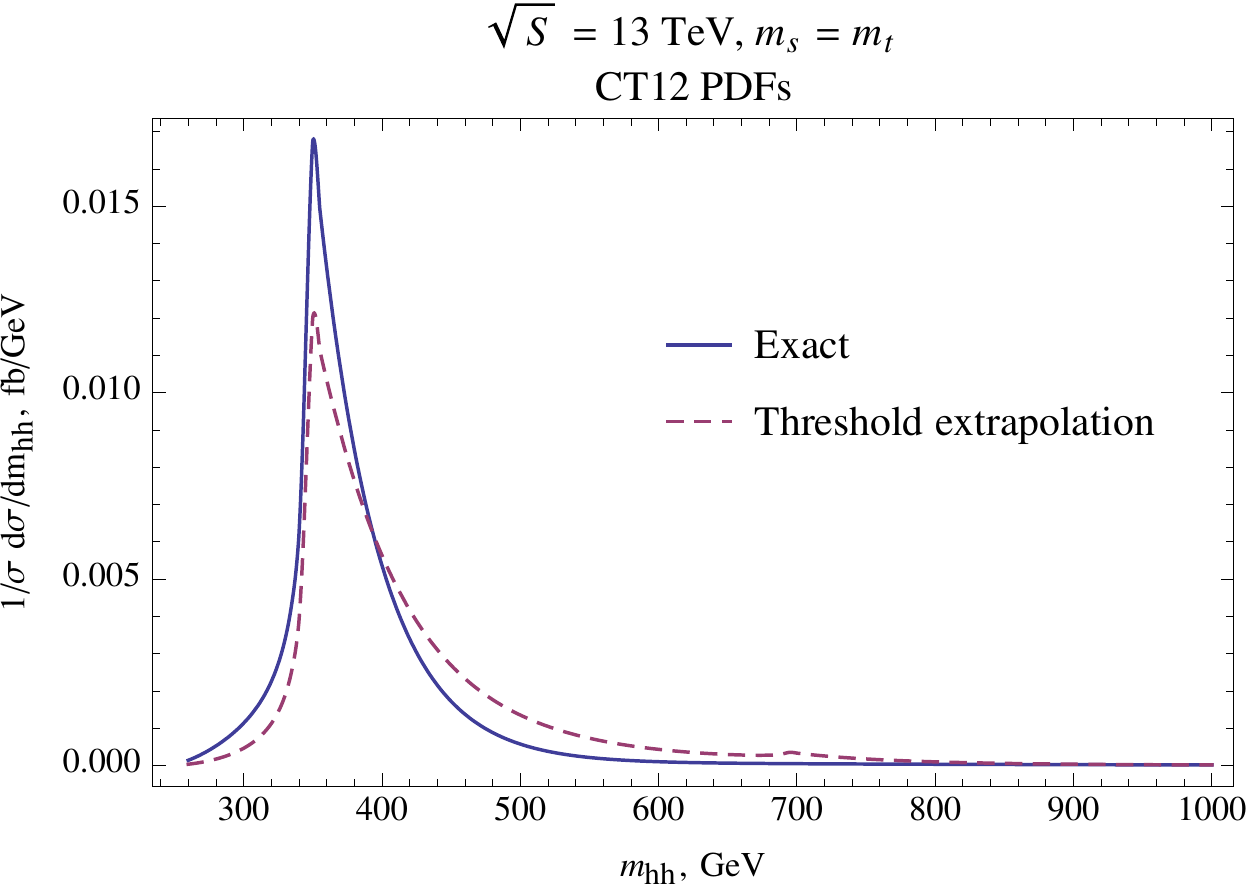}
\includegraphics[width=0.495\textwidth]{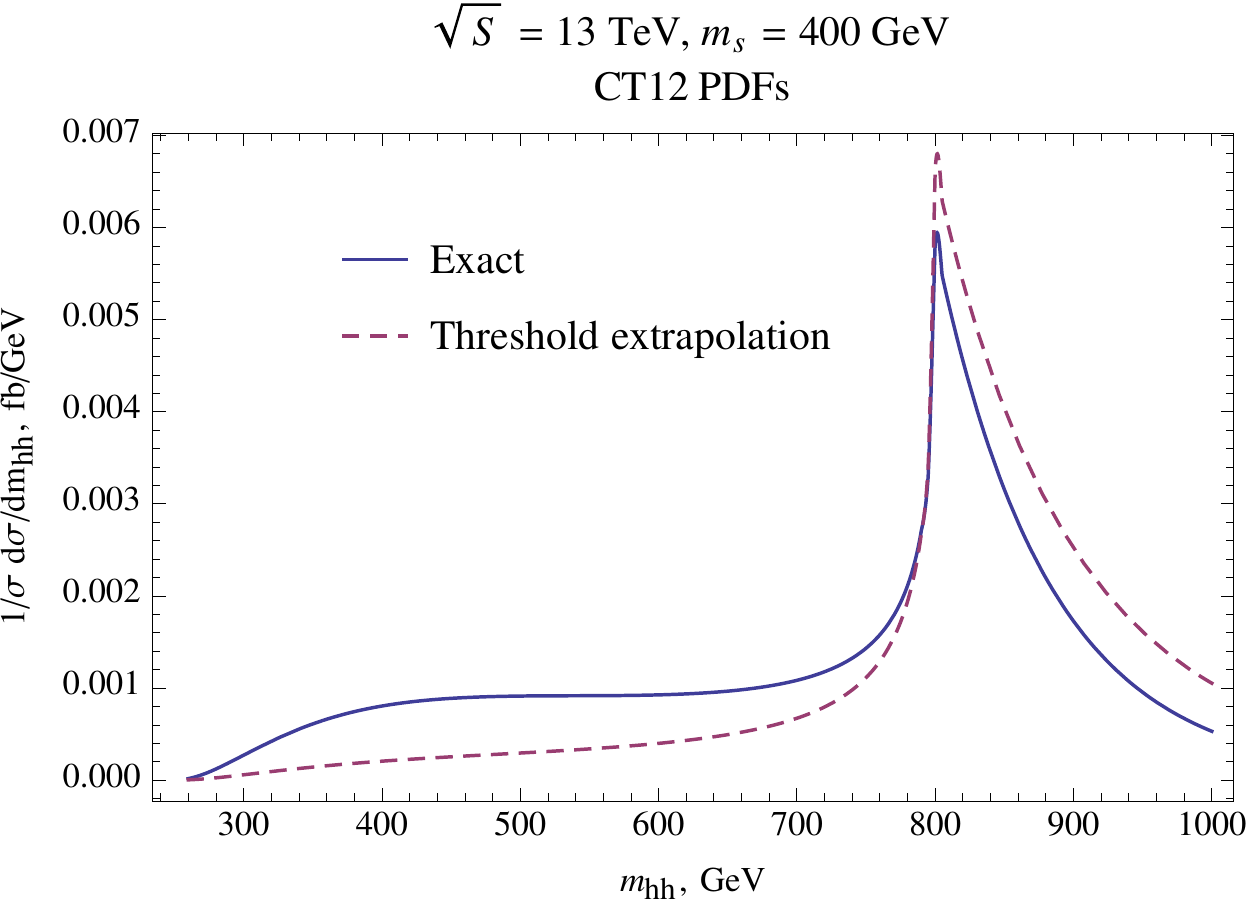}
\par\end{centering}
\caption{\em Normalized invariant mass distributions for $2h$ production with a single loop particle. The upper panels show fermions of masses 173 GeV (left) and 400 GeV (right), while the lower panels show scalars of masses 173 GeV (left) and 400 GeV (right). The solid and dashed lines show the exact distributions and the approximation of Eq. \ref{eq:thrapprox}, respectively.
}
  \label{fig:thrapprox}
\end{figure}
\begin{figure}[t]
\begin{centering}
\includegraphics[width=0.5\textwidth]{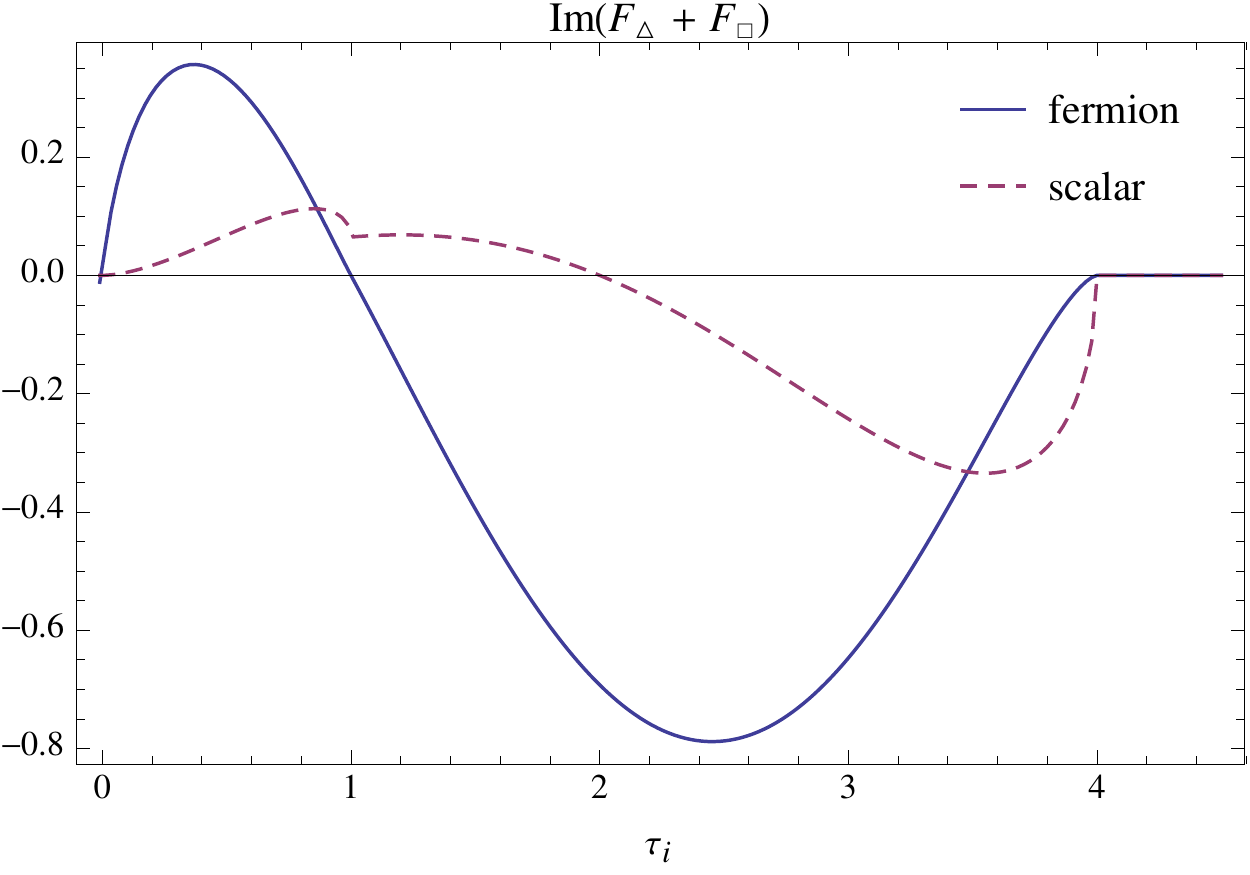}
\par\end{centering}
\caption{\em The imaginary parts of the $2h$ threshold amplitudes with loop fermions (solid) and scalars (dashed). Discontinuities occur when loop propagators go on shell. Appendix \ref{scalarcuts} derives the imaginary part of the threshold amplitude with a loop scalar directly using cut techniques.
}
  \label{fig:imaginary}
\end{figure}

It is interesting to consider how well our closed form expressions for the threshold $2h$ amplitudes approximate the full amplitudes. In the threshold amplitudes $F^{(f,s)}\mid_{th}$ above, the Higgs mass may be considered as a proxy for the CM energy, $\hat{s} = 4 m_h^2$, except in the triangle diagram where it appears in an $s$-channel propagator. This motivates the closed form approximation,
\bea
F_\triangle^{(i)}(\hat{s}, \hat{t}, m_h^2, m_f^2) &\approx& F^{(i)}_\triangle\mid_{th}(\tau_i \to 16 m_i^2 / \hat{s}) \nonumber \\
F_\Box^{(i)}(\hat{s}, \hat{t}, m_h^2, m_f^2) &\approx& F^{(i)}_\Box\mid_{th}(\tau_i \to 16 m_i^2 / \hat{s})\, ,
\label{eq:thrapprox}
\eea
for the $2h$ amplitude beyond threshold. Fig. \ref{fig:thrapprox} shows the result of using this approximation in Eq. \ref{eq:ampgghh} for loop fermions and scalars of various masses. While the total cross section predicted by the expression in Eq. \ref{eq:thrapprox} is not close to the true cross section, the normalized invariant mass distribution is fairly well reproduced. In particular, for $m_{hh} \geq 2 m_i$, two 
different loop particle propagators may go on shell, causing a nonzero imaginary piece of the exact amplitude that is visible as a feature in the invariant mass distribution. Similarly, the threshold approximation in Eq. \ref{eq:thrapprox} includes a term proportional to $\arcsin^2 \left( \frac{\sqrt{\hat{s}}}{2 m_i} \right)$. We note that there is in principle also a discontinuity in the invariant mass distribution at $m_{hh} \geq 4 m_i$, captured by terms proportional to $\arcsin^2 \left( \frac{\sqrt{\hat{s}}}{4 m_i} \right)$ in the threshold approximation. However, there is no visible corresponding feature in the invariant mass distributions of Fig. \ref{fig:thrapprox}. The discontinuities at $m_{hh} \geq 2 m_i$ and $m_{hh} \geq 4 m_i$ correspond to the discontinuities in the threshold amplitudes at $\tau_i = 4$ and $\tau_i = 1$, respectively. Fig. \ref{fig:imaginary} shows the imaginary parts of the threshold amplitudes, allowing us to compare the discontinuities. The unphysical discontinuity at $\tau_i = -1$, which may be traced back to the $\arcsinh$ terms in the threshold amplitudes of Eqs. \ref{eq:ferthr} and \ref{eq:scthr}, is not shown. For both fermions \cite{Li:2013rra} and scalars, we observe a much larger discontinuity at $\tau_i = 4$ than at $\tau_i = 1$, due to the lack of any imaginary piece of the amplitude for $\tau_i > 4$. The much smaller discontinuities in the imaginary part of the threshold amplitudes explain the lack of any visible features at $m_{hh} = 4 m_i$.

\section{Examples}
\label{example}
The previous section demonstrated how the threshold cancellations between triangle and box diagrams render $2h$ production 
extremely sensitive to non-SM couplings.
In this section we consider modifications of the $2h$ distributions from anomalous fermionic Yukawa couplings, from colored scalar loops,
and from fermionic top partners.  Effects of anomalous $t{\overline t}hh$ couplings on the kinematic distributions have been examined in \cite{Chen:2014xra}. In some cases the allowed new interactions are severely restricted by the requirement
that $1h$ production  occur at the observed rate.   
In addition, we are interested in whether $2h$ production distributions can distinguish between fermion and 
scalar loop contributions.

In all of our numerical results we use CT12 NLO PDFs \cite{Owens:2012bv,Gao:2013xoa} with the associated NLO values for $\alpha_s$, and take $m_t=173$~GeV, $m_b=4.3$~GeV, and $m_h=125$~GeV.
For $1h$ production we take $\mu=\mu_R=\mu_F=m_h$, while for $2h$ production we set $\mu=\mu_R=\mu_F=m_{hh}$.  We use the LO $1$-loop predictions
 for both $1h$ and $2h$ production. In addition, the 1-loop functions are evaluated using the software {\tt LoopTools} \cite{Hahn:1998yk},
 as well as independent in-house routines.
 
\subsection{Anomalous Yukawa Couplings} 
\begin{figure}[t]
\begin{centering}
\includegraphics[scale=0.45]{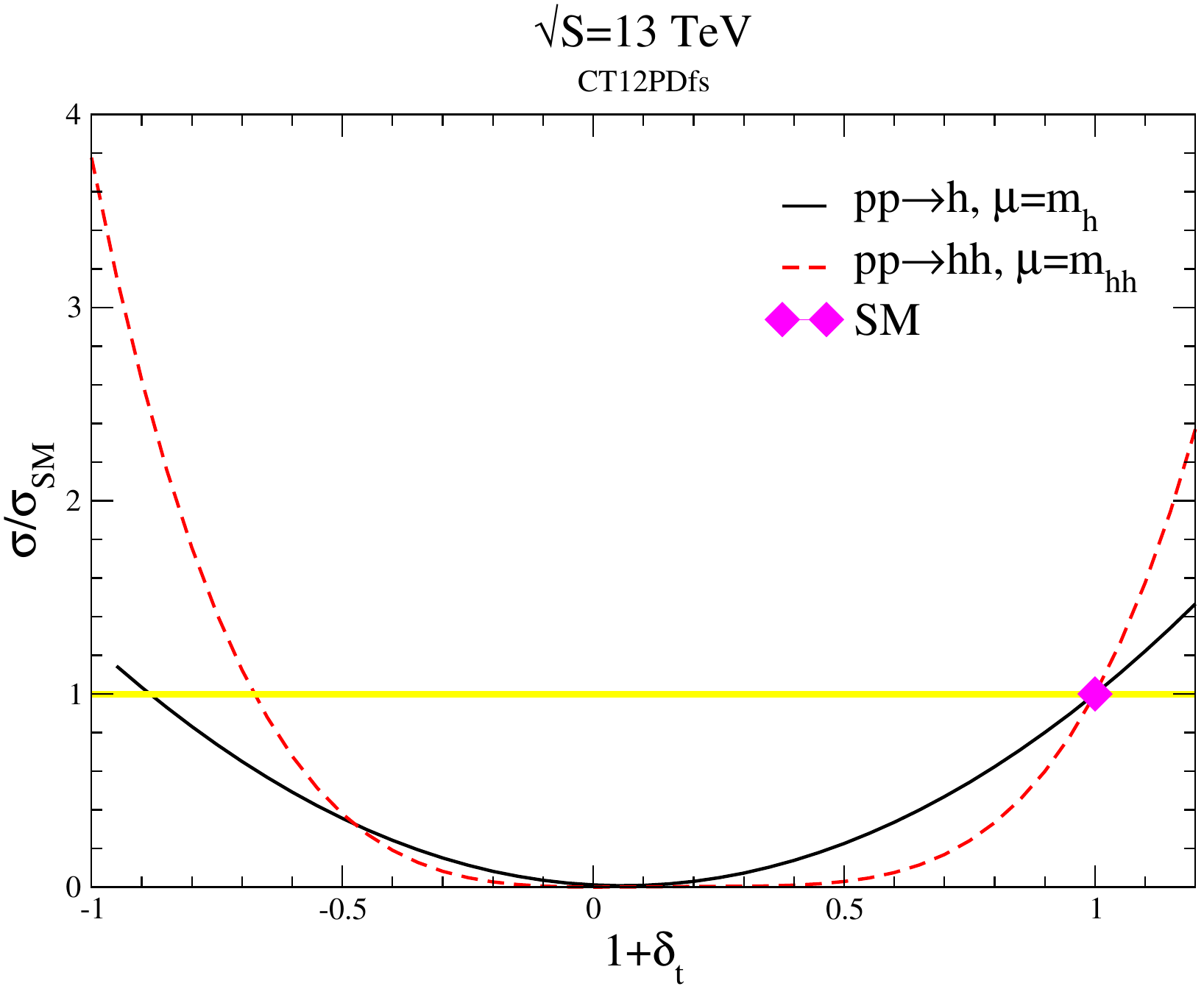}
\includegraphics[scale=0.45]{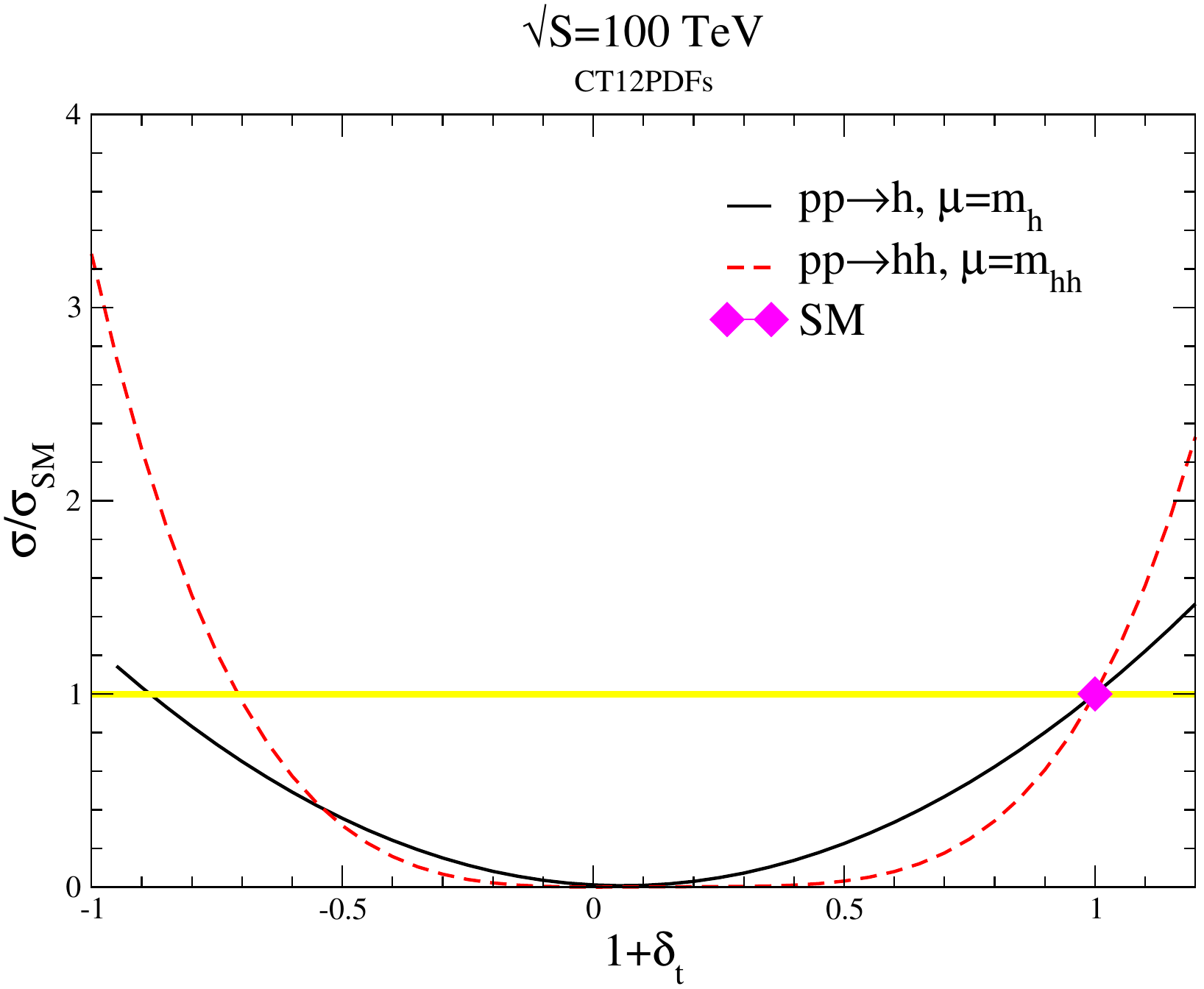}
\par\end{centering}
\caption{\em Total cross sections for $1h$ and $2h$ production with anomalous top-Higgs couplings, normalized to the $1$-loop SM prediction.  The SM rate corresponds
to $\delta_t=0$. All other couplings are assumed to be SM-like.}
  \label{fig:tyuk}
\end{figure}

\begin{figure}[t]
\begin{centering}
\includegraphics[scale=0.45]{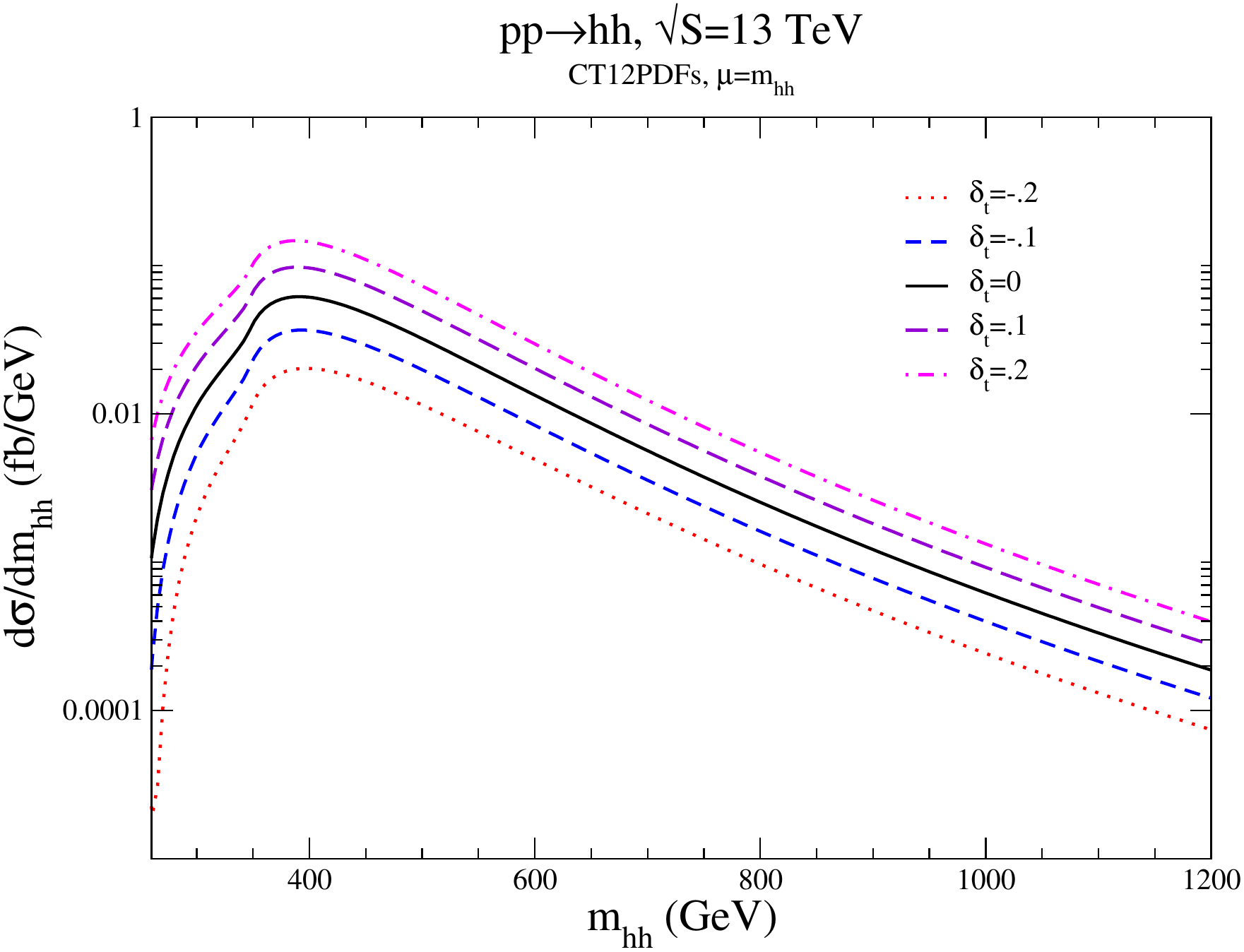}
\includegraphics[scale=.45]{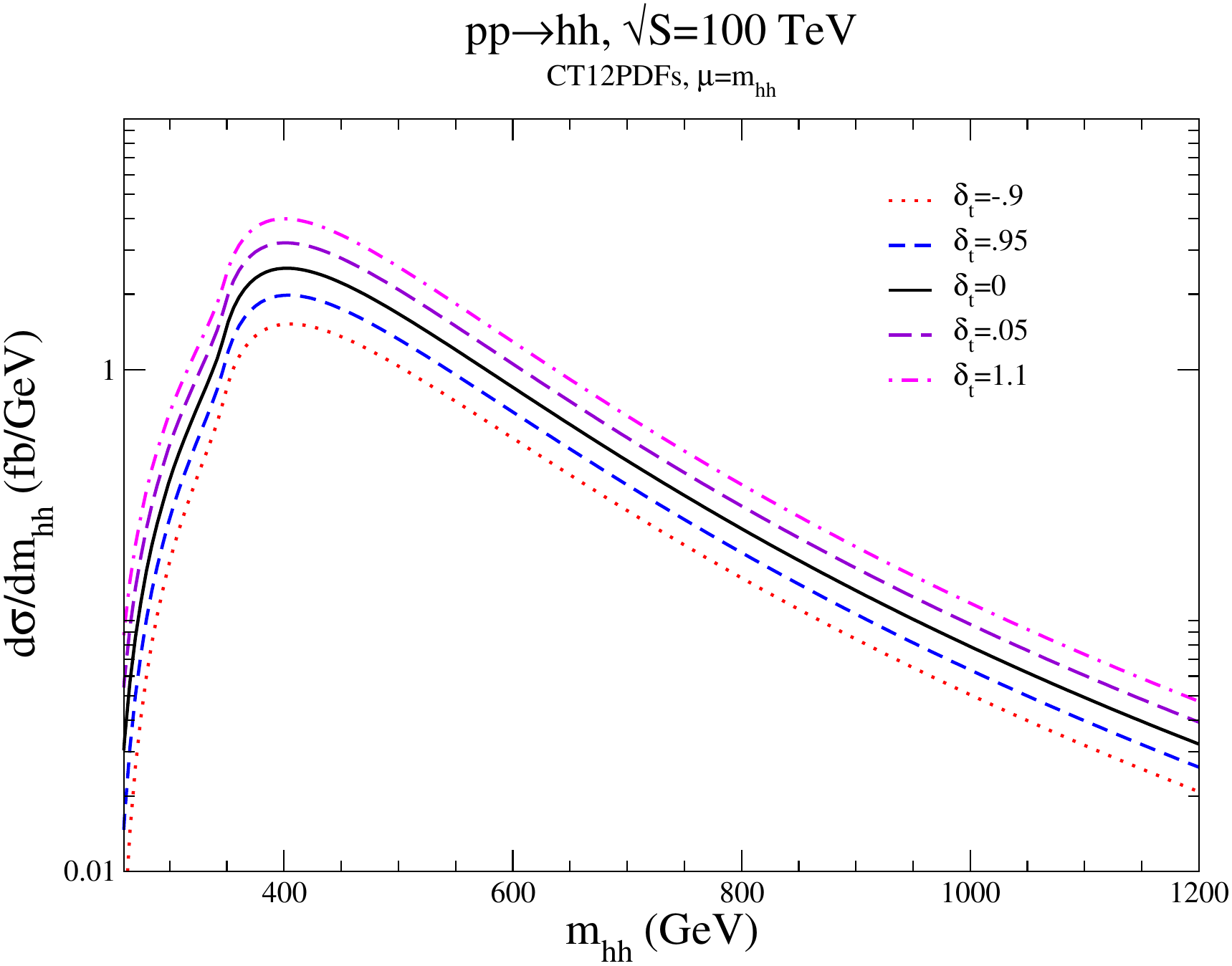}
\par\end{centering}
\caption{\em Invariant mass distributions for $2h$ production with anomalous top Yukawa coupling.  The SM rate corresponds
to $\delta_t=0$. All other couplings are assumed to be SM-like.}
  \label{fig:tyuk_dist}
\end{figure}

\begin{figure}[t]
\begin{centering}
\includegraphics[scale=0.45]{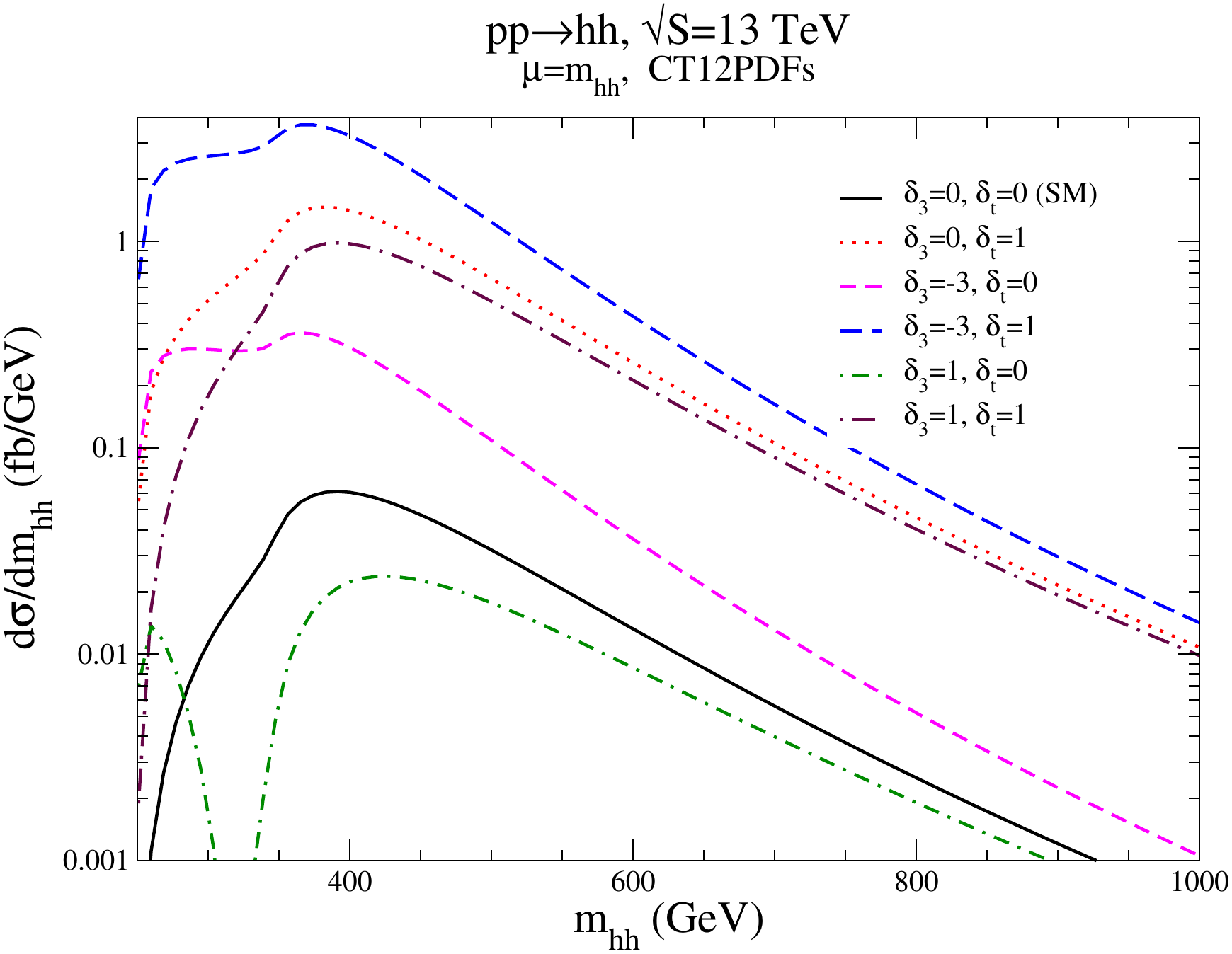}
\includegraphics[scale=.45]{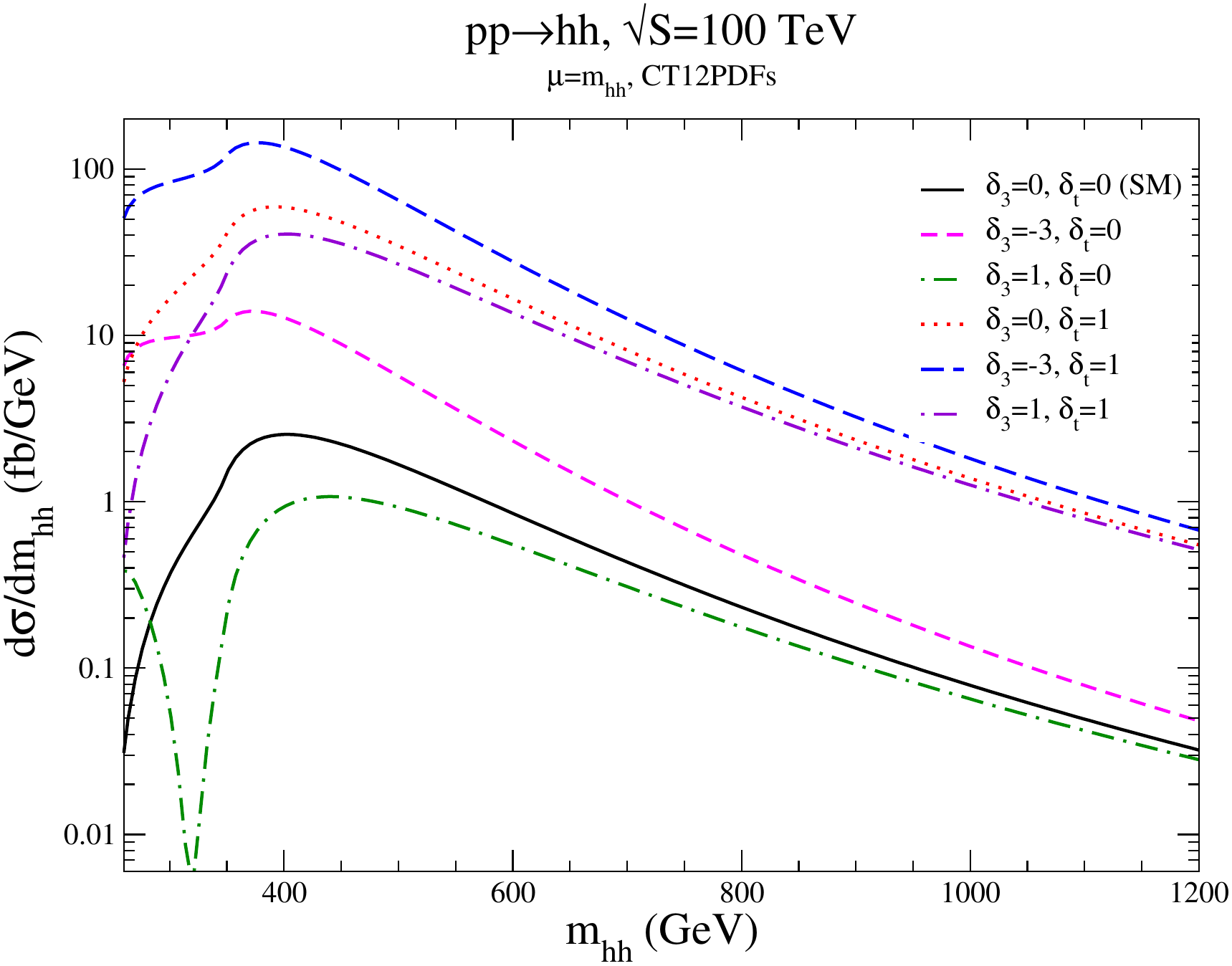}
\par\end{centering}
\caption{\em Invariant mass distributions for $2h$ production with anomalous top Yukawa coupling, $\delta_t$, and anomalous tri-linear
Higgs couplings, $\delta_3$.  The SM corresponds to $\delta_t=\delta_3=0$. All other couplings are assumed to be SM-like. }
  \label{fig:eft_top}
\end{figure}

We begin by considering the effects of anomalous top Yukawa couplings in Eq. \ref{effective}, assuming all other couplings are SM-like.  In Fig. \ref{fig:tyuk}, we show the both
the $1h$ and $2h$ rates, normalized to the one-loop SM rate as a function of the top quark Yukawa.  For positive $\delta_t$,
the requirement that $|R_h-1| \le .20$  only allows a $\sim 40\%$ deviation in $\sigma(gg\rightarrow hh)/\sigma(gg\rightarrow hh)_{SM}$.\footnote{We note that negative
$\delta_t$ is now excluded by global fits to Higgs couplings \cite{Khachatryan:2014jba,atlasfits}.}  As measurements of the $1h$ rate become more precise, the allowed
deviations for the $2h$ rate due to an anomalous top Yukawa coupling will also become smaller.  An important assumption throughout this work is that there are no light particles which could allow for
resonant production of two Higgs bosons, in which case the $m_{hh}$ spectrum would exhibit a clear peak at the mass of the resonance.
The effect of a non-SM top Yukawa coupling on the invariant mass distribution is shown in Fig. \ref{fig:tyuk_dist} for both $\sqrt{S}=13$ and $100$ TeV.\footnote{There is not  much difference in the invariant mass spectra between $\sqrt{S}=13$ and 100 TeV.  This feature has also been observed in Ref. \cite{Chen:2014xra}.}
When $\delta_t\ne 0$, the cancellation between box and triangle diagrams described in the previous section is spoiled  and the resulting cross sections
vary by up to a factor of two \cite{Baur:2002rb,Baur:2002qd}.  This same variation is seen at both $\sqrt{S}=13$ TeV and $100$ TeV. 
The effect of changing the top Yukawa coupling and the tri-linear Higgs coupling in a correlated manner can be quite dramatic, as shown in Fig. 
\ref{fig:eft_top}.  Note the interesting cancellation  for large and positive $\delta_3$.

\begin{figure}[t]
\begin{centering}
\includegraphics[scale=0.45]{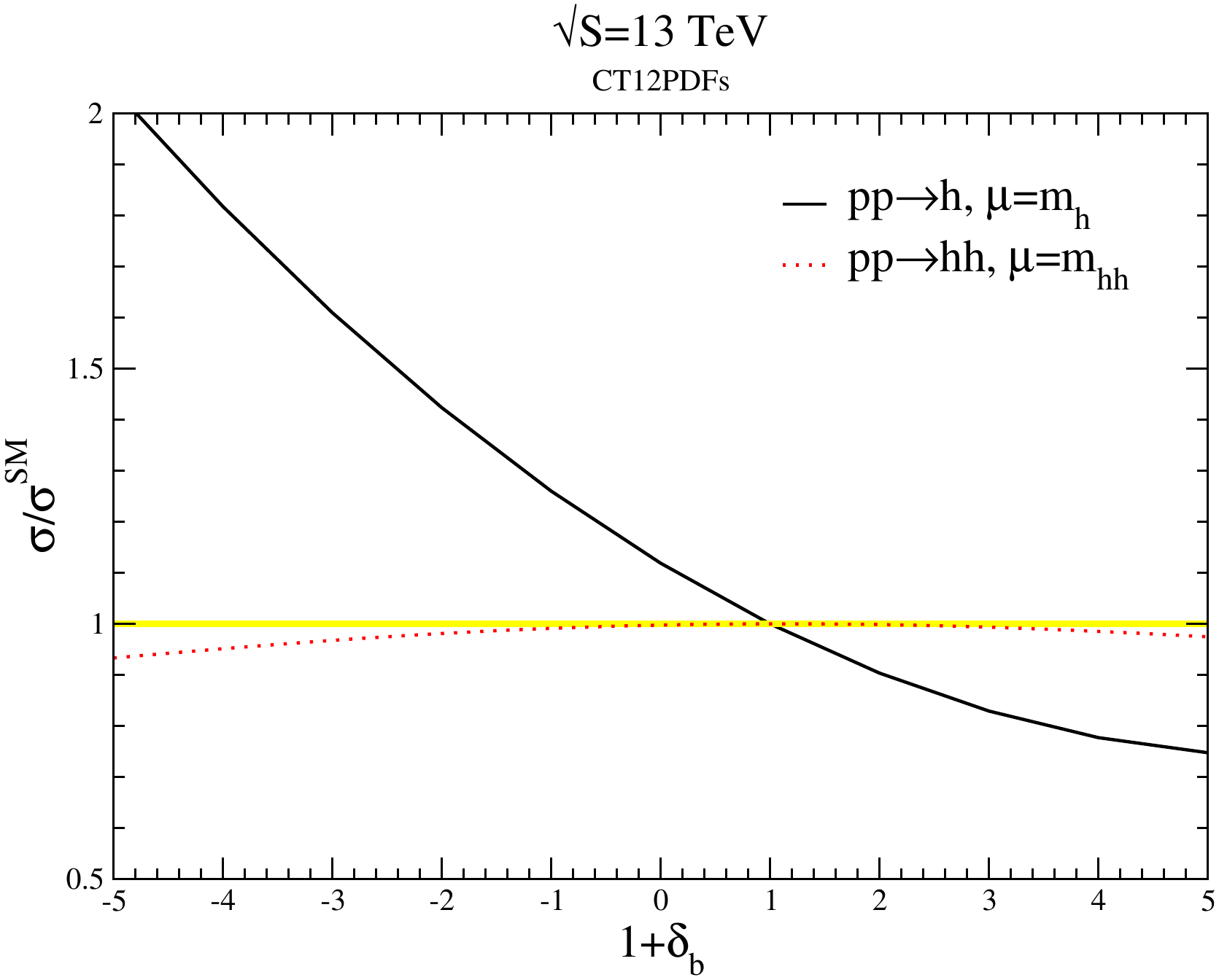}
\includegraphics[scale=0.45]{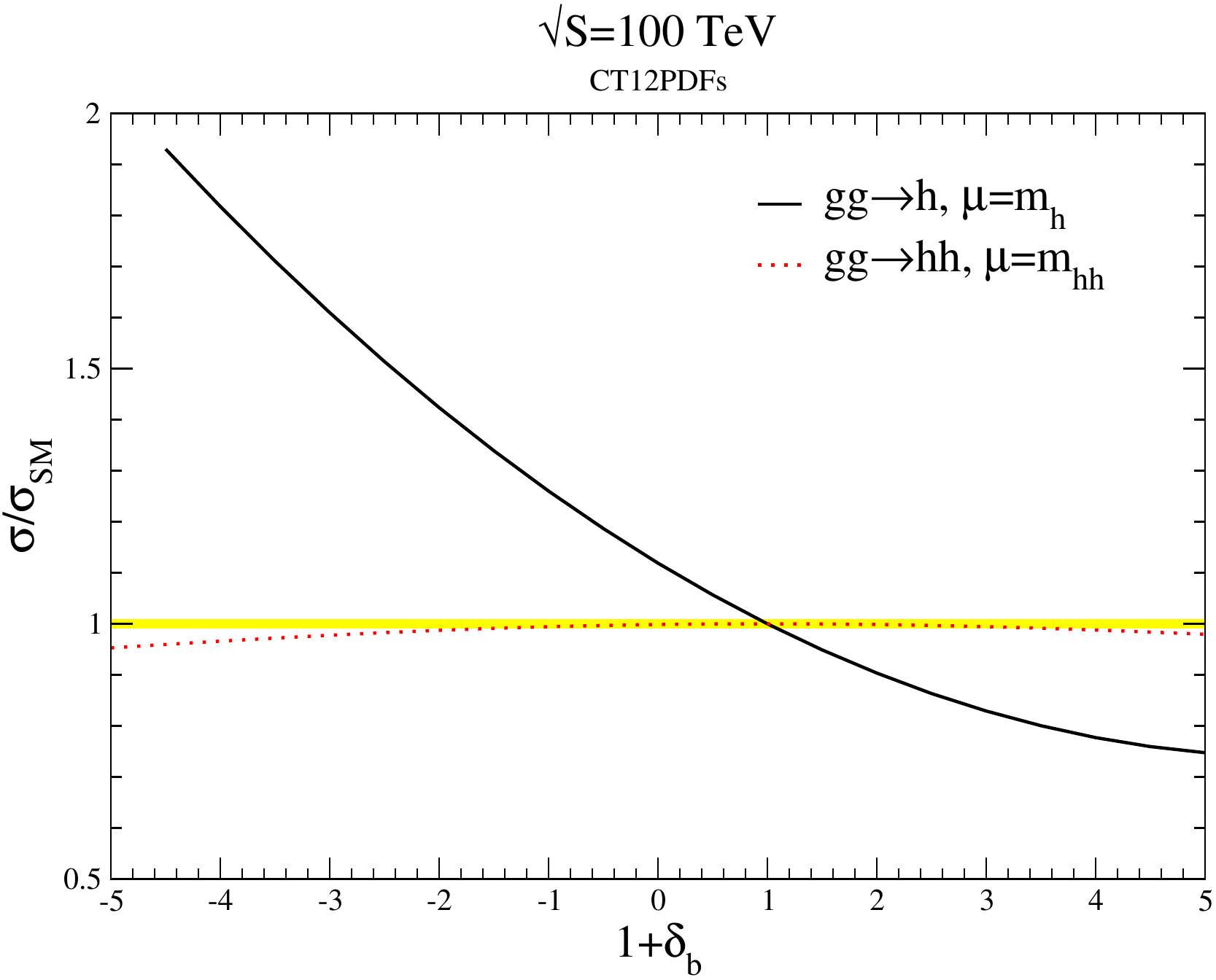}
\par\end{centering}
\caption{\em Total cross sections for $1h$ and $2h$ production with anomalous bottom Yukawa couplings, normalized to the LO SM prediction.  The SM rate corresponds
to $\delta_b=0$. All other couplings are assumed to be SM-like.
}
  \label{fig:byuk}
\end{figure}

In the SM, the contribution of the $b$ quark is small for both the $1h$ and $2h$ production \cite{Anastasiou:2011pi}, although for large
enough $\delta_b$ the production from $b$ quark initial states becomes important \cite{Dicus:1988cx,Dawson:2006dm}.  As the $b$ quark Yukawa is increased, the
rate for $gg\rightarrow h$ is substantially altered, while the $2h$ rate is rather insensitive to the $b$ Yukawa as seen in Fig. \ref{fig:byuk}.
Note that ${\Gamma(h\rightarrow b {\overline b})\sim (1+\delta_b)^2\Gamma(h\rightarrow b{\overline b})_{SM}}$ and the LHC experiments limit the total Higgs
width\cite{Aad:2015xua,Khachatryan:2014iha}, 
 $\Gamma_{h,tot}<5 \Gamma_{h,tot}^{SM}$, so this implies a rough limit $\delta_b \le {\cal{O}}(1)$.

\subsection{Fermionic Top Partners}

In this subsection and the following, we consider fermion and scalar contributions to $gg\rightarrow hh$ and pose the
question:  
\begin{itemize}
\item Can we determine the nature of the loop particle, by examining the properties of the scattering amplitude?
\end{itemize}
We use the 
analytic properties of the amplitude discussed in Sec. \ref{analytic}  to draw some conclusions.  

\begin{figure}[t]
\begin{centering}
\includegraphics[scale=0.45]{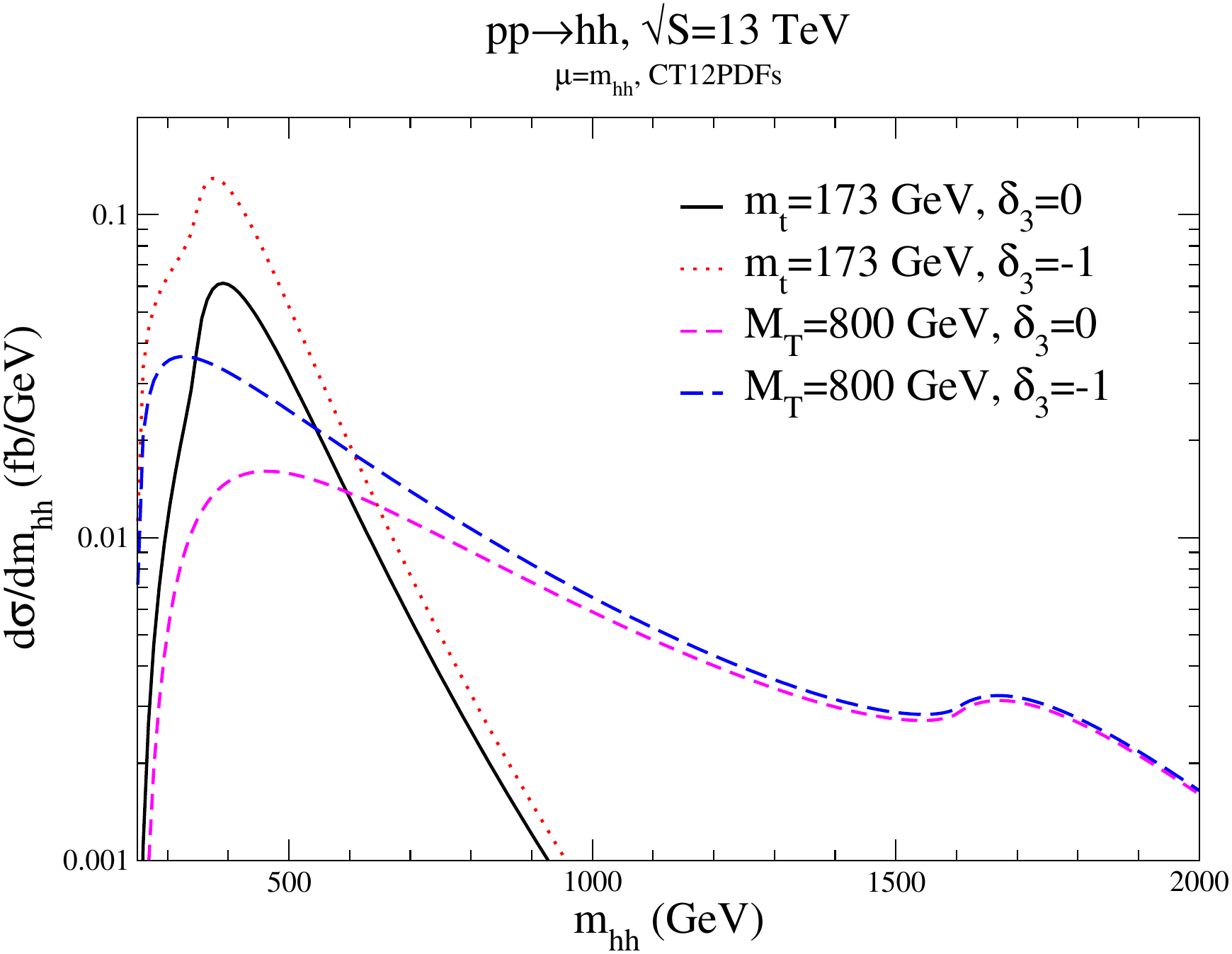}
\includegraphics[scale=0.45]{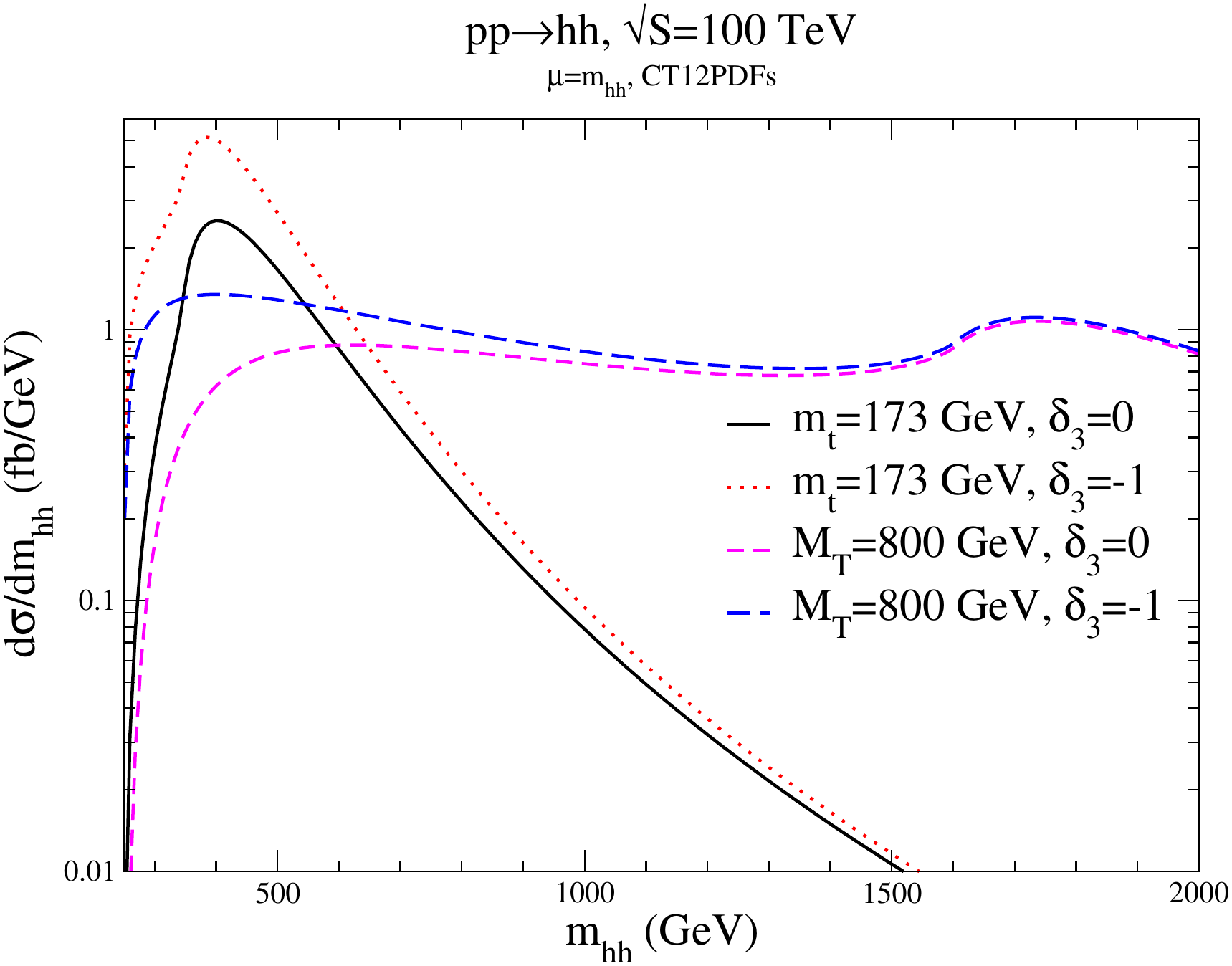}
\par\end{centering}
\caption{\em Invariant mass distribution for a heavy fermion with mass, $M_T$,
 and SM-like Yukawa couplings.  The tri-linear
Higgs coupling is allowed to vary from the SM value of $\delta_3=0$.}
  \label{fig:topdist}
\end{figure}
We begin by considering the effects of a heavy color triplet fermion, with mass $M_T$, with a SM-like
Yukawa coupling to the Higgs boson.
The invariant mass distribution for this heavy fermion is compared with that from $m_t=173$ GeV in Fig. \ref{fig:topdist}.  The distribution
has an interesting dip near $m_{hh}\sim 2 M_T$ due  to the presence of a cut in the amplitude.   This dip persists even when
the Higgs tri-linear coupling is allowed to have a non-SM value, $\delta_3\ne 0$.  At $\sqrt{S}=100$ TeV, the invariant mass spectrum has a more significant support at large $m_{hh}$, compared to that at $\sqrt{S}=13$ TeV.

A 4th generation of chiral fermions would increase the rate for $1h$ production by roughly a factor of $9$ above the SM prediction,
 far above the allowed region from current data \cite{Khachatryan:2014jba,atlasfits}.
So we consider the addition of a vector-like fermonic top partner. In the simplest example, a top partner singlet model, there exists a charge-${2\over 3}$ $SU(2)_L$ singlet particle, ${\cal T}_{L,R}^2$, which mixes with the SM-like top
quark, ${\cal T}^1$.  The Yukawa couplings in the top partner sector are \cite{AguilarSaavedra:2002kr,Dawson:2012di,Aguilar-Saavedra:2013qpa},
\begin{eqnarray}
-L_Y&\sim &
\lambda_2 {\overline{\psi}}^1_L {\tilde H}{\cal T}^1_R
+\lambda_3 
{\overline {\psi}}^1_L
{\tilde {H}} 
{\cal {T}}^2_R+
\lambda_4
{\overline{{\cal {T}}}}^2_L 
{\cal {T}}^1_R+\lambda_5
{\overline{{\cal{T}}}}^2_L
{\cal {T}}^2_R+h.c. \, ,
\label{tplag}
\end{eqnarray}
where
the Standard Model-like particles
are denoted as 
\begin{equation}
\psi_L=\left(\begin{matrix}
{\cal {T}}_L^1\\b_L\end{matrix}\right), \quad {\cal T}_R^1, \quad b_R
\, .
\end{equation}
The addition of the $\lambda_5$ Dirac fermion mass term in Eq. \ref{tplag} means that
the fermion masses are not completely determined by electroweak symmetry breaking.  We can always rotate ${\cal T}^2$ such that $\lambda_4=0$ 
and so there are 3 independent
parameters in the top sector, which we take to be the physical charge-${2\over 3}$ quark masses, $m_t$ and $M_T$, 
along with the mixing angle, $\theta_L$. In the following, we will abbreviate $s_L\equiv \sin\theta_L$, 
$c_L\equiv \cos\theta_L$. 
The couplings of the physical heavy charge-${2\over 3}$ quarks 
to the Higgs boson are,
\begin{eqnarray}
-{\cal{L}}_H&=&{m_t\over v}c_L^2{\overline t}_Lt_Rh
+{M_T\over v}s_L^2{\overline T}_LT_Rh
+s_Lc_L {M_T\over v} {\overline t}_L T_Rh
+s_Lc_L {m_t\over v} {\overline T}_Lt_Rh
+h.c.\, .
\label{higgscoups}
\end{eqnarray}
The parameters of the fermonic top partner model are limited by electroweak precision measurements to 
$\sin\theta_L < .12$ \cite{AguilarSaavedra:2002kr,Dawson:2012di,Aguilar-Saavedra:2013qpa} and
by direct search experiments to 
$M_T>880$ GeV \cite{Chatrchyan:2013uxa,ATLAS:2012qe}.  

\begin{figure}[t]
\begin{centering}
\includegraphics[scale=0.45]{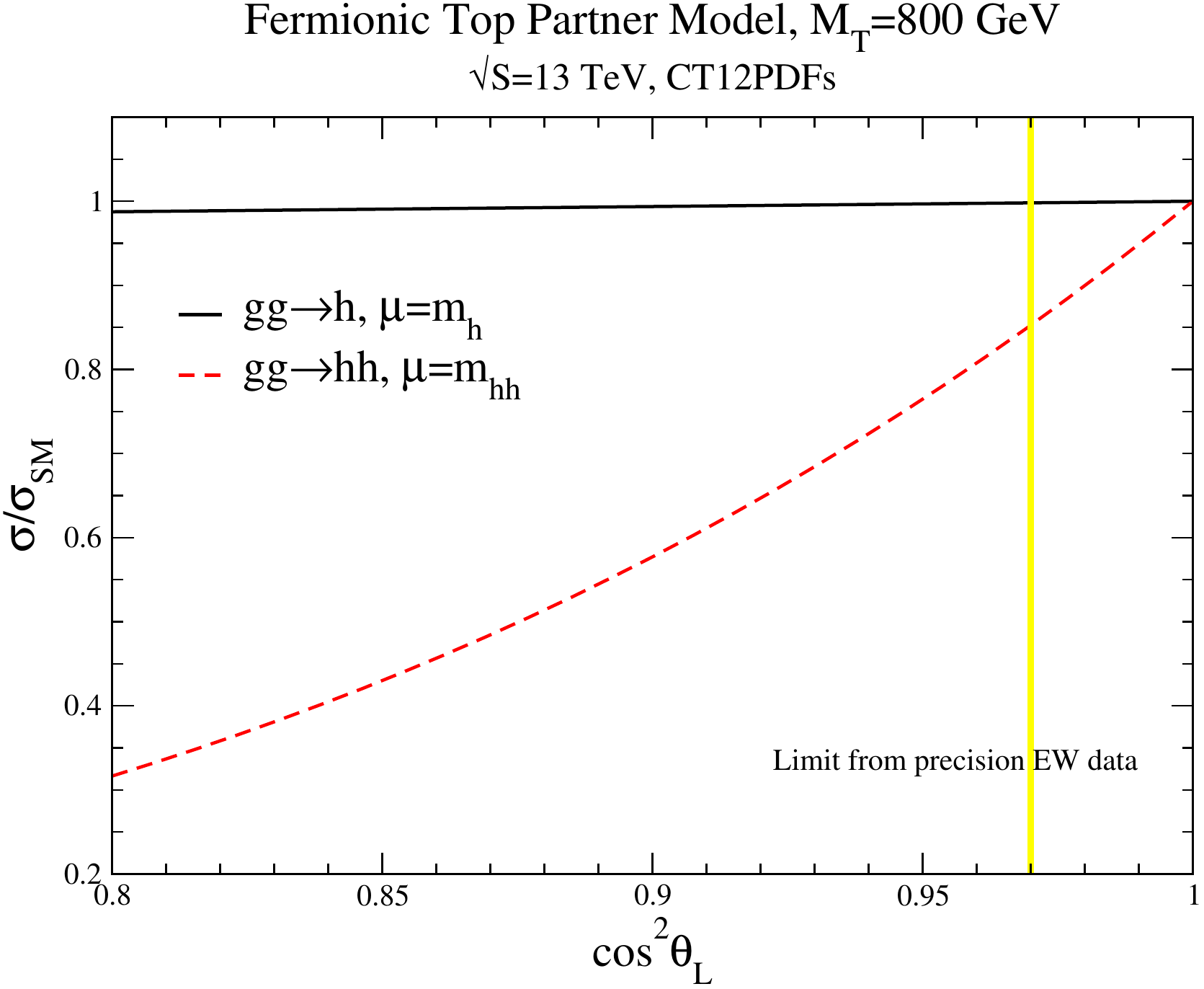}
\includegraphics[scale=0.45]{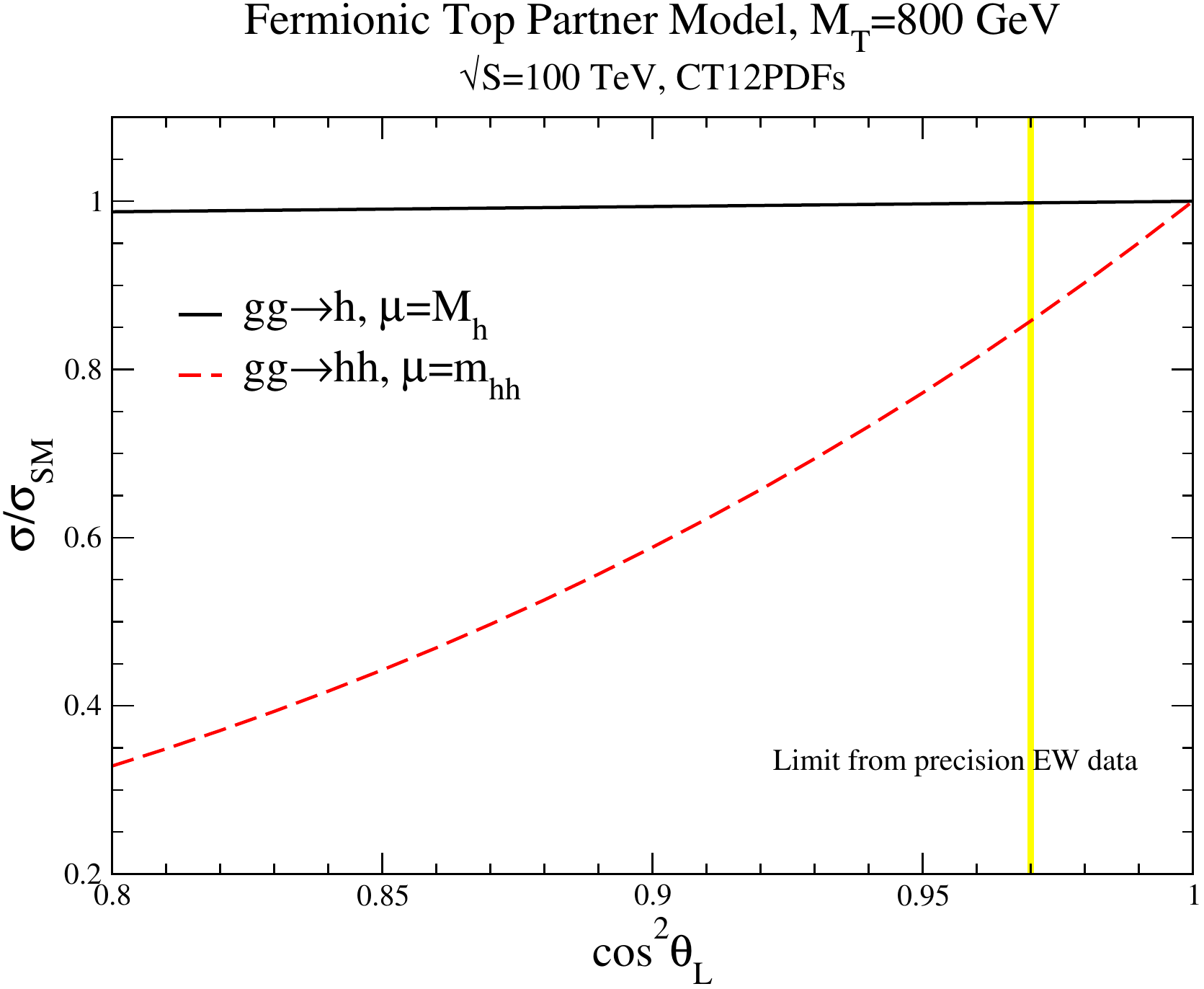}
\par\end{centering}
\caption{\em $1h$ and  $2h$ production, normalized to the SM rate, in the top partner singlet model
with a heavy top partner mass of $M_T=800$ GeV.}
  \label{fig:toppart_sum}
\end{figure}

\begin{figure}[t]
\begin{centering}
\includegraphics[scale=0.45]{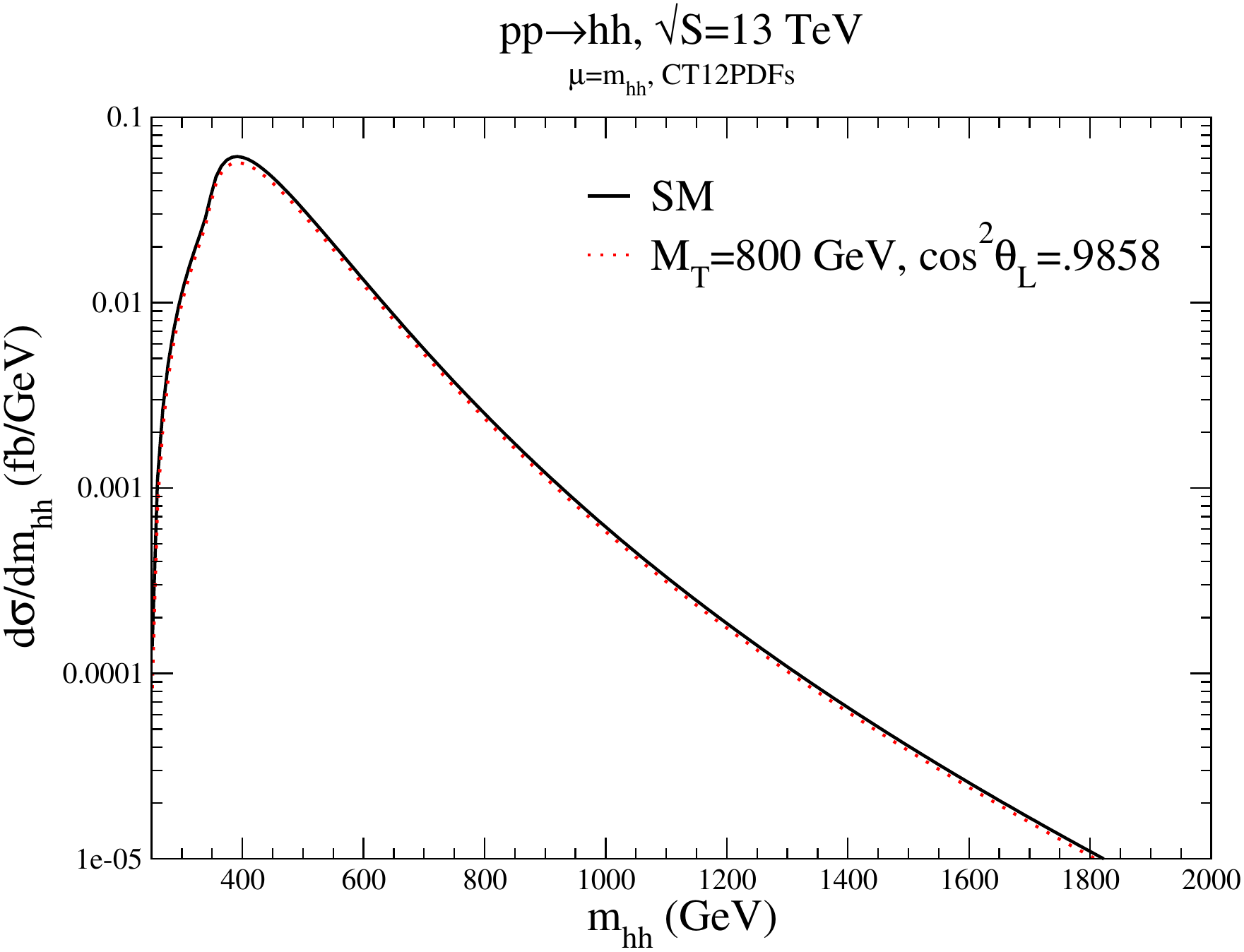}
\includegraphics[scale=0.45]{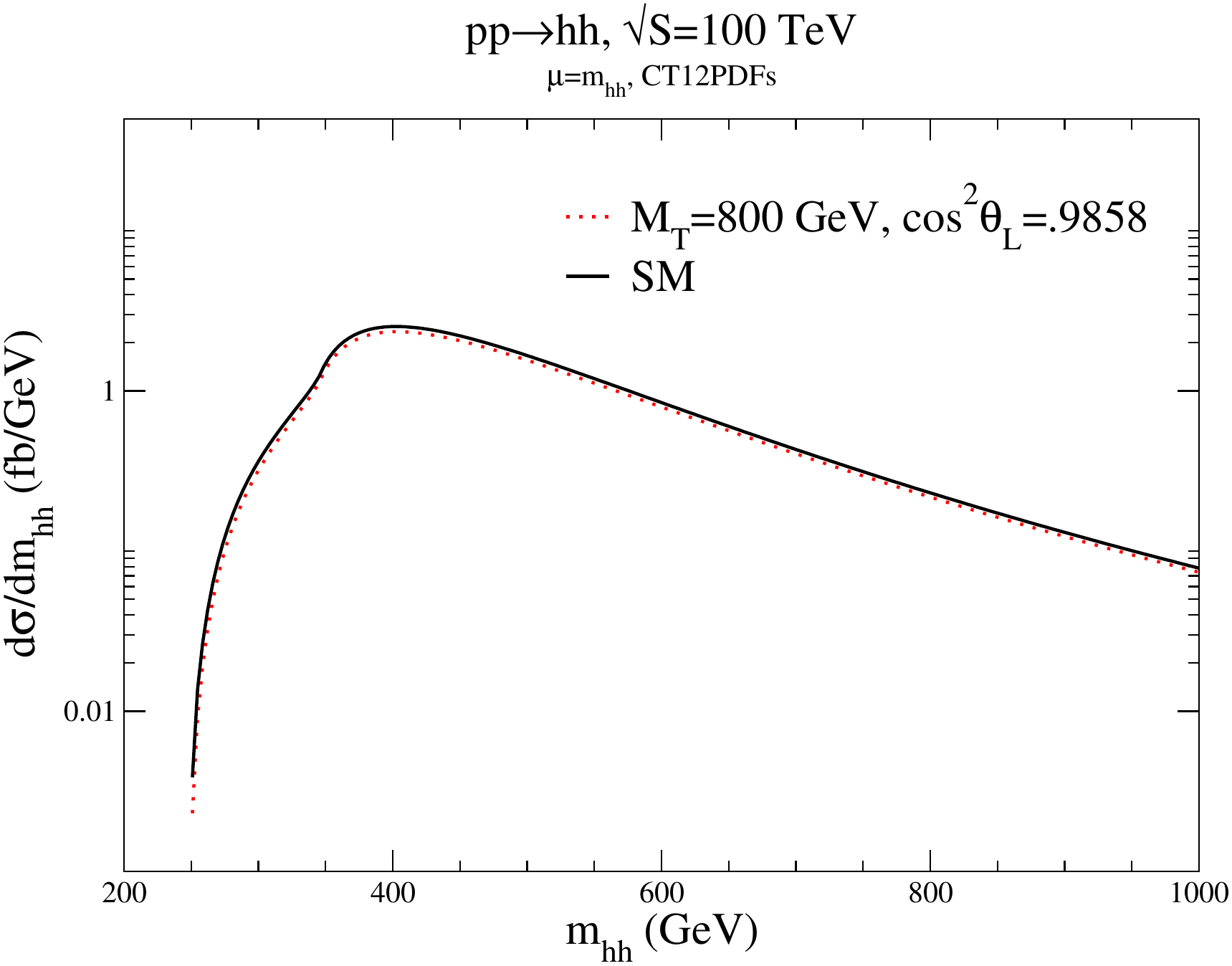}
\par\end{centering}
\caption{\em Invariant mass distribution in the SM (solid) and in the top partner
model with $c_L^2=.9858$ and $M_T=800$  GeV.}
  \label{fig:toppart_dist}
\end{figure}

In the $m_t, M_T\rightarrow\infty$ limit, single
Higgs production in the top partner model is virtually identical to that in the 
SM \cite{Azatov:2011qy,Dawson:2012di,Low:2009di},\footnote{This is  simply a statement of the  decoupling limit.}
\begin{equation}
\sigma_{1h}^{top~ partner}=\sigma_{1h}^{SM}+{\cal O}\biggl({m_h^2\over m_t^2}, {m_h^2\over M_T^2}\biggr)\, .
\end{equation}
 In Fig. \ref{fig:toppart_sum}, we compare the rate for $1h$ and $2h$ production in the singlet top
 partner model with $M_T=800$ GeV, as a function of the mixing angle, $c_L$.  For values of $c_L$
allowed by precision EW measurements, the $1h$ rate can be seen to be indistinguishable
from that of the SM.  On the other hand, $2h$ production receives contributions from the mixed $tTh$ couplings 
of Eq. \ref{higgscoups} and in general can be quite different from the SM prediction, as shown in
Fig. \ref{fig:toppart_sum}. Even imposing the restrictions from precision EW data, $2h$ production can be
reduced by up to $20\%$ from the SM prediction \cite{Chatrchyan:2013uxa,ATLAS:2012qe}, although the rate cannot be increased in this class of model.
The relative reduction of the $2h$ rate is roughly the same at $\sqrt{S}=13$ TeV and $100$ TeV.
The invariant mass spectrum are shown in Fig. \ref{fig:toppart_dist} for the top partner model.
  Because the EW precision constraints require that the mixing angle be very small, the
$m_{hh}$ distribution in indistinguishable from that of the SM.  It would be interesting to investigate slightly less simple models by including 
$t{\overline t}hh$ and $T{\overline T}hh$ couplings, which arise in models where the Higgs arise as a pseudo-Nambu-Goldstone boson \cite{Cheng:2005as,Low:2009di}.


\subsection{Scalar  Top Partners}
\begin{figure}
\includegraphics[scale=0.7]{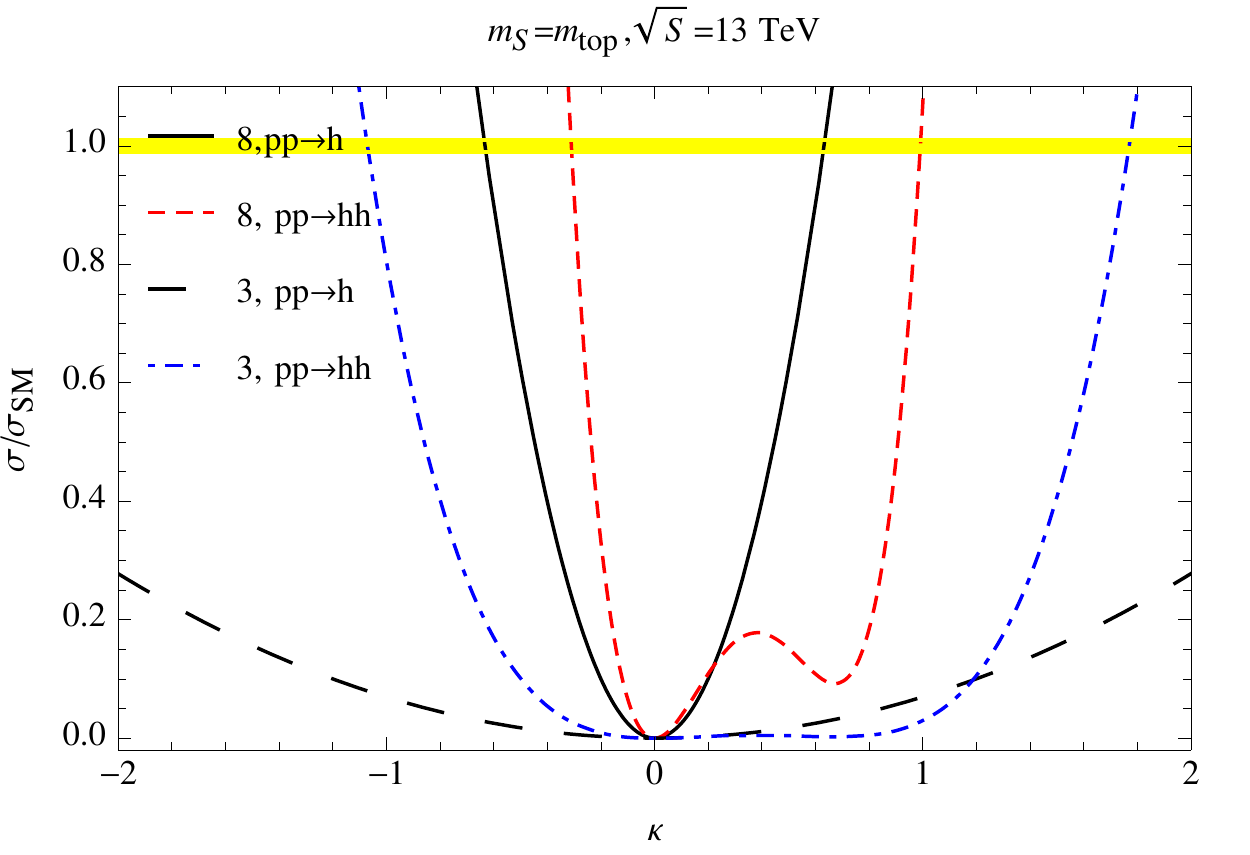}
\caption{\em Comparison of $1h$ (dashed black) and $2h$ production (blue dot-dash) to the SM rate, when the SM top quark is replaced by a color
triplet scalar with mass, $m_s=173 $ GeV.  The solid black (red dashed) curves correspond to the ratios  to the SM predictions for $1h$ and $2h$ with a color
octet 
scalar replacing the top quark. 
}
  \label{fig:scal1}
\end{figure}

\begin{figure}
\includegraphics[scale=0.7]{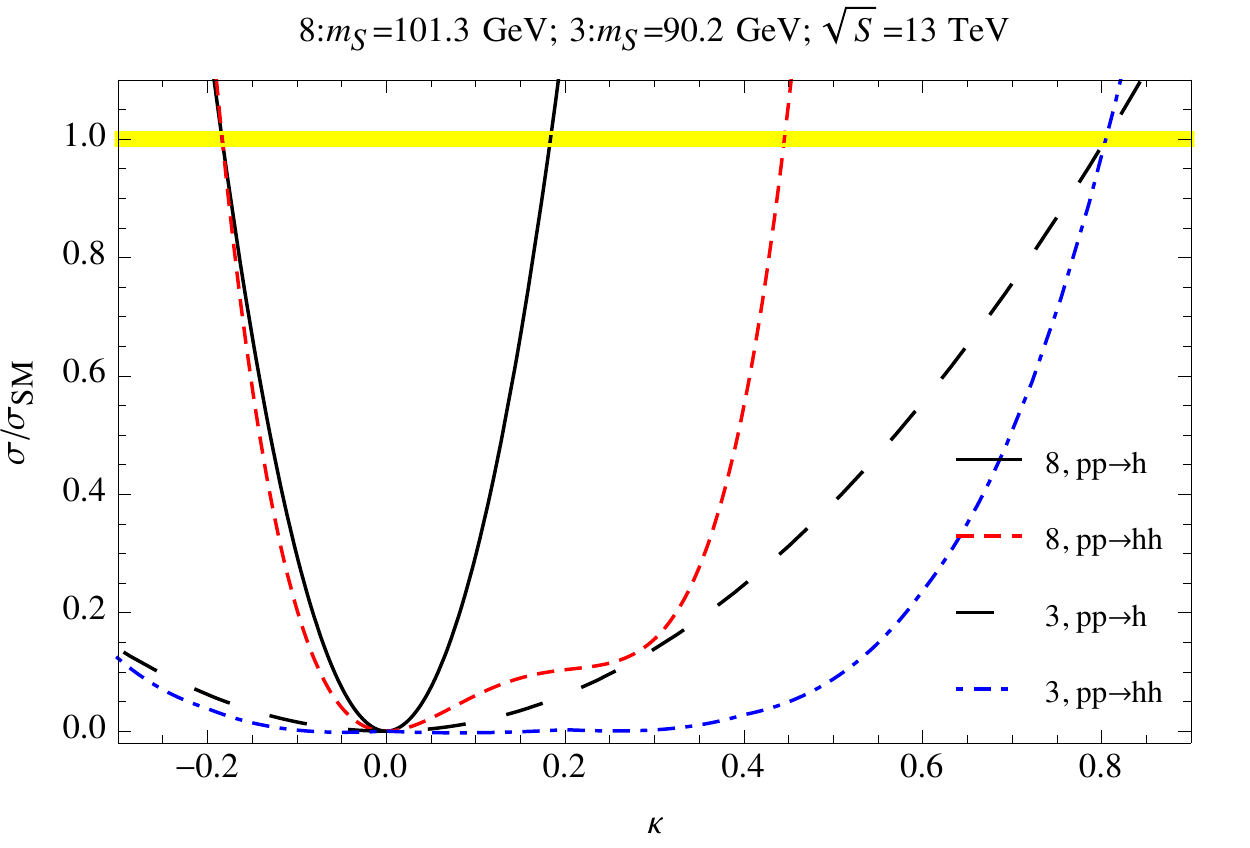}
\caption{\em Ratio of $1h$ production to the SM prediction when the top quark is replaced by a color triplet scalar of mass $m_s=90.2$ GeV (black dash) and by
a color octet scalar of mass, $m_s=101.3 $ GeV (black solid), compared with  the ratio of  $2h$ production to the SM prediction when the top quark is
 replaced by a color triplet scalar of mass $m_s=90.2$ GeV (blue dot- dash) and by
a color octet scalar of mass, $m_s=101.3 $ GeV (red dash).
}
  \label{fig:scal2}
\end{figure}

\begin{figure}
\includegraphics[scale=0.7]{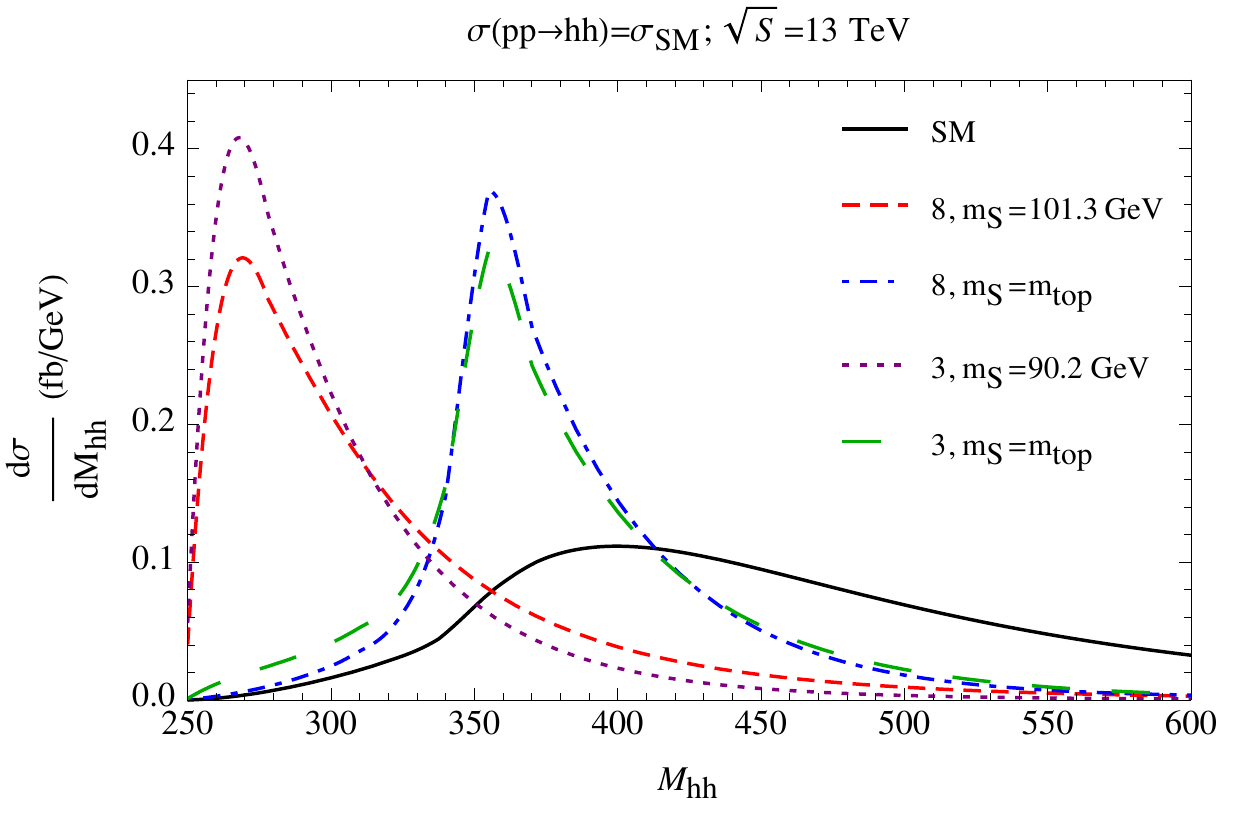}
\caption{\em Distributions for $2h$ production when the parameters are tuned to give the SM total cross sections for $1h$ and $2h$ production.
}
  \label{fig:scal3}
\end{figure}

In this section, we compare the results of the previous section where the loop particles are fermions with $m_{hh}$ distributions
 where instead of
a fermion, there is a colored scalar in the loops. 
We begin by replacing the SM top quark in the $gg\rightarrow h$ triangle diagram and in the $gg\rightarrow hh$ triangle and box diagrams
with a color triplet scalar of the same mass, $m_s=m_t=173$~GeV.  Fig. \ref{fig:scal1} shows the ratio of the total cross
sections for both $1h$ and $2h$ production, normalized to the lowest order SM predictions, in this scenario.  In the case of a color triplet scalar of 
mass $m_s=173$ GeV in the loops, we see that, in order to reproduce the SM rate for $1h$ production (the black dashed line), $\kappa$ need to be quite large, $\kappa \alt 2$.  If $\kappa$ is tuned to obtain $\sigma/\sigma_{SM}=1$ for $gg\rightarrow h$,
then a color octet intermediate particle  replacing the top quark with positive $\kappa $ (the solid black line) would predict a highly suppressed rate
for $2h$ production (the red dashed line).  Alternatively, we can tune both $\kappa$ and the scalar mass such that both $1h$ and $2h$ production have
the SM rates, as shown in Fig. \ref{fig:scal2}. Although the total rates are identical to the SM predictions, the kinematic 
distributions from color octet and triplet intermediate states are quite
different than those from the SM top, as plotted in Fig. \ref{fig:scal3}.   The scalar needs to be quite light to reproduce the SM rates, and the distribution is
peaked at much lower $m_{hh}$ than the SM prediction.

\begin{figure}
\includegraphics[scale=0.7]{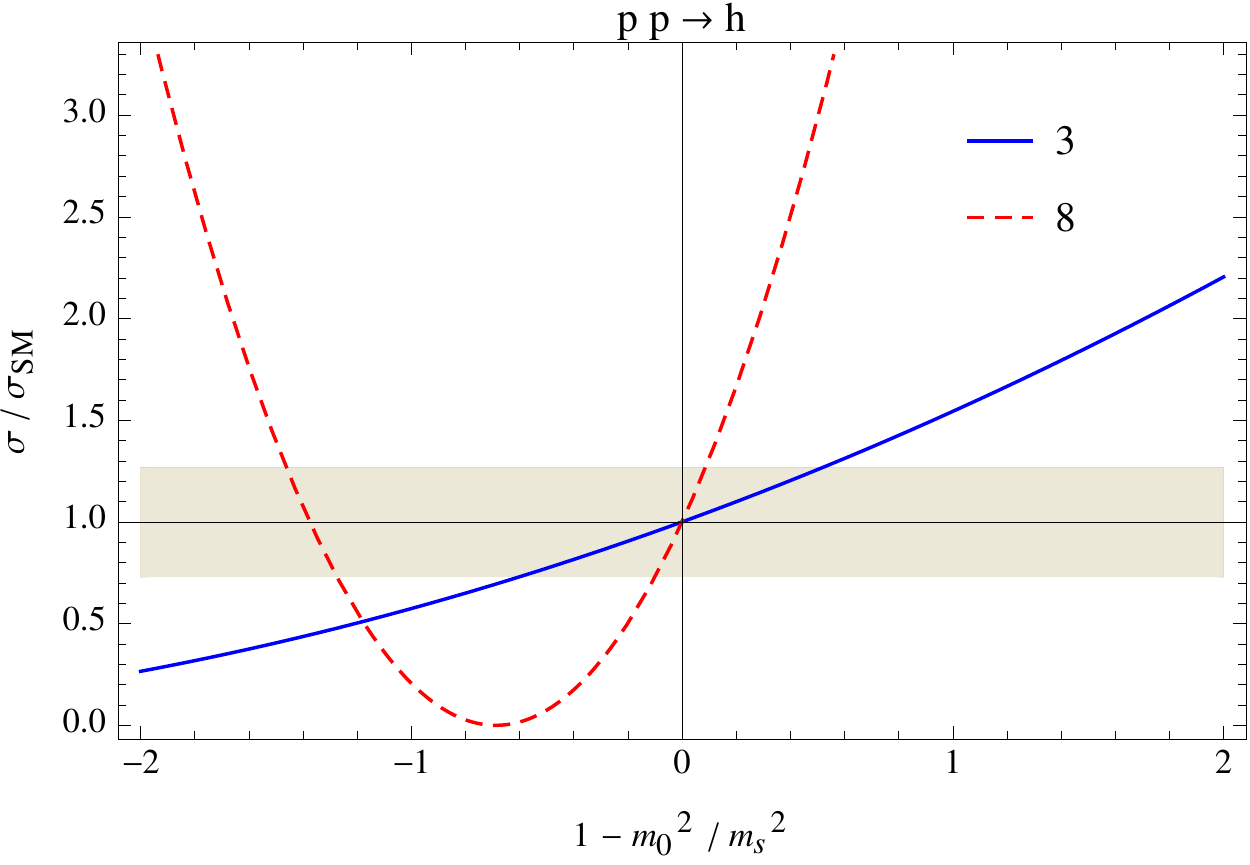}
\caption{\em Ratio of $1h$ production to the SM prediction with a colored triplet (blue solid) or octet (red dashed) scalar in addition to the SM top quark. The shaded band is the $2\sigma$ limit on the rate obtained from an average of current LHC results.
}
  \label{fig:gghscalar}
\end{figure}

We also consider Higgs production in the presence of the SM top quark and a colored scalar\footnote{In supersymmetry there are two colored scalars, the top squarks, mediating in the loop. Such a possibility is beyond the scope of current work and will be pursued elsewhere \cite{myturf}.}. Assuming the top Yukawa is SM-like, an additional scalar receiving all of its mass from electroweak symmetry breaking would give an unacceptably large contribution to the $1h$  production cross section, regardless of its mass and $SU(3)$ representation. This immediately follows from Eq. \ref{eq:sigmalo}:  A heavy color triplet scalar with $\kappa = 2 m_s^2 / v^2$ changes the $1h$ production rate by 54\%. Lighter scalars and scalars in other color representations result in even larger deviations. Fig. \ref{fig:gghscalar} shows the effects of color triplet and octet scalars on $1h$ production in the large mass limit, as functions of the proportion of the scalar mass coming from the Higgs field. Heavy scalars receiving all their masses from the Higgs have $m_0 = 0$, and are not compatible with a simple average of current ATLAS \cite{atlasfits} and CMS \cite{Khachatryan:2014jba} limits on the $1h$ production rate from gluon fusion, which is drawn as a shaded band in Fig. \ref{fig:gghscalar}.

\begin{figure}[t]
\begin{centering}
\includegraphics[scale=0.6]{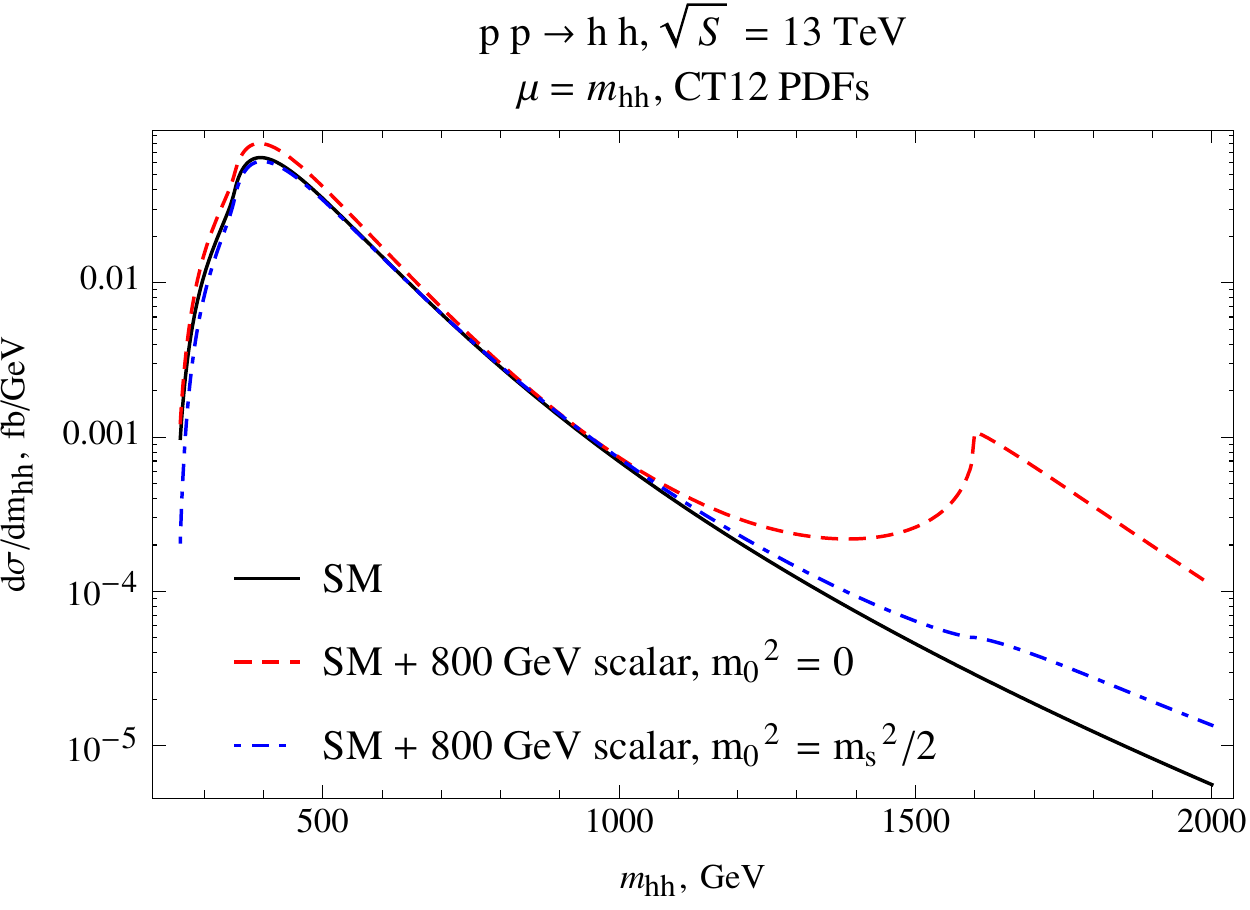}
\includegraphics[scale=0.6]{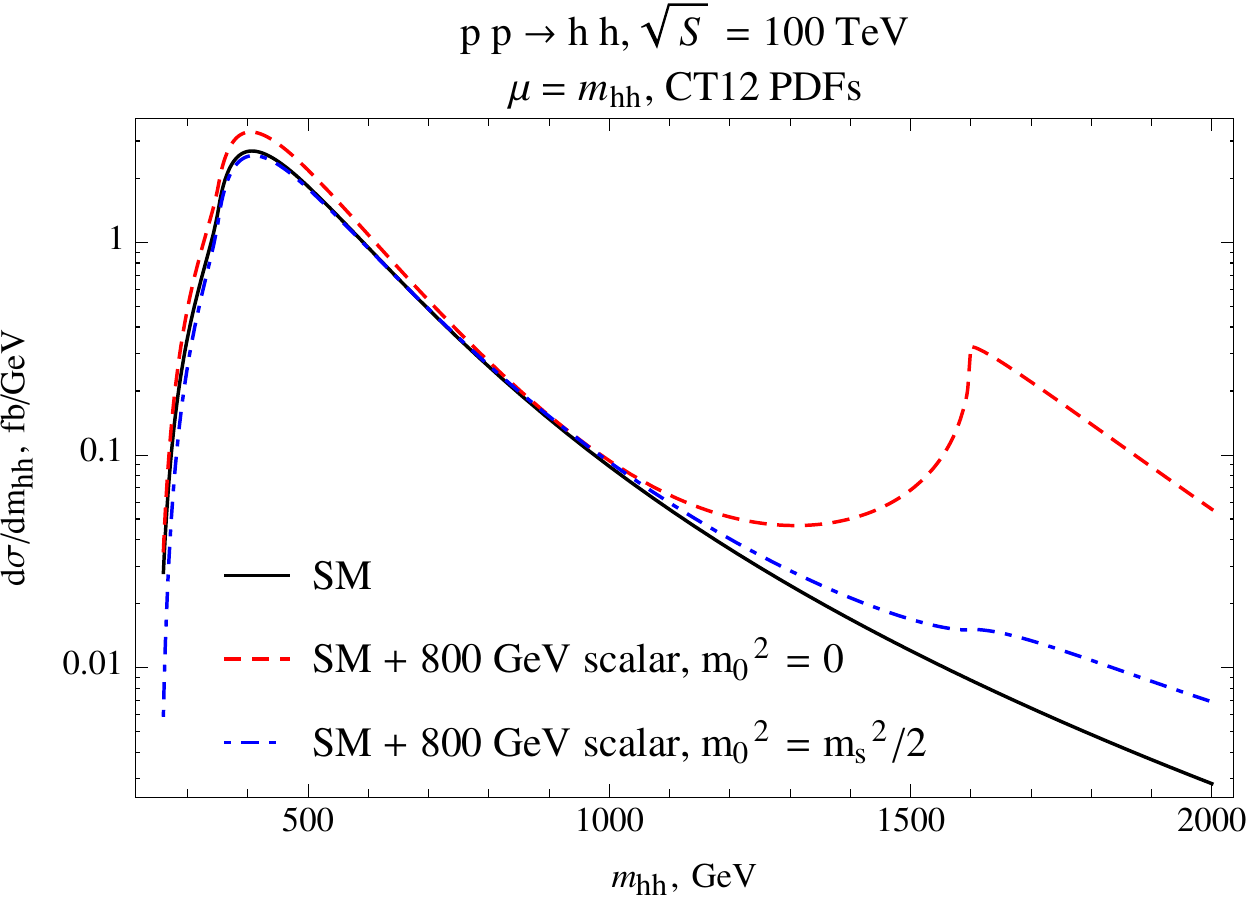}
\par\end{centering}
\caption{\em Invariant mass distribution in $2h$ production with the SM top quark in addition to an 800 GeV color triplet scalar that gets all (red dashed) or half (blue dot-dashed) of its mass from the Higgs. The SM (black solid) is shown for comparison.}
  \label{fig:topsc_dist}
\end{figure}

The limits imposed from $1h$ production constrains the sensitivity of $2h$ measurements to reveal new scalars, especially those with masses close to the weak scale. This is because a light scalar, in addition to the SM top, modifies the $1h$ rate significantly, unless its coupling to the Higgs is small, which at the same time diminishes its impact in $2h$ kinematic distributions. However, heavy scalars will decouple quickly in the $1h$ rate and may show up in the high $m_{hh}$ tail of the $2h$ distribution.  The invariant mass distributions for $2h$ production are shown in Fig. \ref{fig:topsc_dist} assuming a SM-like top quark and an additional 800 GeV color triplet scalar. If the scalar receives half of its mass squared from electroweak symmetry breaking, $m_0^2 = m_s^2 / 2$, the $1h$ rate is in roughly $2 \sigma$ tension with the current measurement, and the $2h$ distribution deviates from the SM expectation starting at $2m_s$, roughly speaking. For comparison, if the entire mass of the scalar was due to the Higgs, the feature at $m_{hh} = 2 m_s$ would be quite significant.

\section{Conclusion}
The observation of double Higgs production will be an important milestone in understanding the nature of electroweak symmetry 
breaking.  In the past the focus of this channel has been on extracting the Higgs tri-linear self-coupling. In this work we showed that the goal can be much broader and  encompass understanding the nature of the UV physics giving rise to Higgs coupling to two gluons, which is otherwise difficult to probe in single Higgs production.

  In Section \ref{analytic}, we examined the differences in the threshold behavior of double Higgs production
resulting from intermediate scalar and fermion loops and in Section \ref{example}, we demonstrated that even if the parameters in a model
with colored scalars are tuned to reproduce the SM rates for single and double Higgs production, the resulting invariant mass distributions
can be significantly different from the SM.  These distributions are also very sensitive to whether an additional scalar gets all of its mass from electroweak
symmetry breaking.  While Higgs plus jet production is also sensitive to the spin of loop particles and has a greater cross section, it does not enjoy the same large amplitude cancellation present in double Higgs production \cite{Bonciani:2007ex}. We also investigated the effects of anomalous top and bottom Yukawa couplings and showed that the resulting
changes in single and double Higgs production relative to the SM rates are roughly the same at $\sqrt{S}=13$ and $100$ TeV.

Clearly, it will be an important experimental question on how to extract the wealth of information contained in the double Higgs production. Our work provides strong motivation to pursue this  issue experimentally.

\begin{acknowledgments}
We thank A. Martin for useful discussions of Ref. \cite{Kribs:2012kz}. A.I.  acknowledges R. Boughezal and F. Petriello for discussions.
The work of S.D. is supported by the U.S.~Department of Energy under grant
No.~DE-AC02-98CH10886. Work at ANL is supported by the U.S.~Department of
Energy under grant No.~DE-AC02-06CH11357. A.I. is supported in part by the U.S.~Department of
Energy under grant  DE-FG02-12ER41811. I.L. is supported in part by the U.S.~Department of Energy under grant No. DE-SC0010143.
\end{acknowledgments}

\appendix
\numberwithin{equation}{section}
\section{$2h$  Amplitudes from scalars}
\label{appendix:2hamp}

Here we collect the contributions from virtual scalars computed in Ref.~\cite{Asakawa:2010xj,Kribs:2012kz}.
\begin{eqnarray}
F^{(s)}_\triangle&=&- 2 \frac{m_s^2 }{{\hat {s}}}  \biggl[1+2m_s^2 C_0({\hat {s}},m_s^2)\biggr] \\
F^{(s)}_{\Box}&=& - 4 {m_s^4\over {\hat{s}}}
\biggl[m_s^2\biggl(D_0({\hat{s}},{\hat{t}},{\hat{u}},m_s^2)+
D_0({\hat{s}},{\hat{u}},{\hat{t}},m_s^2)+
D_0({\hat {t}},{\hat{s}},{\hat{u}},m_s^2)\biggr)
\nonumber \\
&&+{m_h^2-{\hat {u}}\over {\hat {s}}}C_0({\hat {u}},m_s^2)
     +{m_h^2-{\hat {t}}\over {\hat {s}}}C_0({\hat{t}},m_s^2)
     +{p_T^2\over 2}D_0({\hat{t}},{\hat{s}},{\hat{u}},m_s^2)\biggr]\, , \\
G^{(s)}_{\Box}&=&-4 {m_s^4\over {\hat{s}}}
\biggl\{ -C_0^\prime({\hat{s}},m_h^2,m_s^2) \nonumber \\
&&+m_s^2\biggl(D_0({\hat{s}},{\hat{t}},{\hat{u}},m_s^2)
+D_0({\hat{s}},{\hat{u}},{\hat{t}},m_s^2)+
D_0({\hat{t}},{\hat{s}},{\hat{u}},m_s^2)\biggr)
\nonumber \\
&&
+{1\over 2({\hat{u}}{\hat{t}}-m_h^4)}
\biggl[-2{\hat{u}}({\hat{u}}-m_h^2)C_0({\hat{u}},m_s^2)-
2 {\hat{t}} ({\hat{t}}-m_h^2)C_0({\hat{t}},m_s^2)+
{\hat{s}}({\hat{s}}-2m_h^2)C_0({\hat{s}},m_s^2)
\nonumber \\
&&
+{\hat{s}}{\hat{u}}^2D_0({\hat{s}},{\hat{u}},{\hat{t}},m_s^2)+
{\hat{s}}{\hat{ t}}^2D_0({\hat{s}},{\hat{t}},{\hat{u}},m_s^2)+
{\hat{s}}({\hat{s}}-4m_h^2)C_0^\prime({\hat{s}},m_h^2,m_s^2)\biggr]\biggr\}\ ,
\end{eqnarray}
where $p_T^2=({\hat{u}}{\hat {t}}-m_h^4)/{\hat{s}}$.  In the above we have
\begin{eqnarray}
C_0({\hat{s}},m_s^2)&=&C(0,0,{\hat{s}},m_s^2)
\nonumber \\
&=&\int {d^nk\over i \pi^2} 
{1\over [k^2-m_s^2]}
{1\over [ (k+p_1)^2-m_s^2]}
{1\over [(k+p_1+p_2)^2-m_s^2]}\nonumber \\
C_0^\prime({\hat {s}},m_h^2,m_s^2)&=&C(m_h^2,m_h^2,{\hat {s}},m_s^2)
\nonumber \\
&=&
\int {d^nk\over i \pi^2} 
{1\over [k^2-m_s^2]}
{1\over [(k+k_1)^2-m_s^2]}
{1\over [(k+k_1+k_2)^2-m_s^2]}
\nonumber \\
D_0({\hat{s}},{\hat{t}},{\hat{u}},m_s^2)&=& D(0,0,m_h^2,m_h^2,{\hat{s}},{\hat{u}},m_s^2)\nonumber \\
&=&
\int {d^nk\over i \pi^2} 
{1\over [k^2-m_s^2]}
{1\over [(k+p_1)^2-m_s^2]}
{1\over [(k+p_1+p_2)^2-m_s^2]}
{1\over [(k+p_1+p_2-k_1)^2-m_s^2]}
\nonumber \\
D_0({\hat {s}},{\hat {u}},{\hat {t}},m_s^2)&=& D(0,0,m_h^2,m_h^2,{\hat{s}},{\hat{t}},m_s^2)\nonumber \\
&=&
\int {d^nk\over i \pi^2} 
{1\over [k^2-m_s^2]}
{1\over [(k+p_1)^2-m_s^2]}
{1\over [(k+p_1+p_2)^2-m_s^2]}
{1\over [(k+p_1+p_2-k_2)^2-m_s^2]}
\nonumber \\
D_0({\hat{t}},{\hat{s}},{\hat{u}},m_s^2)&=& D(0,m_h^2,0,m_h^2,{\hat{t}},{\hat{u}},m_s^2)\nonumber\\
&=&
\int {d^nk\over i \pi^2} 
{1\over [k^2-m_s^2]}
{1\over [(k+p_1)^2-m_s^2]}
{1\over [(k+p_1-k_1)^2-m_s^2]}
{1\over [(k+p_1+p_2-k_1)^2-m_s^2]}
\nonumber \\
\end{eqnarray}
and $(p_1+p_2)^2=(k_1+k_2)^2={\hat s}$, $(p_1-k_1)^2={\hat{t}}$, 
$(p_1-k_2)^2={\hat{u}}$, and 
$k_1^2=k_2^2=m_h^2$.

\label{appa}
\section{Closed Form Amplitudes for $gg \to hh$}
Here, we calculate the imaginary part of the amplitude for $2h$ production from scalar loops at threshold, using cut techniques. This is a new result analogous to the recent computation for fermions \cite{Li:2013rra}. By using the dispersion relation, we may recover the full $2h$ amplitude in closed form, which is then analyzed in Section \ref{analytic}.

We start with the general amplitude in Eq. \ref{eq:ampgghh}. For simplicity, we assume a single scalar that gets all of its mass from the Higgs, so that $m_0 = 0$ and $\kappa = \kappa_0 = 2 m_s^2 / v^2$ in Eq. \ref{eq:lsc}. Our results can easily be generalized to scalars with arbitrary couplings and masses, and we emphasize that they do not assume a heavy loop particle. At threshold, $\hat{s} = 4 m_h^2$, only the spin 0 piece contributes \cite{Dawson:2012mk}. The imaginary parts of the corresponding form factors $F_\triangle^{(s)}, F_\Box^{(s)}$ can be obtained from cutting all possible $g g \to h h$ diagrams, and sending all cut propagators on shell. We will compute $\Im F_\triangle^{(s)}$ and $\Im F_\Box^{(s)}$ separately at threshold.

\begin{figure}[t]
\begin{centering}
\includegraphics[height=0.5in]{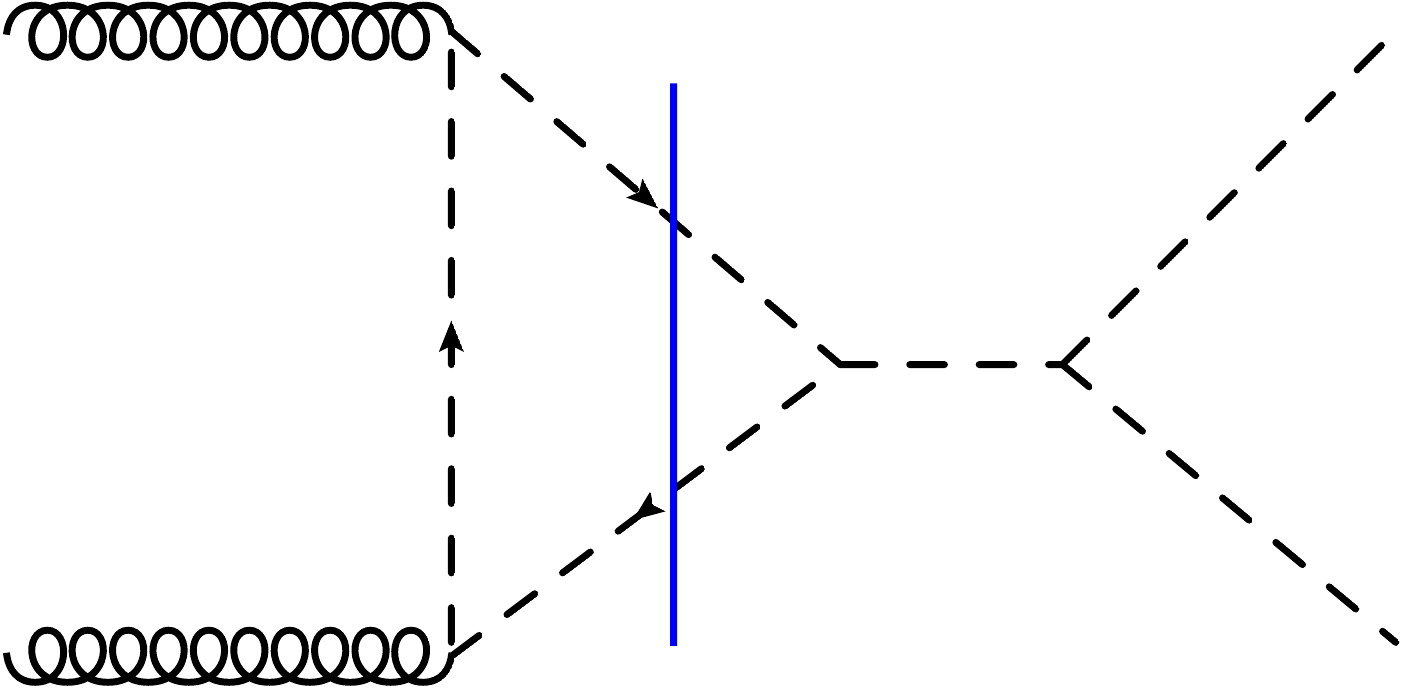} \hspace{0.25in}
\includegraphics[height=0.5in]{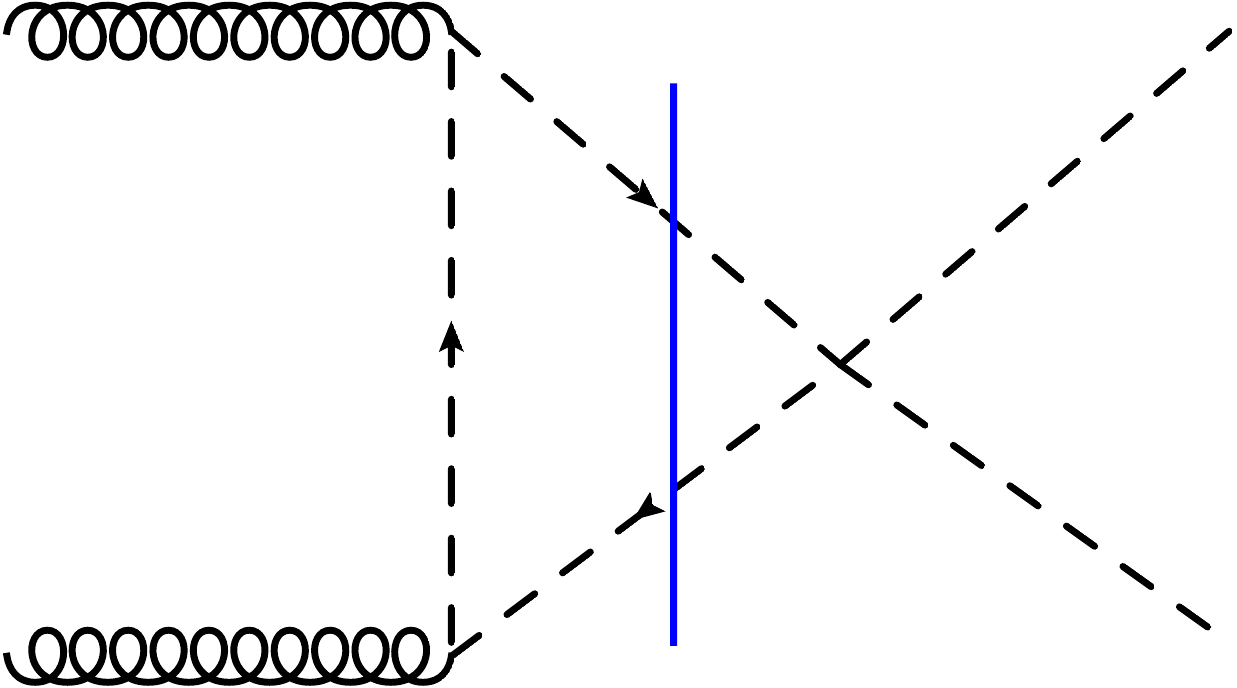} \hspace{0.25in}
\includegraphics[height=0.5in]{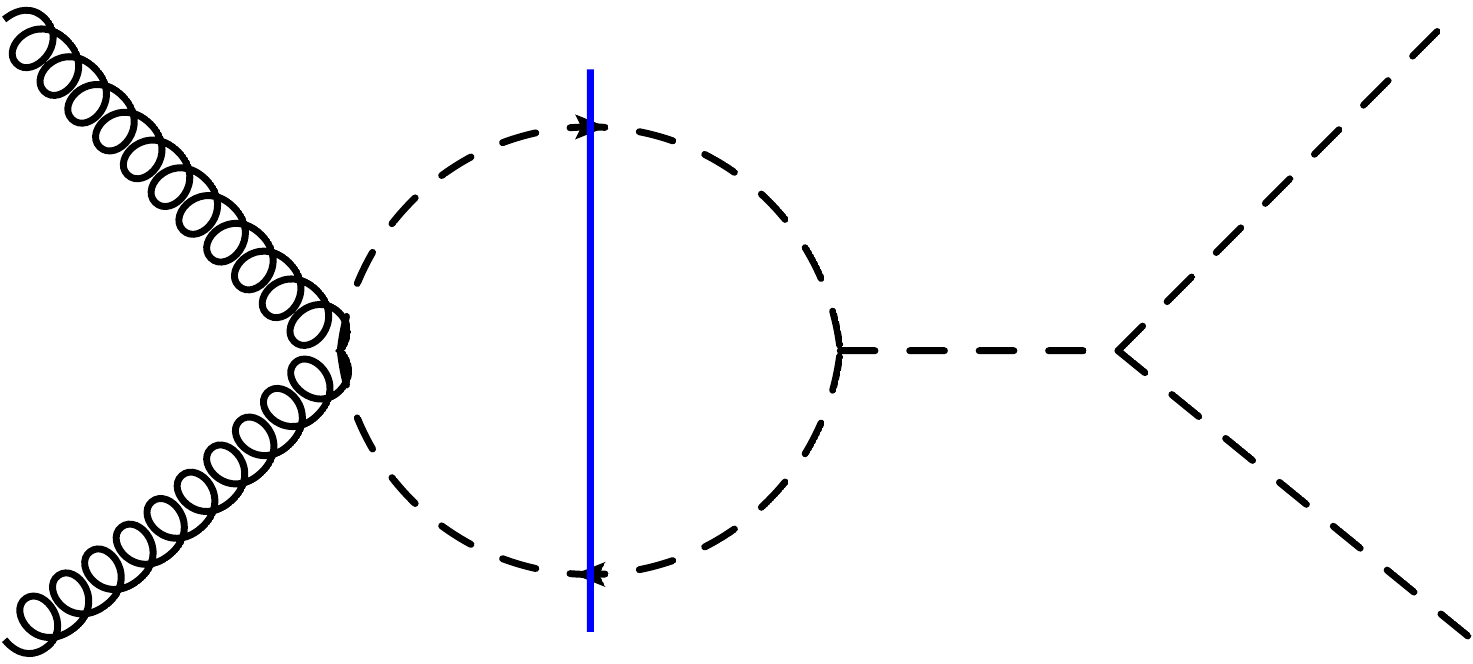} \hspace{0.25in}
\includegraphics[height=0.5in]{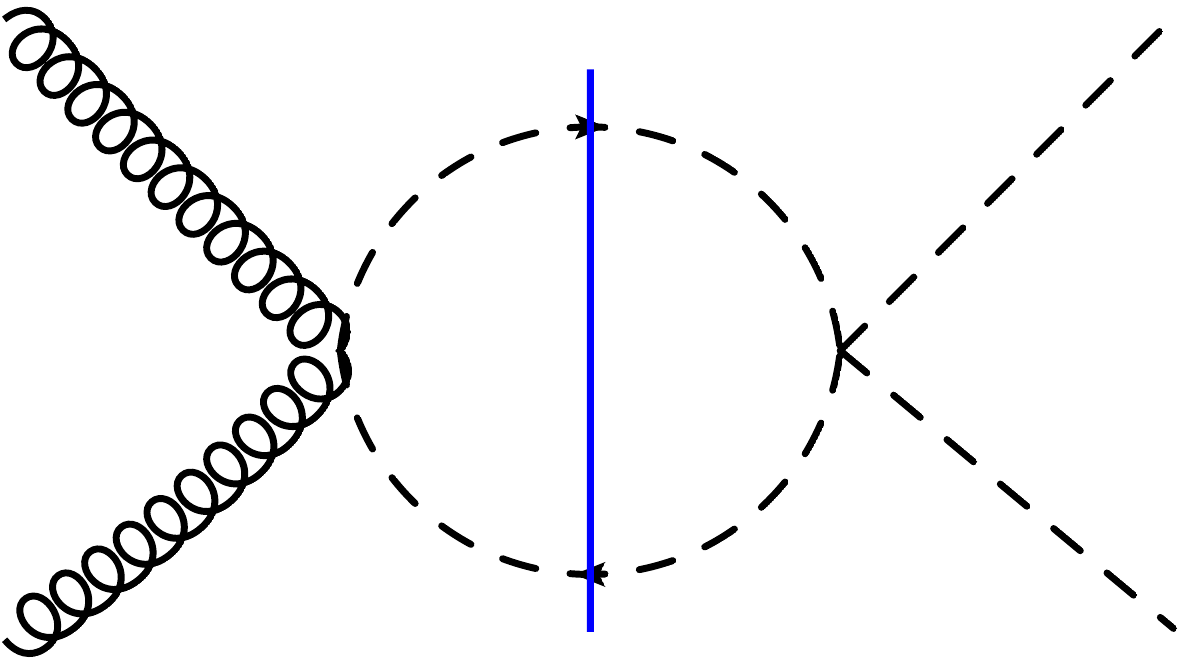}
\par\end{centering}
\caption{\em Diagrams contributing to $F_\triangle^{(s)}$. Bubble diagrams with quartic scalar-gluon vertices are included. The cuts shown are used to calculate $\Im F_\triangle^{(s)}\mid_{th}^{\kappa = \kappa_0}$.
}
  \label{fig:tricuts}
\end{figure}

Fig. \ref{fig:tricuts} shows the diagrams that are responsible for the $F_\triangle^{(s)}$ form factor. In addition to the triangle diagram which may be obtained by replacing the top quark in the SM double Higgs triangle diagram with a scalar, we have the additional $s^* s h^2$ coupling. We also include bubble diagrams with quartic scalar-gluon couplings with the above diagrams in the triangle form factors, as they are related through gauge invariance. Now, the imaginary part of the double Higgs amplitude receives contributions from the cuts shown in the diagrams of Fig. \ref{fig:tricuts}, through
\be
\Im {\cal M} \supset \int d \Pi_2 {\cal M}_L {\cal M}_R
\ee
where ${\cal M}$, ${\cal M}_L$ and ${\cal M}_R$ refer to the full double Higgs amplitude and the left/right halves of a cut diagram. The integral $\int d\Pi_2$ is over the phase space of the cut propagators. Each cut diagram in Fig. \ref{fig:tricuts} contributes separately to $\Im F_\triangle^{(s)}$. The halves of the cut diagrams are simply tree-level amplitudes for $g g \to s^* s$ and $s^* s \to h h$. Furthermore, since we are interested in the amplitude cancellation at threshold, we may project out the spin 0 piece of the amplitude to get $\Im F_\triangle^{(s)}\mid_{th}$. The kinematics of Appendix \ref{appa} simplify considerably for $\hat{s} = 4 m_h^2$, and we are left with
\be
\Im F_\triangle^{(s)}\mid_{th} = \frac{\pi}{32} T(s) \theta \left( \frac{4}{\tau_s} - 1 \right) \tau_s^{2} \log \frac{1 + \sqrt{1 - \tau_s/4}}{1 - \sqrt{1 - \tau_s/4}}
\label{eq:tricuts}
\ee

\begin{figure}[t]
\begin{centering}
\includegraphics[height=0.5in]{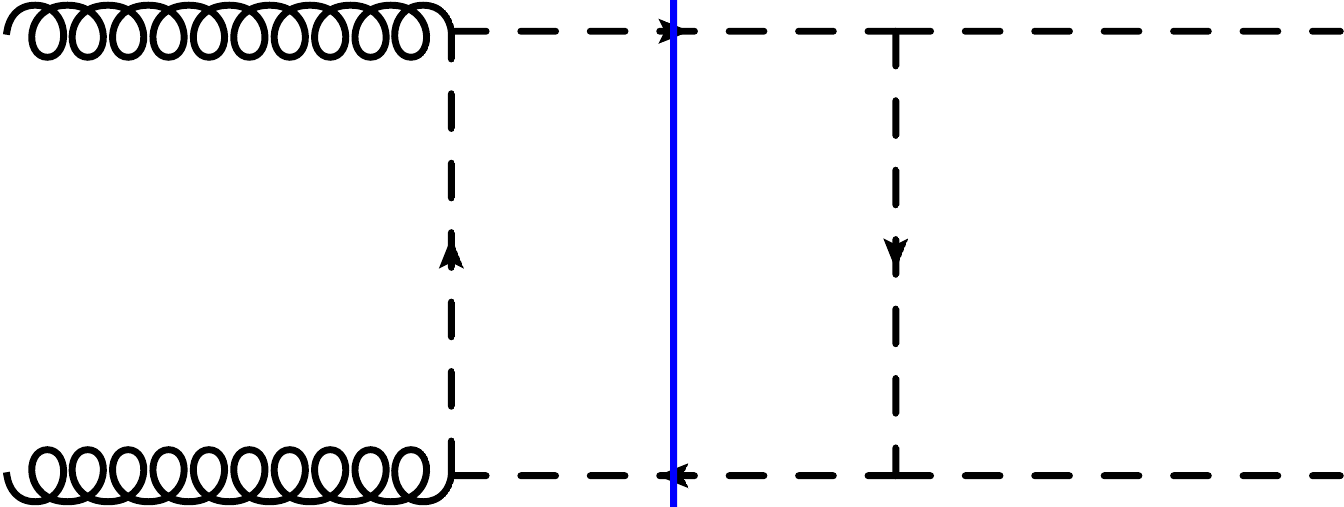} \hspace{0.25in}
\includegraphics[height=0.5in]{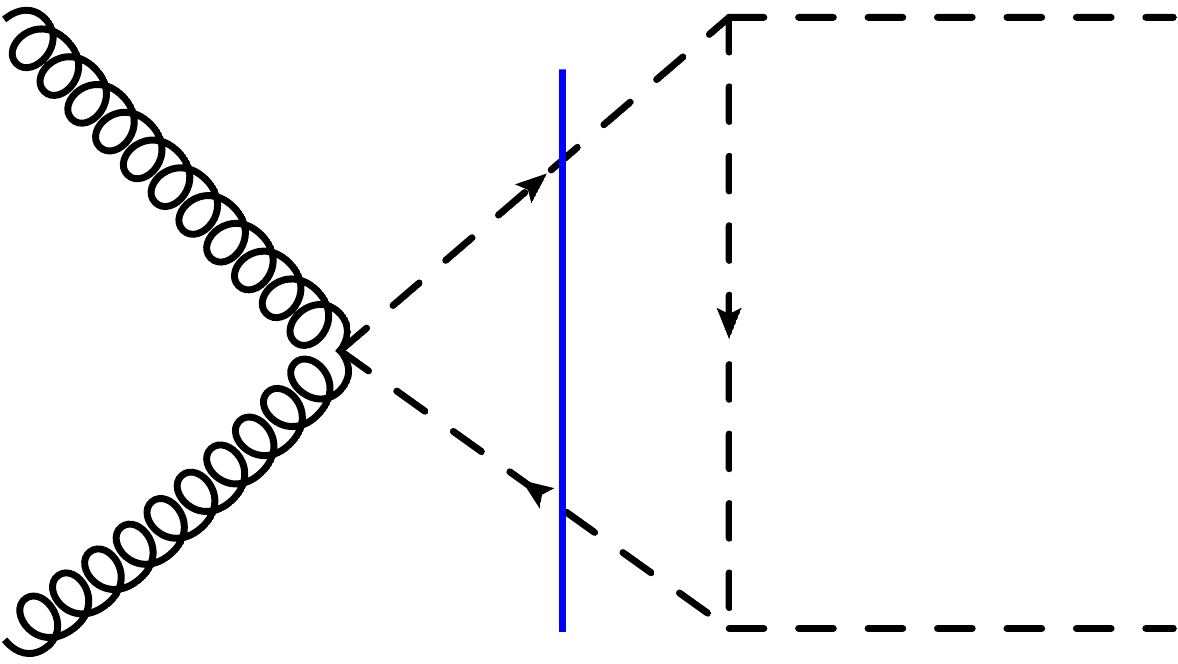} \hspace{0.25in}
\includegraphics[height=0.5in]{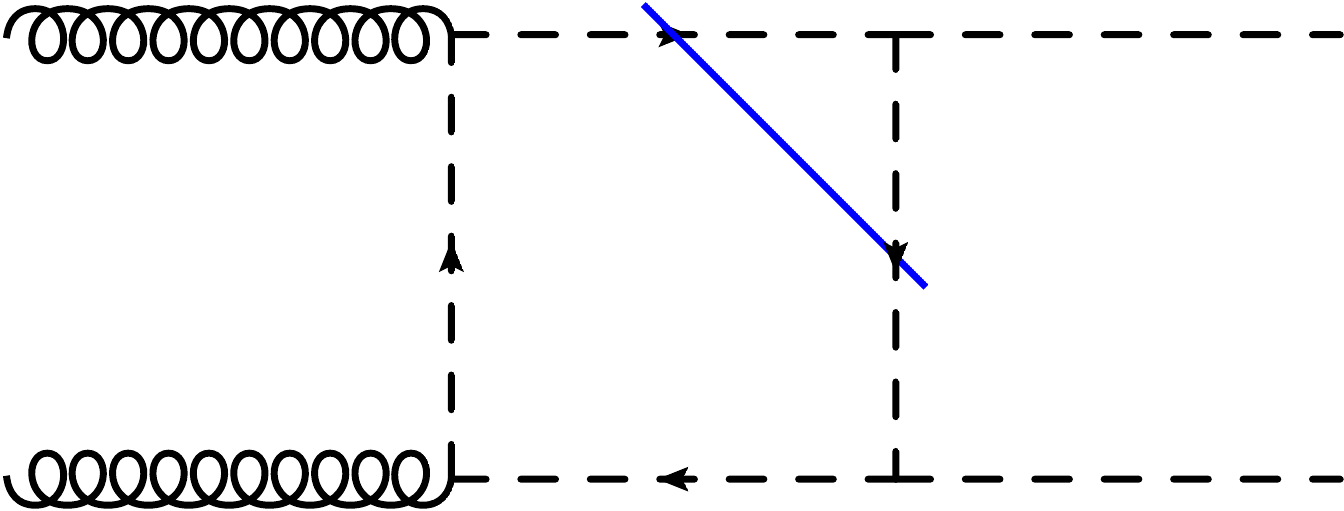} \\[0.25in]
\includegraphics[height=0.5in]{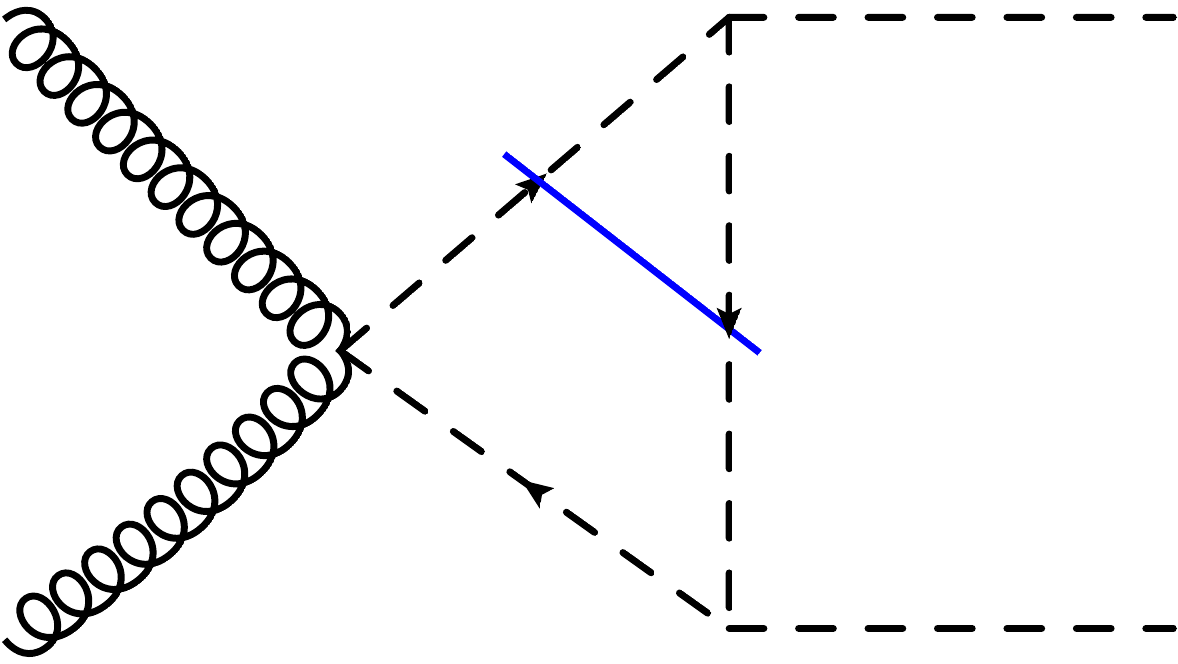} \hspace{0.25in}
\includegraphics[height=0.5in]{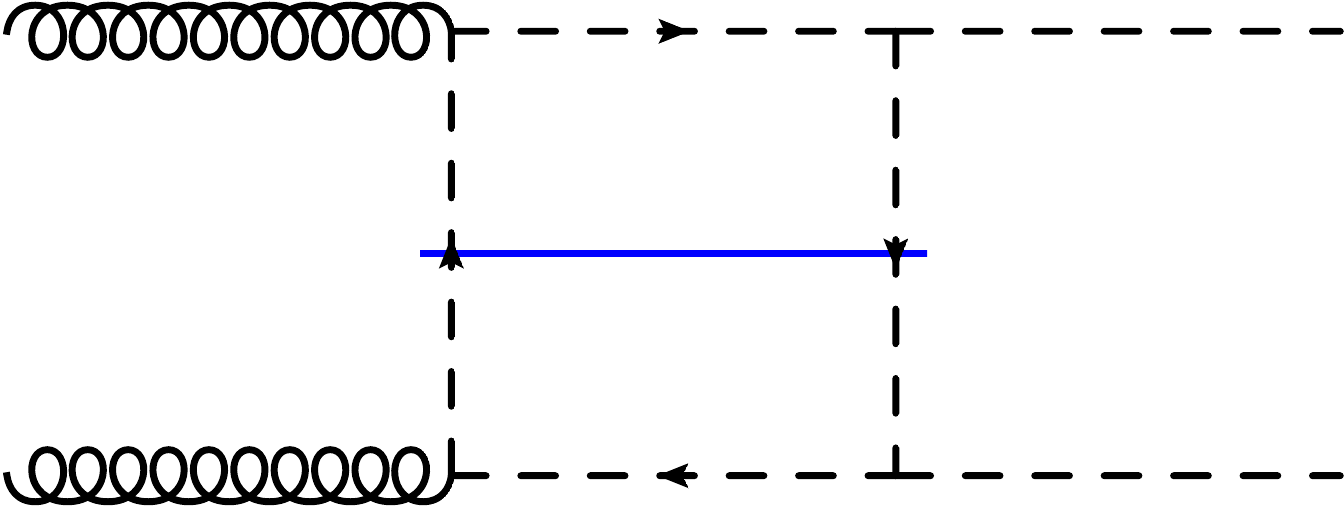}
\par\end{centering}
\caption{\em Diagrams contributing to $F_\Box^{(s)}$ and $G_\Box^{(s)}$. Triangle diagrams with quartic scalar-gluon vertices are included. The cuts shown are used to calculate $\Im F_\Box^{(s)}\mid_{th}^{\kappa = \kappa_0}$.
}
  \label{fig:boxcuts}
\end{figure}

Cut diagrams that contribute to $F_\Box^{(s)}$ and $G_\Box^{(s)}$ are shown in Fig. \ref{fig:boxcuts}. The first two diagrams of Fig. \ref{fig:boxcuts} have identical cuts to the diagrams of Fig. \ref{fig:tricuts}, and the left sides are the same as in the earlier diagrams. At threshold with the cut propagators on shell, comparison of the right sides of these diagrams with those of Fig. \ref{fig:tricuts} immediately gives
\be
\Im F_\Box^{(s)}\mid_{th} \supset -\frac{\tau_s}{2} \Im F_\triangle^{(s)}\mid_{th}
\label{eq:boxcuts1}
\ee
This is the contribution of the top row of Fig. \ref{fig:boxcuts} to $\Im F_\Box^{(s)}\mid_{th}$.

The contributions of the cuts in the third and fourth diagrams of Fig. \ref{fig:boxcuts} to $\Im F_\Box^{(s)}$ at threshold may be computed from the tree-level amplitudes for $g g \to s^* s h$ and $s^* s \to h$. Note that the two adjacent propagators attaching to either external Higgs may be cut, each choice leading to an identical set of contributions to the imaginary amplitude. Only one such set of cuts is shown in these diagrams. Both sets of cuts together yield
\be
\Im F_\Box^{(s)}\mid_{th} \supset -\frac{\pi}{16} T(s) \theta \left( \frac{1}{\tau_s} - 1 \right) \tau_s^2 \left( \sqrt{1 - \tau_s} - \left( 1 + \frac{\tau_s}{2} \right) \log \frac{1 + \sqrt{1 - \tau_s}}{1 - \sqrt{1 - \tau_s}} \right)
\label{eq:boxcuts2}
\ee
Also, there is no contribution to the imaginary part of the $g g \to h h$ amplitude from cutting two adjacent propagators attaching to an external gluon, because the amplitude for $g \to s^* s$ is zero when the scalars are put on shell.

Finally, the last cut diagram of Fig. \ref{fig:boxcuts} gives a contribution to $\Im F_\Box^{(s)}$ that may be calculated at threshold from the $g h \to s^* s$ amplitude. We proceed as before, and find the final contribution
\be
\Im F_\Box^{(s)}\mid_{th} \supset \frac{\pi}{16} T(s) \theta \left( -\frac{1}{\tau_s} - 1 \right) \tau_s^2 \left( \sqrt{1 + \tau_s} - \left( 1 + \frac{\tau_s}{2} \right) \log \frac{1 + \sqrt{1 + \tau_s}}{1 - \sqrt{1 + \tau_s}} \right)
\label{eq:boxcuts3}
\ee

The sum of the right-hand sides of Eqs. \ref{eq:tricuts}, \ref{eq:boxcuts1}, \ref{eq:boxcuts2} and \ref{eq:boxcuts3} gives the full imaginary $2h$ amplitude at threshold. Now, we turn to the limits of the full amplitude as $\tau_s \to 0, \infty$. In the limit $\tau_s \to 0$, the amplitude vanishes since for a scalar that gets all its mass from the Higgs, $\kappa$ is proportional to $\tau_s$ through
\be
\kappa = \frac{m_h^2}{2 v^2} \tau_s
\ee
On the other hand, in the infinite scalar mass limit $\tau_s \to \infty$ we may apply the low-energy theorem. From the effective Lagrangian for the interaction between scalars and gluons \cite{Shifman:1979eb}, we know that the $2h$ amplitude goes as
\bea
A_{hh} &\propto& \left \langle hh \Bigg| \log \left( 1 + \frac{2h}{v} + \frac{h^2}{v^2} \right) \Bigg| 0 \right \rangle \nonumber \\
&=& \left \langle hh \Bigg| \frac{2h}{v} - \frac{h^2}{v^2} \Bigg| 0 \right \rangle
\eea
which vanishes for the SM Higgs self-coupling \cite{Li:2013rra}. Given the limiting behavior of the amplitude combined with full knowledge of its imaginary part, then, the dispersion relation gives the full amplitude in Eq. \ref{eq:scthr}.
\label{scalarcuts}

\newpage
\bibliography{hh}


\end{document}